\documentclass[fleqn,usenatbib]{mnras}

\usepackage{newtxtext,newtxmath}

\usepackage[T1]{fontenc}

\DeclareRobustCommand{\VAN}[3]{#2}
\let\VANthebibliography\thebibliography
\def\thebibliography{\DeclareRobustCommand{\VAN}[3]{##3}\VANthebibliography}

\usepackage{graphicx}	
\usepackage{amsmath}	
\usepackage{color}
\usepackage{comment}
\usepackage{threeparttable}
\usepackage{subfig}

\title[MRI turbulence at large magnetic Prandtl numbers]{MRI turbulence in accretion discs at large magnetic Prandtl numbers}

\author[L. E. Held and G. Mamatsashvili]{
Loren E. Held$^{1}$\thanks{E-mail: loren.held@aei.mpg.de (LEH)} and George Mamatsashvili$^{2,3}$
\\
$^{1}$Max Planck Institute for Gravitational Physics (Albert Einstein Institute), Am M{\"u}hlenberg 1, Potsdam 14476, Germany\\
$^{2}$Helmholtz-Zentrum Dresden-Rossendorf, Bautzner Landstra{\ss}e 400, Dresden 01328, Germany\\
$^{3}$E. Kharadze Georgian National Astrophysical Observatory, Abastumani 0301, Georgia}

\date{Accepted XXX. Received YYY; in original form ZZZ}

\pubyear{2021}

\begin{document}
\label{firstpage}
\pagerange{\pageref{firstpage}--\pageref{lastpage}}
\maketitle

\begin{abstract}
The effect of large magnetic Prandtl number $\text{Pm}$ (the ratio of viscosity to resistivity) on the turbulent transport and energetics of the magnetorotational instability (MRI) is poorly understood, despite the realization of this regime in astrophysical environments as disparate as discs from binary neutron star mergers, the inner regions of low mass X-ray binaries and active galactic nuclei, and the interiors of protoneutron stars. We investigate the MRI dynamo and associated turbulence in the regime $\text{Pm}>1$ by carrying out fully compressible, 3D MHD shearing box simulations using the finite-volume code \textsc{PLUTO}, focusing mostly on the case of Keplerian shear relevant to accretion discs. We find that when the magnetic Reynolds number is kept fixed, the turbulent transport (as parameterized by $\alpha$, the ratio of stress to thermal pressure) scales with the magnetic Prandtl number as $\alpha \sim \text{Pm}^{\delta}$, with $\delta \sim 0.5-0.7$ up to $\text{Pm} \sim 128$. However, this scaling weakens as the magnetic Reynolds number is increased. Importantly, compared to previous studies, we find a new effect at very large $\text{Pm}$ -- the turbulent energy and stress begin to plateau, no longer depending on ${\rm Pm}$. To understand these results we have carried out a detailed analysis of the turbulent dynamics in Fourier space, focusing on the effect of increasing $\text{Pm}$ on the transverse cascade -- a key non-linear process induced by the disc shear flow that is responsible for the sustenance of MRI turbulence. Finally, we find that $\alpha$-$\text{Pm}$ scaling is sensitive to the box vertical-to-radial aspect ratio, as well as to the background shear.
\end{abstract}

\begin{keywords}
accretion, accretion discs -- magnetohydrodynamics -- instabilities, turbulence
\end{keywords}



\section{Introduction}
\label{INTRO}
The transport of angular momentum via the magnetorotational instability (MRI) in accretion discs has been the subject of intense investigation for over three decades \citep{balbus1991powerful,hawley1995}. A key physical parameter controlling the efficiency of angular momentum transport in MRI turbulence is the magnetic Prandtl number $\text{Pm} \equiv \nu / \eta$, the ratio of kinematic viscosity $\nu$ to magnetic diffusivity $\eta$ \citep{balbus1998instability,lesur2007}. This parameter is expected to vary significantly between different types of accretion disc, and even within different regions of the same disc. For example, $\text{Pm} \ll 1$ in the outer regions of X-ray binaries and AGN discs, and in protoplanetary discs (beyond 1 AU), but $\text{Pm} \gg 1$ in the inner regions of X-ray binaries and AGN discs \citep{balbus2008}, and also in the very inner parts of protoplanetary discs \citep{lesur2021magnetohydrodynamics}. In particular, the regime $\text{Pm} \gg 1$ is thought to be relevant (at all radii) to the hyperaccreting neutrino-cooled discs that can form from binary neutron star (BNS) and black hole-neutron star (BHNS) mergers \citep{rossi2008}. Both types of merger have been discovered recently: the first binary neutron star merger (GW170817) was observed both in gravitational waves and via electromagnetic signatures \citep{abbott2017gw170817}, and the first two BHNS mergers (GW200105; GW200115) were detected in gravitational waves only \citep{abbott2021observation}. However, the dynamics of MRI turbulence (in the absence of an imposed mean magnetic field) have only been sparingly studied in numerical simulations at the large magnetic Prandtl numbers relevant to discs in these objects \citep{lesur2007, simon2009, simon2011, potter2017}, and even then the range of magnetic Prandtl numbers explored so far has been relatively small ($\text{Pm} \lesssim 16)$.

Due to its central importance as a source of angular momentum transport (at least in sufficiently ionized discs), the MRI has been studied extensively in local shearing box simulations covering a rich parameter space: both in unstratified  and vertically stratified discs (e.g. \cite{hawley1995,stonehawley1996}), under isothermal and non-isothermal conditions, in ideal and non-ideal simulations \citep{fromang2007,fromang2007mhd}, and with different initial magnetic field configurations. The configuration of the initial magnetic field is particularly important. Three configurations are commonly studied in the literature: net-vertical-magnetic-flux (NVF), net-toroidal-magnetic-flux (NTF), and zero-net-magnetic-flux (ZNF). The zero-net-flux MRI, in particular, is of great interest because it supports a non-linear dynamo that can sustain MHD turbulence even in the absence of an imposed mean magnetic field.\footnote{We use the terms zero-net-magnetic-flux (ZNF) MRI turbulence and MRI dynamo interchangeably in the text.} The net-vertical-flux MRI contains exponentially growing axisymmetric modes known as channel flows, which are characterised by channel-like horizontal streaming motions due to the instability \citep{goodman1994,hawley1995,bodo2008,pessah2009,longaretti2010,latter2009,latter2010mri,pessah2010,mamatsashvili2013,Murphy_Pessah2015,gogichaishvili2018active}. By contrast, such exponentially growing modes of classical modal instability are absent in the net-toroidal-flux and zero-net-flux MRI, which consist instead only of transiently (i.e. nonmodally) growing non-axisymmetric modes \citep{balbus1992,papaloizou1997,herault2011periodic,squire2014,gogichaishvili2017,riols2017magnetorotational}. 

Finally, we note that the ZNF configuration adopted in our simulations is an idealized one: it describes only a local patch of an actual disc. In general any portion of a global disc is bound to contain at least a small net vertical field component, which might be absent in a local patch. This mean field component is conserved under the shearing box boundary conditions and can greatly influence the long-term sustenance dynamics of the turbulence. Thus, one should exercise some caution when extrapolating ZNF shearing box results of MRI turbulence and its associated dynamo to real discs.

\subsection{Self-sustenance mechanism of MRI dynamo}
A significant amount of work has been dedicated to elucidating the self-sustenance mechanism of the MRI dynamo. Early 3D MHD simulations of a vertically stratified disc found that zero-net-flux MRI turbulence is characterized by a large-scale oscillatory axisymmetric azimuthal field \citep{brandenburg1995dynamo}. Later studies (at relatively low Reynolds numbers \text{Re} and magnetic Reynolds numbers \text{Rm}, i.e. in the tens and hundreds) discovered that this axisymmetric azimuthal field -- provided that it is modulated in the radial direction -- can continuously seed non-axisymmetric shear waves that interact non-linearly to produce an axisymmetric electromotive force (EMF). This EMF is responsible for the reversal of the azimuthal magnetic field and also induces a large-scale radial field. The radial field is then turned back into an azimuthal field by the background shear, thus completing the cycle \citep{rincon2007,rincon2008subcritical,lesur2008localized,lesur2008self,herault2011periodic, rincon2019}.

Recently \cite{gogichaishvili2017} and \cite{mamatsashvili2020zero} have studied the dynamics of fully developed net-toroidal-flux and zero-net-flux MRI turbulence, respectively, at much higher Reynolds and magnetic Reynolds numbers (up to $\sim 10^4$) in Fourier space. In contrast to the models mentioned above, in this case a larger number of dynamically active modes (degrees of freedom) are excited in the flow and hence the turbulent dynamics is much more complex. It was shown that a principal agent in the self-sustenance of MRI turbulence is {\it the non-linear transverse cascade} -- a generic, topologically new type of non-linear process in shear flows (such as disc flow) which arises from anisotropy of the non-linear dynamics due to shear \citep{horton2010,mamatsashvili2014}. The transverse cascade plays a key role in replenishing (seeding) new non-axisymmetric modes whose radial magnetic field then leads to amplification of the azimuthal field through the MRI, thereby maintaining the turbulence. This self-sustaining scheme, based on the interplay between linear MRI growth and the non-linear transverse cascade, is a generalization of the one at lower Reynolds numbers proposed in earlier studies. However, at low magnetic Prandtl numbers (${\rm Pm} \leq 1$) turbulent diffusion of large-scale (`active') modes by small-scale (`passive') modes via the direct cascade interferes with this cycle, because it prevails over the transverse cascade which is essential for the self-sustenance mechanism, and hence makes it hard to excite sustained MRI-turbulence in this regime \citep{riols2015dissipative,riols2017magnetorotational, mamatsashvili2020zero}.

\subsection{Role of explicit dissipation coefficients in MRI turbulence simulations}
\label{INTRO_RoleOfExplicitDissipationCoeffs}
Another important question is how the saturation of MRI turbulence depends on dissipation coefficients (viscosity $\nu$ and magnetic diffusivity $\eta$). Early studies in an unstratified shearing box found that in the absence of explicit dissipation coefficients (ideal simulations) the MRI was not converged, i.e. the turbulent transport (parametrized by the stress-to-thermal pressure ratio $\alpha$) decreases monotonically as the resolution is increased \citep{fromang2007}.\footnote{However, \cite{shi2016} find that this non-converegence of ideal MHD MRI with increasing resolution disappears in sufficiently tall unstratified boxes, i.e. when the vertical-to-radial aspect ratio exceeds 2.5.} This non-convergence has also been reported in stratified shearing boxes \citep{bodo2014,ryan2017}, thus necessitating the inclusion of explicit dissipation coefficients to ensure convergence of the results with resolution, at least in small boxes \citep{fromang2007mhd,fromang2010}.\footnote{By small boxes we mean of size one scale height $H$ in the radial and vertical directions. These are sometimes also referred to as `standard' boxes in the literature.}

\cite{lesur2007} have found in incompressible unstratified shearing box simulations with net-vertical-magnetic-flux that turbulent transport depends linearly on the magnetic Prandtl number (up to $\text{Pm} = 8$), and \cite{simon2009, simon2011} have found similar results for fully compressible simulations initialized with a net-toroidal magnetic field. In the presence of a mean magnetic field (NTF or NVF) and sufficiently high magnetic Reynolds numbers ($10^4-10^5$), the MRI is sustained even in the small magnetic Prandtl limit $\text{Pm}\leq1$ \citep{simon2009,meheut2015}), with \cite{meheut2015} finding that the $\alpha$ turbulent transport parameter converges to a finite value as $\text{Pm}\rightarrow0$. The zero-net-flux case is more complicated: \cite{fromang2007mhd} found that MRI dies when $\text{Pm} < 2$. This has been confirmed in incompressible unstratified simulations at much larger resolution by \cite{mamatsashvili2020zero}, who also find that turbulence decays for $\text{Pm} < 2$ even at relatively large Re and Rm ($\text{Re} = 6000, \text{Rm} = 12000$). For $\text{Pm} \geq 2$, however, unstratified ZNF simulations show that $\alpha$ increases with the magnetic Prandtl number \citep{simon2009,simon2011,potter2017}.

\subsection{Astrophysical applications}
\label{INTRO_AstrophysicalApplications}
The Reynolds and magnetic Reynolds numbers in many astrophysical discs tend to be extremely large,\footnote{An exception is in protoplanetary discs at around $1\,$AU at the mid-plane, where $\text{Rm} \sim 1$.} and the corresponding dissipation scales small compared to the characteristic length scales of the systems. In particular, microphysical viscosity is far too small to facilitate accretion on time-scales of order those inferred from observations \citep{lin1985}, and therefore turbulent transport in discs is often characterized by an effective or turbulent (`alpha') viscosity \citep{1973shakura}. Similarly, mean-field models often employ a turbulent resistivity to characterize the turbulent diffusion of magnetic fields at large scales, such as in global disc winds, or the radial transport of large-scale (poloidal) flux through the disc. While the concept of `effective' or turbulent diffusion coefficients is useful in these contexts, actual dissipation of turbulent energy is facilitated by microphysical dissipation processes (viscosity and resistivity). Reconnection of magnetic fields, which is integral for determining the saturation level of the MRI dynamo, occurs in thin current sheets whose length scale is comparable to the microphysical resistive scale, and not at a length scale associated with some effective resistivity. In any case, it is the \textit{ratio} of dissipation scales that is the important factor in determining the saturation level of MRI turbulence. If the viscous scale is much larger than the resistive scale (i.e. the $\text{Pm} \gg 1$ regime) then velocity fluctuations are damped at the resistive scale, thus impeding magnetic reconnection. A build-up of magnetic energy at the resistive scale ensues, which can then cascade back up to larger scales \citep{balbus1998instability,balbus2008, simon2009}. 

The magnetic Prandtl number varies considerably between different astrophysical systems (see Figure 6 of \cite{rincon2019} for a handy summary). Accretion discs, somewhat uniquely among astronomical objects, occupy a very wide range in \text{Pm} space, as demonstrated by several studies that have calculated this number by taking into consideration the microphysics of the disc. \cite{balbus2008} carried out 1D calculations of discs around compact objects (both low mass X-ray binaries and active galactic nuclei) and found $\text{Pm} \gg 1$ in the hot inner regions of the disc, with a transition to $\text{Pm} < 1$ occurring at around 50 Schwarzchild radii ($R_S$) from the central black hole. The authors assumed microphysical viscosity and resistivity are given by their classical \cite{spitzer1962} values (viscosity arising due to non-relativistic Coulomb interactions between protons, and resistivity due to non-relativistic Coulomb interactions between electrons and protons). A similar study investigating discs from binary neutron star mergers was carried out by \cite{rossi2008} (but taking into account the vastly different microphysics characteristic of these hyperaccreting discs, such as relativistic effects and electron degeneracy). They found that in merger remnant discs $\text{Pm} \gg 1$ everywhere. In particular, for an accretion rate of $0.2\, \text{M}_\odot\,\text{s}^{-1}$, Pm ranges from  $\text{Pm} \sim 10$ at around $5\,R_S$ to around $\text{Pm} = 70$ at $60\,R_S$. Finally, the regime $\text{Pm}\gg1$ is also relevant to the MRI-unstable neutrino-cooled interiors of protoneutron stars \citep{guilet2015neutrino}. In these systems the combination of very large Rm ($\sim 10^{19}$) and relatively low Re ($\sim 10^5$-$10^6$), the latter on account of the large neutrino viscosity, result in a magnetic Prandtl number as high as $\sim 10^{13}$.

In addition to its influence on the saturation of the MRI, the expected variation of Pm in certain types of accretion discs might have consequences for thermal instability and thus short-term variability. A sufficiently steep dependence of turbulent stress on the magnetic Prandtl number can render the disc thermally unstable. This was demonstrated for X-ray binaries, first using 1D alpha disc models \citep{takahashi2011,potter2014}, and then in full 3D numerical simulations of the MRI \citep{potter2017}. More recently, \cite{kawanaka2019} have also used 1D alpha disc models with a neutrino-cooling prescription to show that a temperature-dependent magnetic Prandtl number leads to viscous and thermal instability in discs around binary neutron star merger remnants.

\subsection{Motivation and outline}
Our aim in this paper is to explore the saturation and energetics of zero-net-flux MRI turbulence in the regime of large magnetic Prandtl numbers. We do so by means of fully compressible numerical simulations. In order to isolate the effect of large ${\rm Pm}$, only, on the turbulence, we omit complications such as stratification, and employ an isothermal equation of state. Our results show that the saturation of MRI turbulence, in terms of both turbulent transport and magnetic field strength, scales as a power-law with respect to ${\rm Pm}$ for moderately large ${\rm Pm}$. A novel result is that at very large ${\rm Pm}$, we find that this power-law begins to plateau. To understand this behaviour we carry out a detailed analysis of the turbulent dynamics in Fourier space, concentrating on the effect that increasing ${\rm Pm}$ has on the non-linear transfers (transverse cascade) responsible for sustaining the MRI dynamo. Finally, we also find that the scaling is sensitive to box vertical-to-radial aspect ratio, as well as to the shear parameter, the latter suggesting that accretion discs (from binary neutron star mergers, for example) are likely to exhibit weaker scaling of turbulent saturation with ${\rm Pm}$ than the interiors of protoneutron stars. This dependence of the magnetic energy - Pm scaling on the strength of the background shear suggests quantitative differences between accretion discs and protoneutron stars in terms of the growth of magnetic fields via the MRI dynamo.

The structure of the paper is as follows. In Section \ref{METHODS} we discuss our set-up, initial conditions, parameters, and diagnostics. We present our key results in Section \ref{RESULTS_RealSpace}, in particular how $\alpha$ scales with the magnetic Prandtl number, focusing on dynamics in real space. In Section \ref{RESULTS_SpectralAnalysis} we perform a spectral analysis of our results in order to elucidate the energetics of the MRI dynamo as the magnetic Prandtl number is increased. In Section \ref{RESULTS_Discussion} we discuss how the $\alpha$-Pm scaling depends on the shear parameter and on the vertical-to-radial box aspect ratio, and also compare our results to previous work. Finally, we present our conclusions in Section \ref{CONCLUSIONS}.

\section{Methods}
\label{METHODS}

\subsection{Governing equations}
\label{METHODS_GoverningEquations}
We work in the shearing box approximation
\citep{goldreich1965,hawley1995,latter2017local},
which treats a local region of a disc as a Cartesian box located at some fiducial radius $r = r_0$ and orbiting with the angular frequency of the disc at that radius $\Omega_0 \equiv \Omega(r_0)$. A point in the box has
Cartesian coordinates $(x, y, z)$ along the radial, azimuthal/toroidal, and vertical directions, respectively. In this rotating frame, the equations of non-ideal MHD are

\begin{equation}
\partial_t \rho + \nabla \cdot (\rho \mathbf{u}) = 0, \label{SB1}
\end{equation}
\begin{multline}
\partial_t \mathbf{u} + \mathbf{u}\cdot\nabla \mathbf{u} = -\frac{1}{\rho} \nabla P - 2\Omega_0 \mathbf{e}_z \times \mathbf{u} + 2q\Omega_0^2x\mathbf{e}_x+ \\\frac{1}{\mu_0 \rho}(\nabla\times\mathbf{B})\times\mathbf{B}+\frac{1}{\rho}\nabla \cdot \mathbf{T}, \label{SB2}
\end{multline} 
\begin{equation}
\partial_t \mathbf{B} = \nabla\times(\mathbf{u}\times\mathbf{B})+\eta\nabla^2\mathbf{B}, \label{SB4}
\end{equation}
with the symbols taking their usual meanings. We close the system with the equation of state for an isothermal gas $P = c_s^2 \rho$ where $c_s^2$ is the (constant) speed of sound.
 
All our simulations are unstratified and the effective gravitational potential is embodied in the radial tidal acceleration $2q\Omega_0^2x$ (third term on the right-hand side of Equation \ref{SB2}), where $q$ is the dimensionless shear parameter $q \equiv -\left.d\ln{\Omega}/d\ln{r}\right\vert_{r=r_0}$. For Keplerian discs $q=3/2$, a value we adopt throughout this paper except in Section \ref{RESULTS_EffectOfChangingShearParameter} where we vary $q$ in order to investigate the dependence of the $\alpha-\text{Pm}$ scaling on the shear parameter. 

To control the magnetic Prandtl number (see Section \ref{Methods_Parameters}) we employ explicit diffusion coefficients. The viscous stress tensor is given by $\mathbf{T} \equiv 2\rho \nu \mathbf{S}$, where $\nu$ is the kinematic viscosity, and $\mathbf{S} \equiv (1/2)[\nabla \mathbf{u} + (\nabla \mathbf{u})^\text{T}] - (1/3)(\nabla\cdot\mathbf{u})\mathbf{I}$ is the traceless shear tensor \citep{landau1987}. The explicit magnetic diffusivity is denoted by $\eta$: it is related to the resistivity $\xi $ via $\eta \equiv \xi/\mu_0$, where $\mu_0$ is the permeability of free space (note that from now on we will use the terms resistivity and magnetic diffusivity interchangeably). We keep the viscosity and resistivity fixed in space and time in any given simulation.

\subsection{Important parameters}
\label{Methods_Parameters}
The magnetic Reynolds number compares inductive to resistive effects and is given by

\begin{equation}
\text{Rm} = \frac{c_s H}{\eta},
\end{equation}
where $c_s$ is the isothermal speed of sound, $H=c_s/\Omega_0$ is the scale height (see Section \ref{METHODS_Units} for definitions), and $\eta$ is the magnetic diffusivity. The Reynolds number compares inertial to viscous forces and is given by

\begin{equation}
\text{Re} = \frac{c_s H}{\nu},
\end{equation}
where $\nu$ is the kinematic viscosity. Finally, the ratio of Rm to Re defines the magnetic Prandtl number,

\begin{equation}
\text{Pm} \equiv \frac{Rm}{Re} =  \frac{\nu}{\eta},
\end{equation}
which serves as the key control parameter in our simulations.

\subsection{Numerical set-up}
\label{METHODS_NumericalSetUp}

\subsubsection{Code}
\label{Methods_Codes}
For our simulations we use the conservative, finite-volume code \textsc{PLUTO} \citep{mignone2007}. We employ the HLLD Riemann solver, 2nd-order-in-space linear interpolation, and the 2nd-order-in-time Runge-Kutta algorithm. In addition, in order to enforce the condition that $\nabla\cdot\mathbf{B}=0$ we employ Constrained Transport (CT), and use the UCT-Contact algorithm to calculate the EMF at cell edges. To allow for longer time-steps, we take advantage of the \textsc{FARGO} scheme \citep{mignone2012}. When explicit resistivity $\eta$ and viscosity $\nu$ are included, we further reduce the computational time via the Super-Time-Stepping (STS) scheme \citep{alexiades1996super}. Ghost zones are used to implement the boundary conditions.

We use the built-in shearing box module in \textsc{PLUTO}
\citep{mignone2012}. Rather than solving Equations \eqref{SB1}-\eqref{SB4} (primitive
form), \textsc{PLUTO} solves the governing equations in conservative form.

\begin{figure}
\centering
\includegraphics[scale=0.23]{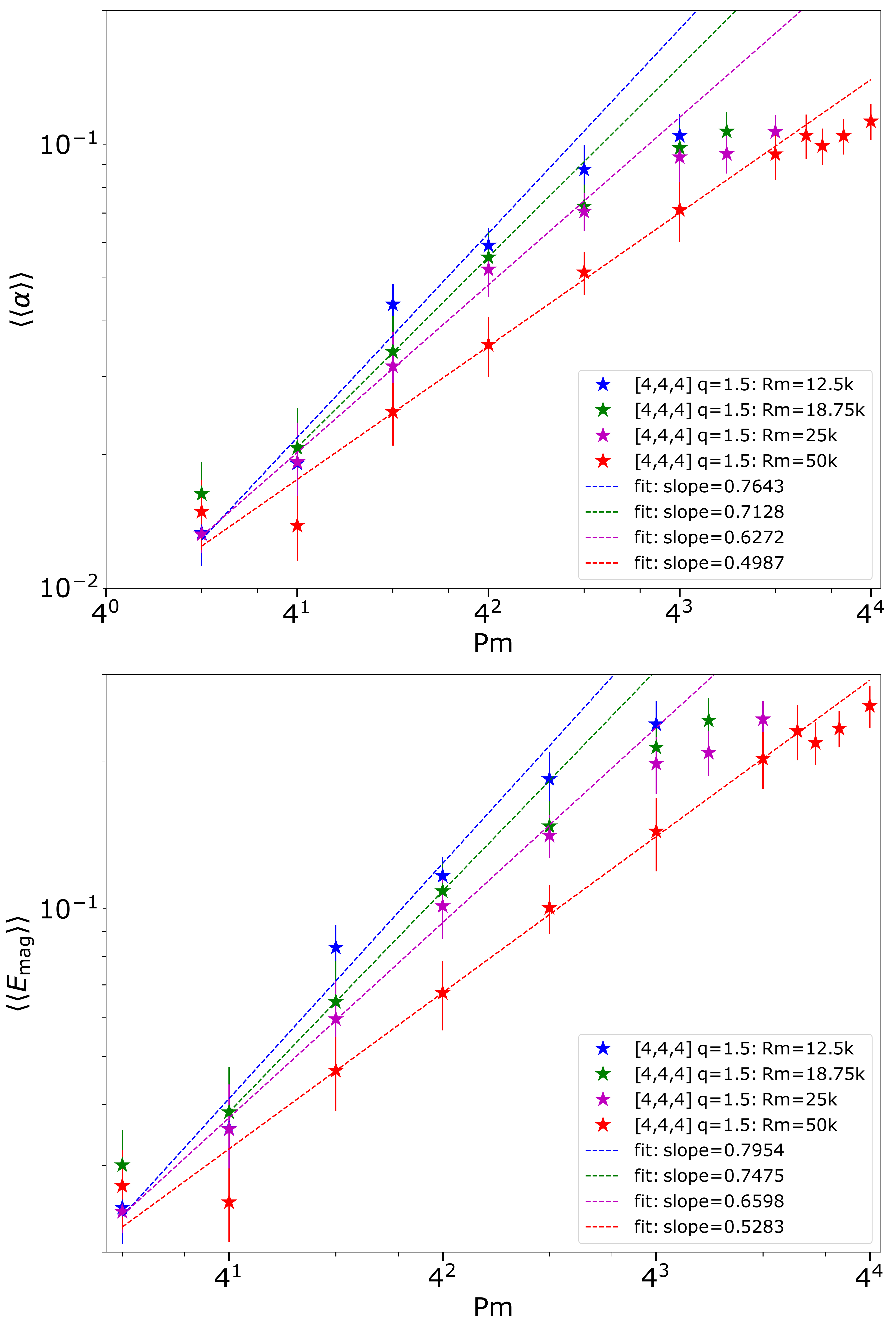}
\caption{Comparison of $\alpha$-Pm scaling (top) and $E_\text{mag}$-Pm scaling (bottom) between four sets of simulations with different magnetic Reynolds numbers of $\text{Rm}=12500$ (blue), $\text{Rm}=18750$ (green), $\text{Rm}=25000$ (magenta), and $\text{Rm}=50000$ (red), respectively (the abscissa is in log scale to base 4). The dashed lines show fits to the data. Error bars denote the standard deviation measured in each simulation. All simulations were run in cubic boxes of size $4H\times4H\times4H$ with a shear parameter of $q = 1.5$ (Keplerian shear).}
\label{FIGURE_AlphaPmRelationship}
\end{figure}

\subsubsection{Initial conditions}
\label{METHODS_InitialConditions}
All our simulations are initialized from an equilibrium with a uniform
density profile. The background velocity is given by $\mathbf{u} = -q \Omega_0 x \,
\mathbf{e}_y$. At initialization we usually perturb all the
velocity components with random noise exhibiting a flat power
spectrum. The perturbations $\delta \mathbf{u}$ have maximum
amplitude of about $5\times10^{-2}\,c_{s0}$, unless stated otherwise. Here $c_{s0}$ is the sound speed at initialization. All simulations are initialized with commonly used \textit{zero-net-flux} (ZNF) magnetic field configuration $\mathbf{B}_0 = B_0\sin{(2\pi x/L_x)}\mathbf{\hat{e}}_z$. The field strength $B_0$ is controlled through the ratio of gas pressure to magnetic pressure $\beta_0 \equiv 2P /B_0^2$, which we set to $\beta_0 \equiv 1000$. 

\subsubsection{Units}
\label{METHODS_Units}
Note that from this point on, all quantities are given in terms of dimensionless code units. Time units are selected so that $\Omega_0 = 1$. The length unit is chosen so that the initial sound speed $c_{s0} = 1$, which in turn defines a reference scale height $H_0\equiv c_{s0} / \Omega_0=1$. Finally the mass unit is set by the initial density which is $\rho_0 = 1$. From now on we drop subscripts on the angular frequency and scale height.

\subsubsection{Box size and resolution}
\label{METHODS_BoxSizeAndResolution}
The majority of our simulations are run at a resolution of $N_x\times N_y\times N_z=512\times512\times512$ in a box of size $L_x\times L_y\times L_z = 4H\times4H\times4H$ (i.e. $128$ cells per scale height). We refer to runs with this box size as our \textit{cubic box} runs. To test convergence we have also run select cubic box simulations at resolutions of $128^3$, $256^3$, and $1024^3$, corresponding to $32, 64,$ and $256$ cells per scale height, respectively (see Appendix \ref{APPENDIX_ConvergenceStudy}). To investigate the dependence on box vertical-to-radial aspect ratio we have also repeated select runs in boxes of size $H\times3H\times 3H$ and $H\times4H\times 4H$ (keeping the resolution per scale height the same as in the cubic box runs, i.e. 128 cells per $H$). We refer to these as our \textit{vertical slab} or \textit{tall box} runs. Finally to compare to previous work we have repeated select runs in boxes of size $H\times4H\times H$, which we refer to as \textit{finger box} runs. All the simulations described in this paper (with corresponding values of the main parameters, i.e. resolution, box size, ${\rm Re}$, ${\rm Rm}$, and ${\rm Pm}$, as well as time- and volume-averages of key diagnostics, e.g. magnetic energy density, $\alpha$, and Maxwell and Reynolds stresses (see Section \ref{METHODS_Diagnostics} for definitions)), are listed in Tables \ref{TABLE_LargeBoxSimulations} and \ref{TABLE_TallBoxSimulations} in the appendix.

\subsubsection{Boundary conditions and mass source term}
\label{METHODS_BoundaryConditions}
We use standard shear-periodic boundary conditions in the $x$-direction
\cite[see][]{hawley1995} and periodic boundary conditions in the
$y$- and $z$-directions. Thus mass should not escape the box, and we expect the total mass to be conserved in these runs.

\subsection{Diagnostics}
\label{METHODS_Diagnostics}
Below we define various diagnostics in real space, only. For diagnostics in Fourier space the reader should refer to Section \ref{RESULTS_GoverningEquationsInSpectralSpace}.

\subsubsection{Averaged quantities}
\label{METHODS_AveragedQuantities}
The volume-average of a quantity $X$ is denoted $\langle X \rangle$ and is defined as 
\begin{equation}
\langle X \rangle(t) \equiv \frac{1}{V} \int_V X(x, y, z, t) dV
\end{equation}
where $V$ is the volume of the box.

We are also interested in averaging certain quantities (e.g. the Reynolds stress) over time. The temporal average of a quantity $X$ is denoted $\langle{X}\rangle_t$ and is defined as
\begin{equation}
\langle X \rangle_t (x, y, z) \equiv \frac{1}{\Delta t} \int_{t_i}^{t_f} X(x, y, z, t) dt,
\end{equation}
where we integrate from some initial time $t_i$ to some final time $t_f$ and $\Delta t \equiv t_f - t_i$.

The horizontal average of a quantity $X$ is denoted $\langle{X}\rangle_{xy}$ and is defined as

\begin{equation}
    \langle X \rangle_{xy}(z,t) \equiv \frac{1}{A} \int_A X(x,y,z,t) dA,
\end{equation}
where $A$ is the horizontal area of the box. Horizontal averages over different coordinate directions (e.g. over the $y$- and $z$-directions) are defined in a similar manner.

\subsubsection{Reynolds and magnetic stresses, and alpha}
\label{METHODS_ReynoldsAndMagneticStressesAndAlpha}
In accretion discs, the radial transport of angular momentum is
related to the $xy$-component of the total stress
\begin{equation}
\Pi_{xy} \equiv R_{xy} + M_{xy},
\label{totalstress}
\end{equation}
in which $R_{xy} \equiv \rho u_x \delta u_y$ is the Reynolds stress, where $\delta u_y \equiv u_y + q\Omega x$ is the fluctuating part of the y-component of the total velocity $u_y$, and $M_{xy} \equiv -B_x B_y$ is the magnetic stress. The total stress is related to the classic dimensionless angular momentum transport parameter $\alpha$ by
\begin{equation}
\alpha \equiv \frac{\langle \Pi_{xy} \rangle}{\langle P \rangle},
\label{alpha}
\end{equation}
where $P$ is the gas pressure. Note that some definitions of alpha include the dimensionless shear parameter $q$ (see Section \ref{METHODS_GoverningEquations}) in the denominator. To compare our results more easily with the literature, we drop this factor.

\subsubsection{Energetics}
\label{METHODS_EnergyDensities}
\begin{figure}
    \centering
    \includegraphics[scale=0.24]{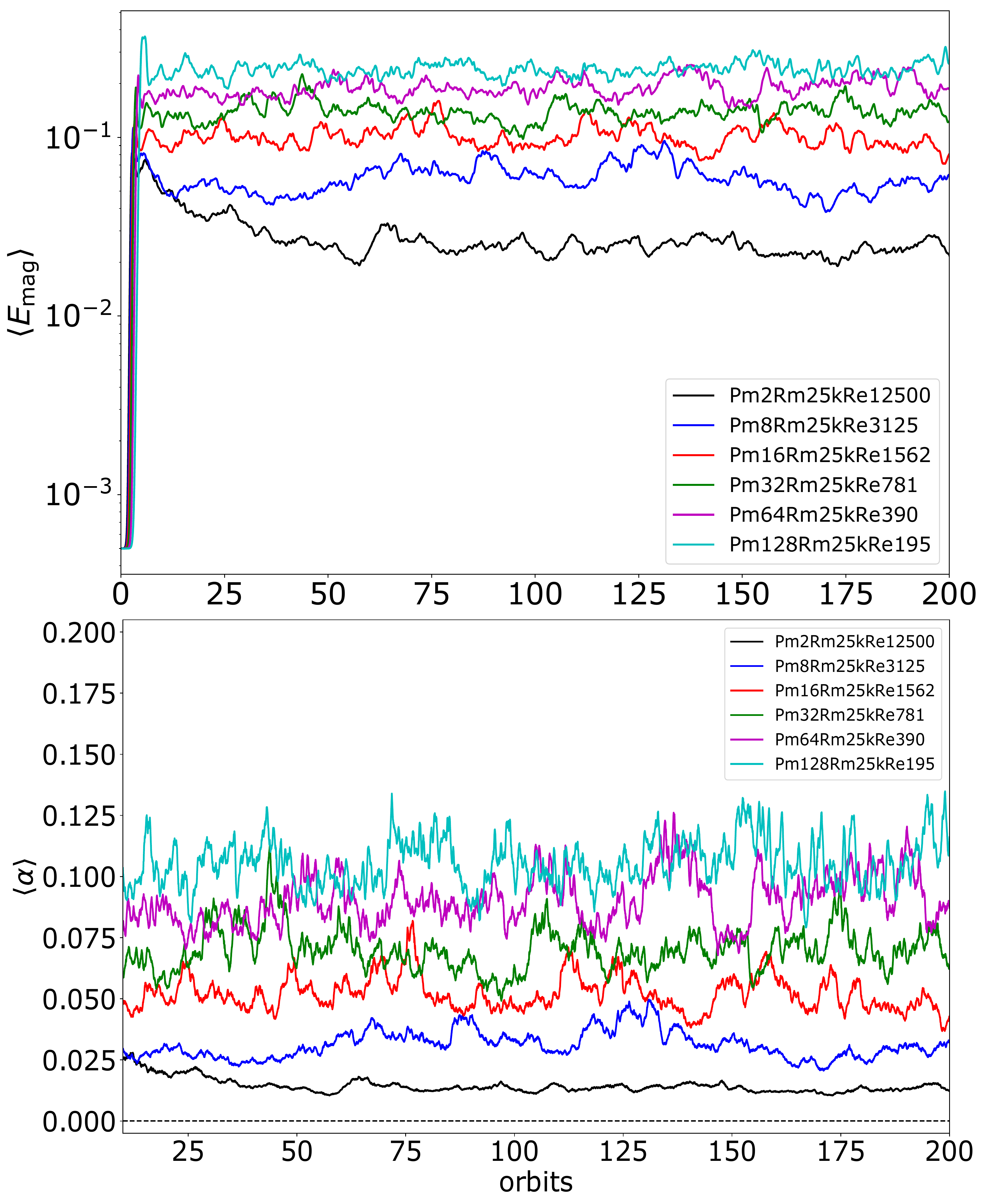}
    \caption{Time evolution of volume-averaged magnetic energy density (upper panel) and turbulent stress (lower panel) from select simulations with different magnetic Prandtl number. The magnetic Reynolds number is fixed at $\text{Rm}=2.5\times10^4$ in all runs.}
    \label{FIGURE_TimeSeriesRes128Rm25k}
\end{figure}

The total energy density is given by
\begin{equation}
E_\text{total} = \frac{1}{2}\rho u^2 + \frac{1}{2}B^2 + \rho \Phi + \rho e,
\end{equation}
where the terms on the right-hand side correspond to the kinetic $E_{\text{kin}}$, magnetic $E_{\text{mag}}$, gravitational potential and thermal $E_{\text{th}}$ energy densities, respectively. Here $\Phi =-\frac{3}{2}\Omega_0^2x^2$ is the effective gravitational potential in the shearing box approximation for a Keplerian disc. 

In some of our simulations we employ `tall' boxes (i.e. $L_z/L_x > 1$). These tend to exhibit a coherent, large-scale, vertically-varying toroidal field \citep{lesur2008self,shi2016}. A useful diagnostic for quantifying this large-scale toroidal field is to decompose the azimuthal magnetic energy density into a mean part (due to the large-scale field) and contributions due to smaller-scale fluctuations \citep[see Equations 9-10 of][]{shi2016}, thus

\begin{equation}
    E_{\text{mag,y}} = E_{\text{mag},y,\text{avg}} + E_{\text{mag},y,\text{pert}},
    \label{EQUN_TotalByField}
\end{equation}
where the magnetic energy density of the background toroidal field is defined as

\begin{equation}
   E_{\text{mag},y,\text{avg}} = \frac{1}{2} \langle (\langle B_y\rangle_{xy})^2 \rangle_z,
   \label{EQUN_EmagyMean}
\end{equation}
while the magnetic energy density in the small-scale toroidal field is defined as

\begin{equation}
   E_{\text{mag},y,\text{pert}} = \frac{1}{2} \langle (B_y-\langle B_y \rangle_{xy})^2 \rangle.
   \label{EQUN_EmagyPerturbed}
\end{equation}

\subsubsection{Correlation lengths}
\label{METHODS_CorrelationLengths}

To quantify the size of structures observed in the flow field, we compute the 1D correlation length \citep{fromang2007}. The 1D (vertically-averaged) correlation length in the $x$-direction of the $j$th-component of a vector $\mathbf{X}$, where $j \in {x,y,z}$ and $\mathbf{X} \in {\mathbf{u},\mathbf{B}}$ is defined as

\begin{equation}
    L_x(X_j) = \left\langle \frac{\int \int X_j(x,y=0,z) X_j(x',y=0,z) dx' dx}{\int X_j^2(x,y=0,z) dx}\right\rangle_z,
    \label{EQUN_CorrelationLength}
\end{equation}
where $x'$ and $x$ denote two points along the $x$-axis. We use a similar definition for computing the correlation length in the $z$-direction $L_z(X_j)$ (in which case the average is taken over the $x$-direction).

\section{Turbulent dynamics at large magnetic Prandtl numbers}
\label{RESULTS_RealSpace}

In this section, we investigate the dynamics of MRI turbulence in the large magnetic Prandtl number regime $({\rm Pm} > 1)$ using various common diagnostics (time-series, flow field, correlation lengths, spacetime diagrams, etc.). A deeper, spectral analysis of the turbulent dynamics is presented separately in Section \ref{RESULTS_SpectralAnalysis}. We consider four sets of simulations: each set is carried out at \textit{fixed} magnetic Reynolds number with $\text{Rm} \in\{12500, 18750, 25000, 50000\}$. We present a convergence study of the results at $\text{Pm}=4$ and $\text{Rm}=18750$ in Appendix \ref{APPENDIX_ConvergenceStudy}.  Within any one set we increase the magnetic Prandtl number $\text{Pm} \equiv \text{Rm}/\text{Re}$ (starting at $\text{Pm}=2$) by increasing the explicit viscosity (i.e. \textit{decreasing} Re), keeping the resistivity (i.e. Rm) fixed. The four sets of simulations are listed in Table \ref{TABLE_LargeBoxSimulations}. Note that the Reynolds and magnetic Reynolds numbers in our simulations (as in all simulations of discs with explicit diffusion coefficients) are far smaller than those expected in most real discs (e.g. in discs from binary neutron star mergers $\text{Rm}\sim 10^{19}$ and $\text{Re} \sim 10^{16}$-$10^{18}$ \citep{rossi2008}). However, in terms of the key control parameter (the magnetic Prandtl number), our simulations investigate $\text{Pm} = 2$ to $\text{Pm} = 256$ and thus significantly overlap with the Pm-regime of discs around merger remnants, as discussed in Section \ref{INTRO_AstrophysicalApplications}.

All simulations described in this section were run in boxes of size $4H\times4H\times4H$ at a resolution of $512^3$ (128 cells per scale height) and in the regime of Keplerian shear ($q = 1.5$). We defer a discussion of the effect of vertical-to-radial box size $L_z/L_x$, and of the shear parameter $q$ on the turbulent dynamics to Section \ref{RESULTS_Discussion}.

\subsection{Alpha-Pm relationship}
\label{RESULTS_AlphaPmRelationship}

\begin{figure*}
\centering
\includegraphics[scale=0.35]{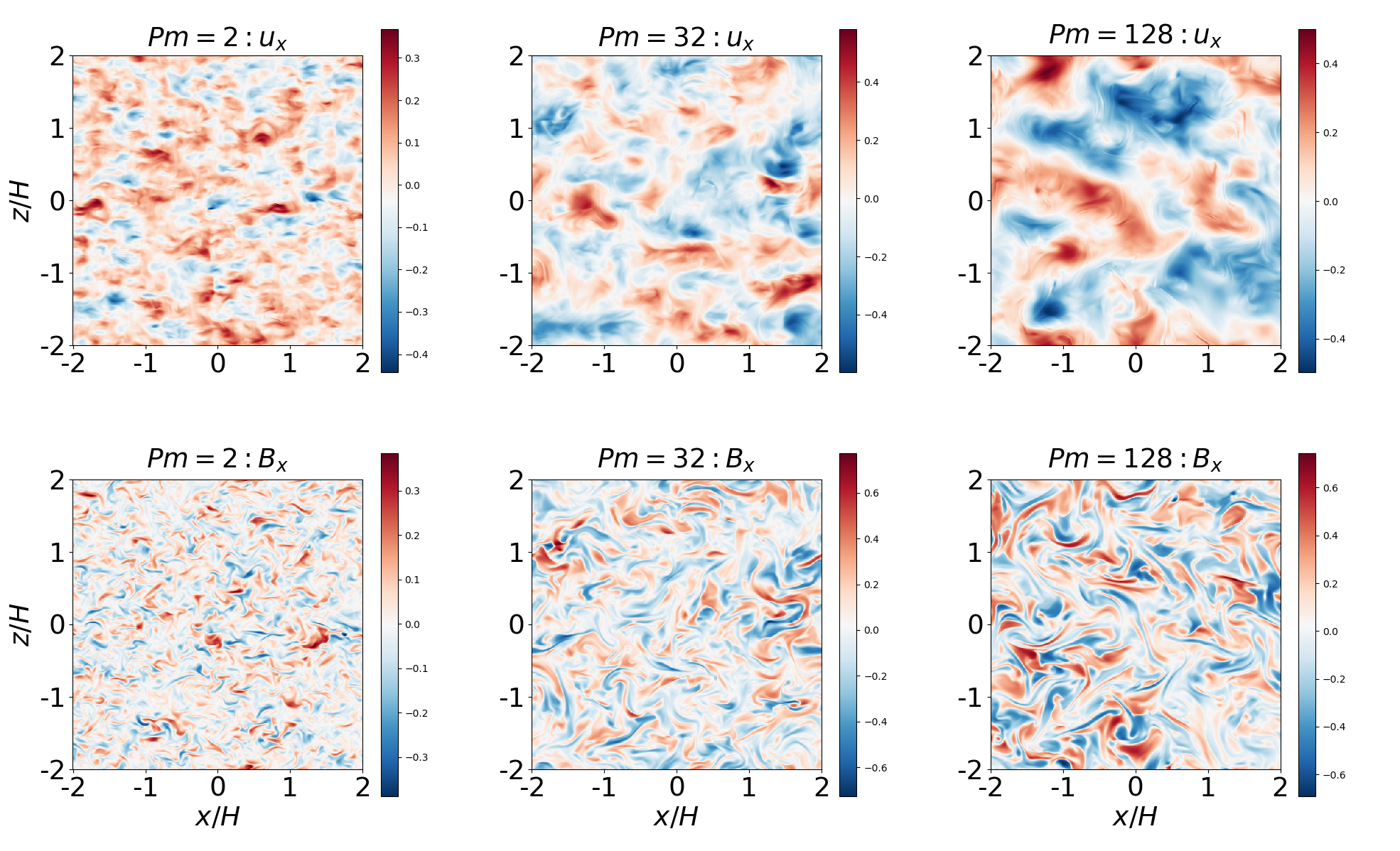}
\caption{Comparison of $u_x$ velocity component (top row) and $B_x$ magnetic field component (bottom row) from three simulations at $\text{Pm}=2$ (left), $\text{Pm}=32$ (middle), and $\text{Pm}=128$ (right). The magnetic Reynolds number is fixed at $\text{Rm}=2.5\times10^{4}$ in all simulations.}
\label{FIGURE_Res128Rm25kSeriesFlowFieldComparison}
\end{figure*}

Our key results are shown in Figure \ref{FIGURE_AlphaPmRelationship}. The top panel shows the $\alpha$-Pm relationship from our runs, while the bottom panel shows the relationship between total magnetic energy $E_\text{mag}$ and Pm. Three salient features emerge from this figure:

First, we find a clear power-law scaling relationship between $\alpha$ and the magnetic Prandtl number, i.e. $\alpha \propto \text{Pm}^{\delta}$ in the interval between $\text{Pm}=2$ and $\text{Pm}\sim 64$ regardless of the magnetic Reynolds number. 

Second, the scaling appears to become weaker as the magnetic Reynolds number ${\rm Rm}$ is increased: at $\text{Rm} = 12500$ we find $\delta \sim 0.76$, at $\text{Rm} =18750$ we find $\delta \sim 0.71$, at $\text{Rm} = 25000$ we measure a scaling of $\delta \sim 0.63$, and at $\text{Rm}=50000$ we measure a scaling of $\delta \sim 0.5$.\footnote{We fit data between $\text{Pm}=2$ and $\text{Pm}=16,32$, and 64 for the $\text{Rm}=12500, 25000$, and 50000 series, respectively. In the ($\text{Pm}=2$, $\text{Rm}=18750$) simulation the stress and magnetic energy increase slightly over the last 100 orbits, which result in the data points for $\alpha$ and $E_\text{mag}$ for this simulation being outliers. Thus for the $\text{Rm}=18750$ series we fit data only between $\text{Pm}=4$ and $\text{Pm=16}$.} We find similar scaling laws when considering the increase of magnetic energy density with ${\rm Pm}$ (bottom panel of Figure \ref{FIGURE_AlphaPmRelationship}), though in this case the slopes are $3\%-5\%$ steeper.

Third, the new effect that we observe here compared to previous related studies \citep{lesur2007,simon2009,simon2011,meheut2015} is that the turbulent stress and magnetic energy begin to plateau at large ${\rm Pm}$, with the onset of the plateau occurring at smaller ${\rm Pm}$ the larger ${\rm Rm}$, e.g. the onset occurs at around $\text{Pm}=32$ for $\text{Rm}=12500$ and $\text{Rm}=18750$, at around $\text{Pm}=64$ for $\text{Rm}=25000$ and at around $\text{Rm}=128$ for $\text{Rm}=50000$. Moreover, within the plateau we observe little to no dependence of the results on ${\rm Rm}$ (or on ${\rm Re}$).

\subsection{Change in turbulent dynamics with Pm}
\label{RESULTS_TimeSeriesAndFlowField}
In order to understand the increase in the saturation level of the turbulence with ${\rm Pm}$, we next turn to various diagnostics in real space. For brevity we restrict the bulk of the discussion in this section to the set of simulations at $\text{Rm} = 25000$ (which serve as our fiducial simulations).

\subsubsection{Time evolution of averaged quantities}
\label{SECTION_TimeEvolutionOfAveragedQuantities}

In Figure \ref{FIGURE_TimeSeriesRes128Rm25k}, we show the time series of the volume-averaged magnetic energy density (top panel) and turbulent stress $\alpha$ (bottom panel) from select runs ranging from $\text{Pm}=2$ (black curve) to $\text{Pm}=128$ (light blue curve), all run at a \textit{fixed} magnetic Reynolds number of $\text{Rm}=25000$. As the magnetic Prandtl number is increased, the peak magnetic energy at the onset of non-linear saturation also increases. Afterwards, the flow settles into quasi-steady equilibrium faster as ${\rm Pm}$ increases: at $\text{Pm}=2$ the energy decays over 50 orbits or so before reaching quasi-steady equilibrium, compared to the $\text{Pm}=16$ runs and above where the energy reaches equilibrium within a few orbits of non-linear saturation. However, the MRI-turbulence is maintained at all Prandtl numbers that we investigated. Finally there is a marked increase in the the magnetic energy in the quasi-steady state as Pm is increased. For example, $\langle E_\text{mag} \rangle \sim 0.02$ at $\text{Pm}=2$ compared to $\langle E_\text{mag} \rangle \sim 0.24$ at $\text{Pm}=128$, an increase of an order of magnitude, which is consistent with the scaling-law behaviour mentioned above (Figure \ref{FIGURE_AlphaPmRelationship}). The same behavior is reflected in the turbulent transport (bottom panel), which increases from $\langle \alpha \rangle \sim 0.01$ at $\text{Pm}=2$ before leveling out at $\langle \alpha \rangle \sim 0.1$ at $\text{Pm}=128$. As the magnetic Prandtl number enters the plateau region shown in Figure \ref{FIGURE_AlphaPmRelationship} ($\text{Pm}\sim64$ and above), the different time-series become bunched up, reflecting the saturation of the magnetic energy and stress at large ${\rm Pm}$, as seen in Figure \ref{FIGURE_AlphaPmRelationship}.

We observe similar behaviour in the individual stresses (Maxwell and Reynolds; not shown). The Maxwell stress increases with Pm until the plateau region, as expected from the behaviour of the magnetic energy. In Table \ref{TABLE_LargeBoxSimulations} we list the ratio of Maxwell stress to Reynolds stress $\langle \langle M_{xy} \rangle \rangle / \langle \langle R_{xy} \rangle \rangle$. This quantity increases with Pm ($\sim 4.6$ at $\text{Pm}=2$ compared to $\sim 7.7$ at $\text{Pm}=128$). Interestingly, the Reynolds stress does not decrease as Pm increases, despite the increase in explicit viscosity (decrease in Reynolds number Re): instead we find that \textit{both} the Maxwell and Reynolds stresses increase as Pm is increased. Thus the increase in the Maxwell stress relative to the Reynolds stress is not due to weakened hydrodynamic activity. 

\subsubsection{Structure of the flow}
\label{SECTION_StructureOfTheFlow}
In Figure \ref{FIGURE_Res128Rm25kSeriesFlowFieldComparison} we compare snapshots of the $x$-components of the velocity and magnetic field (in $xz$-plane) from our $\text{Pm} = 2$,  $\text{Pm} = 32$, and $\text{Pm}=128$ runs. The first two of these simulations lie within the region where we observe power-law scaling in $\alpha$ with Pm, while the last simulation occurs in the plateau region (see Figure \ref{FIGURE_AlphaPmRelationship}). As Pm is increased, structures in the velocity field become noticeably larger, which is not unexpected given that the Reynolds number decreases from $\text{Re}=12500$ (at $\text{Pm} = 2$) to $\text{Re}=195$ (at $\text{Pm} = 128$), and thus that the viscous scale moves to smaller wavenumbers. Of greater interest is that we also observe larger structures in the \textit{magnetic} field as $\text{Pm}$ is increased, despite the magnetic Reynolds number remaining fixed in these simulations. To quantify the size of the structures visible in the plot, we have calculated the correlation lengths for the velocity and magnetic fields in both the $x$- and $z$-directions (see Equation \ref{EQUN_CorrelationLength}). At $\text{Pm}=2$ we find $(L_x[u_x],L_x[B_x], L_z[u_x], L_z[B_x]) \sim (0.34, 0.10, 0.51, 0.07)H$, at $\text{Pm}=32$ we find $(L_x[u_x],L_x[B_x], L_z[u_x], L_z[B_x]) \sim (0.61, 0.17, 0.32, 0.10)H$, and at  $\text{Pm}=128$ we find $(L_x[u_x],L_x[B_x], L_z[u_x], L_z[B_x]) \sim (0.74, 0.18, 0.33, 0.09)H$, which reflects the increase in the size of structures in the flow with Pm that can be seen by eye in Figure \ref{FIGURE_Res128Rm25kSeriesFlowFieldComparison}.

\begin{figure}
\centering
\includegraphics[scale=0.24]{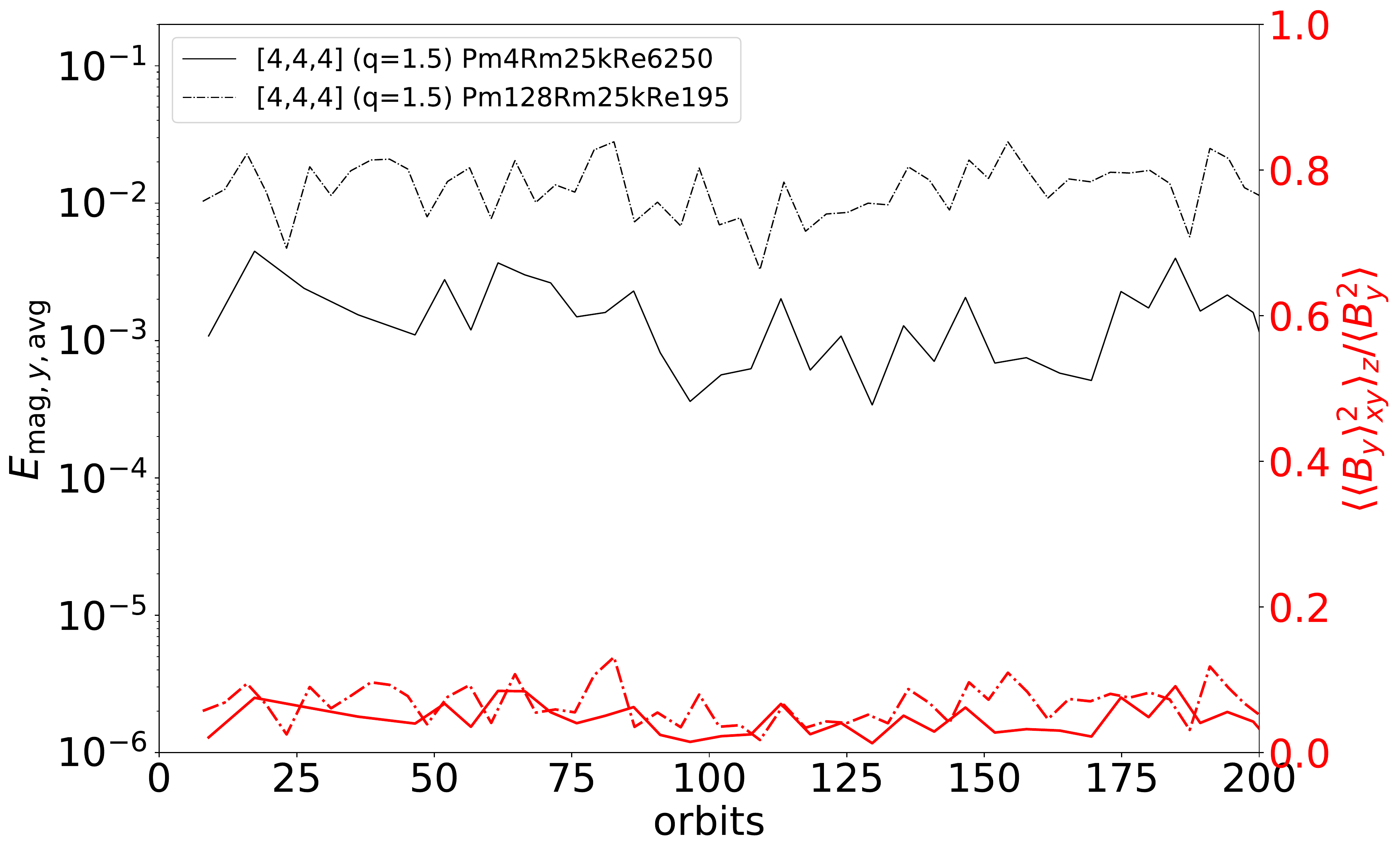}
\caption{Comparison of magnetic energy in the background toroidal field (black curves) and of the ratio of magnetic energy in the background toroidal field to the total energy in the toroidal field (red curves) for simulations at two different Pm. Solid curves: $\text{Pm}=4$ (power-law scaling region). Dash-dot curves: $\text{Pm}=128$ (plateau region). Both simulations were run with $\text{Rm}=25000$.}
\label{FIGURE_EmagyavgTimeSeriesComparisonPm4Pm128}
\end{figure}

\cite{shi2016} report the emergence of a large-scale toroidal magnetic field which undergoes a change of sign as a function of the vertical coordinate in tall boxes ($H\times4H\times4H$): see their Figure 1. In addition, the density field in their tall boxes exhibits only small-scale turbulence. In `finger' boxes ($H\times4H\times H$), on the other hand, they find no such mean toroidal magnetic field,  while the density field exhibits large-scale density waves superimposed on small-scale turbulence. Our cubic boxes are the same size in the vertical direction ($L_z = 4H$) as the non-ideal simulations of \cite{shi2016}, but the vertical-to-radial aspect ratio in our fiducial simulations is unity. We have inspected snapshots of toroidal magnetic field $B_y$ and density $\rho$ in the $xz$-plane in our simulations during the non-linear phase (not shown). We find no sign of a large-scale $B_y$ field. In the density field we observe density waves superimposed on a background of small-scale turbulence. Thus the flow field in our large cubic box simulations ($4H\times4H\times4H$) more closely resembles that found in the runs with small vertical-to-radial aspect ratio of \cite{shi2016} rather than that found in their tall boxes. 

To quantify the presence (if any) of a large-scale mean toroidal field,  we have also calculated the magnetic energy density in the mean toroidal field (see Equation \ref{EQUN_EmagyMean}), both at $\text{Pm}=4$ (which is inside the power-law scaling region in Figure \ref{FIGURE_AlphaPmRelationship}) and at $\text{Pm}=128$ (which is inside the plateau region). See Figure \ref{FIGURE_EmagyavgTimeSeriesComparisonPm4Pm128}. The energy in the mean toroidal magnetic field (black curves) increases by around an order of magnitude as we increase the magnetic Prandtl number from $\text{Pm}=4$ to $\text{Pm}=128$, but this simply reflects the overall increase in total magnetic energy as Pm is increased. The \textit{fraction} of total magnetic energy contained in the mean toroidal magnetic field hardly changes as Pm is increased. Furthermore the energy in the mean toroidal field is negligible compared to the energy of the total toroidal magnetic field ($\sim3\%$ at $\text{Pm}=4$ , $\sim5\%$ at $\text{Pm}=128$). We conclude that (in our large cubic boxes) there is no significant mean toroidal magnetic field in the regime of very large magnetic Prandtl numbers (corresponding to the plateau region in \ref{FIGURE_AlphaPmRelationship}). Note that when we investigate the effect of the vertical-to-radial box aspect ratio on the results, we do observe the emergence of a strong mean magnetic field in tall boxes ($L_z/L_x \gtrsim 2.5$), which we discuss in greater detail in Section \ref{RESULTS_EffectOfChangingBoxAspectRatio}.

\subsubsection{Spacetime diagrams}
\label{SECTION_SpacetimeDiagrams}
Finally we have also compared spacetime diagrams of the vertical profile of the horizontally-averaged total toroidal magnetic field $\langle B_y \rangle_{xy}(z)$ (not shown) at $\text{Pm}=2$ and $\text{Pm}=32$ between orbits 200 and 300. In both cases we observe `patchy' structure in the toroidal magnetic field characteristic of the unstratified MRI dynamo (see, for example, the bottom panel of Figure 16 in \cite{mamatsashvili2020zero}). These coherent patches of  $\langle B_y \rangle_{xy}$ are about a scale height in size, and can be interpreted as a weak, height-dependent toroidal field that changes sign 2-3 times over the vertical extent of the box. Any individual patch also changes sign every 5-10 orbits. We find no appreciable difference in the size of these toroidal flux tubes or the time-scales over which they change sign between $\text{Pm}=2$ and $\text{Pm}=32$, however.

\section{Spectral analysis}
\label{RESULTS_SpectralAnalysis}
This section is devoted to the spectral dynamics of MRI-turbulence in Fourier space and is structured as follows. In Section \ref{RESULTS_GoverningEquationsInSpectralSpace}, we introduce the governing equations and diagnostic terms in Fourier space. We then consider (in Section \ref{RESULTS_EnergySpectra}), the 2D and 1D (shell-averaged) spectra of kinetic and magnetic energy, focusing on how these vary with increasing magnetic Prandtl number ${\rm Pm}$ (at fixed magnetic Reynolds number ${\rm Rm}$). Then, in Section \ref{RESULTS_SpectraOfDynamicalTerms}, we analyze the spectra of dynamical terms (Maxwell stress and nonlinear transfers), and how these relate to the self-sustenance of the MRI-turbulence. We continue our analysis of these dynamical terms in Section \ref{RESULTS_DependenceOfTurbulentDynamicsInSpectralSpaceOnPm}, where we discuss how they change with increasing ${\rm Pm}$ (at fixed ${\rm Rm}$), and how this can explain two key results: the power-law scaling of alpha with ${\rm Pm}$, and the plateau at large ${\rm Pm}$. Finally, in Section \ref{RESULTS_DependenceOnReAtFixedPm}, we investigate how the same diagnostics depend on ${\rm Pm}$ (at fixed Reynolds number ${\rm Re}$), and also on ${\rm Re}$ (at fixed ${\rm Pm}$). The latter helps explain why $\alpha-{\rm Pm}$ scaling weakens with increasing ${\rm Re}$ (or ${\rm Rm}$).

\begin{figure}
\centering
\includegraphics[scale=0.45]{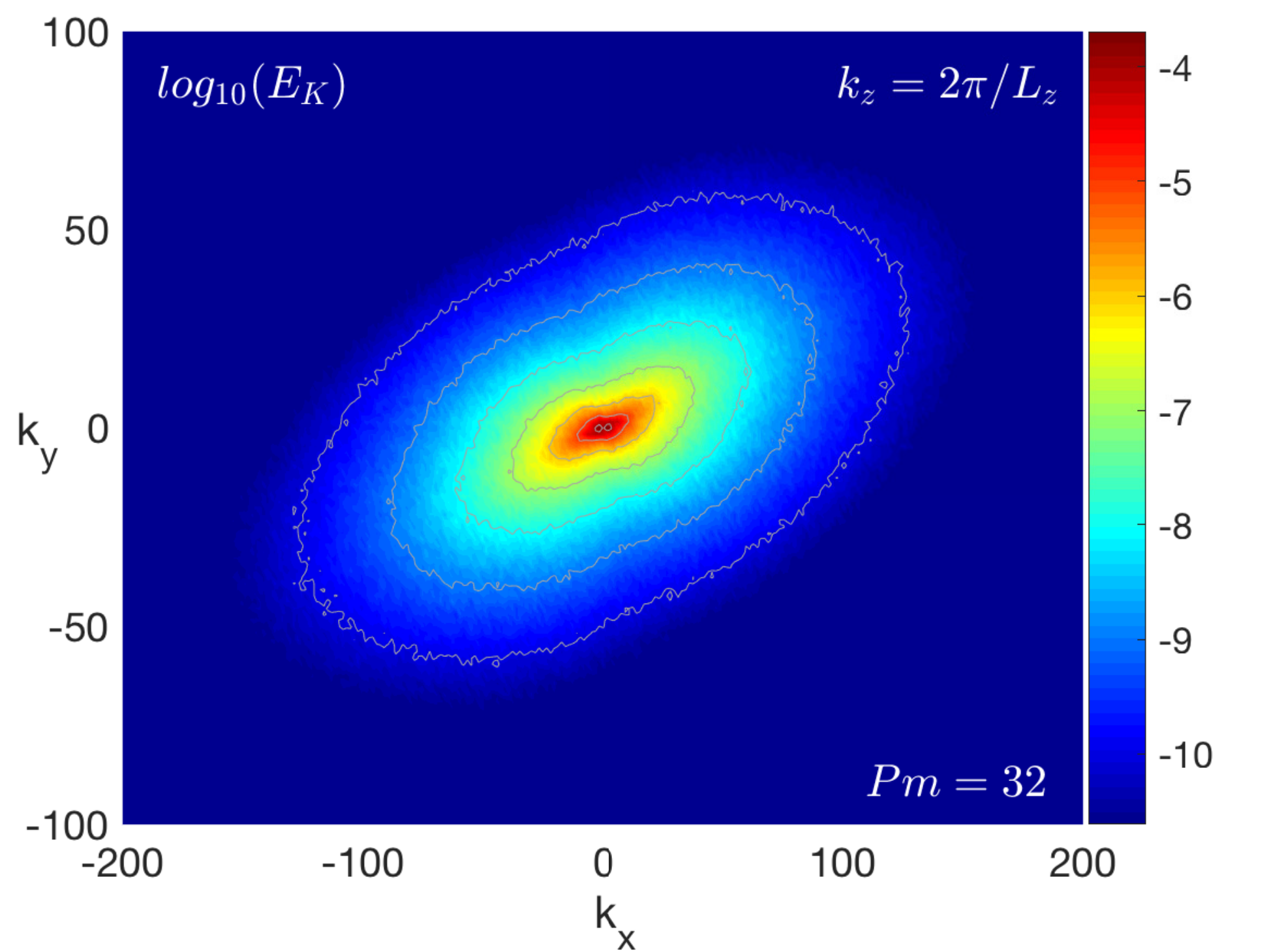}
\includegraphics[scale=0.45]{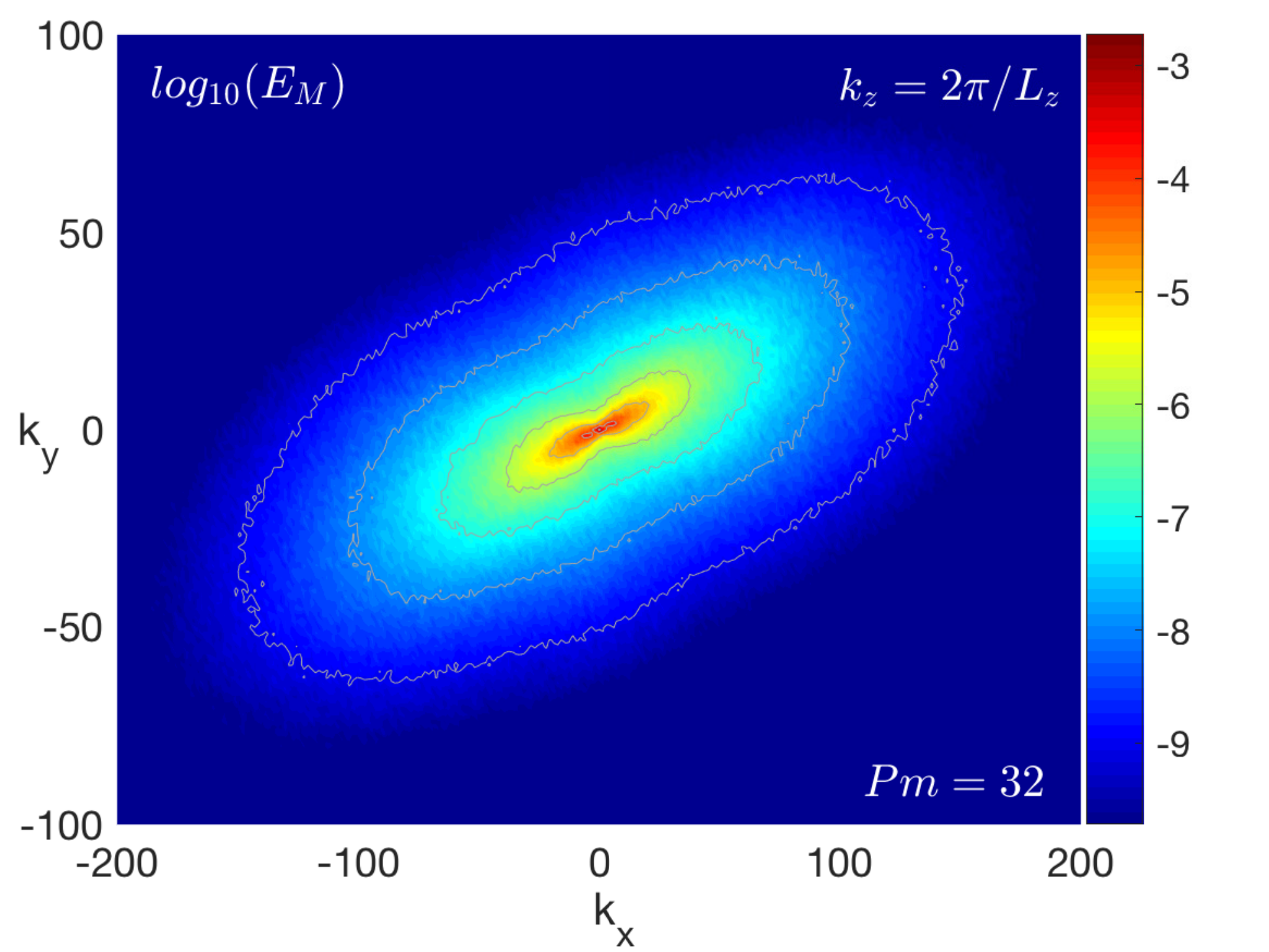}
\caption{Logarithm of the spectral kinetic (top) and magnetic (bottom) energies in $(k_x,k_y)$-slice at $k_z=2\pi/L_z$ for ${\rm Pm}=32$ and ${\rm Rm}=2.5\times10^4$. Due to the background shear, these spectra are noticeably anisotropic, i.e. they depend on the wavevector polar angle, having larger power at $k_x/k_y > 0$ for a given $k_y$.}\label{FIGURE_2DEkin_Emag_Pm32_kz1}
\end{figure}
\begin{figure}
\includegraphics[scale=0.45]{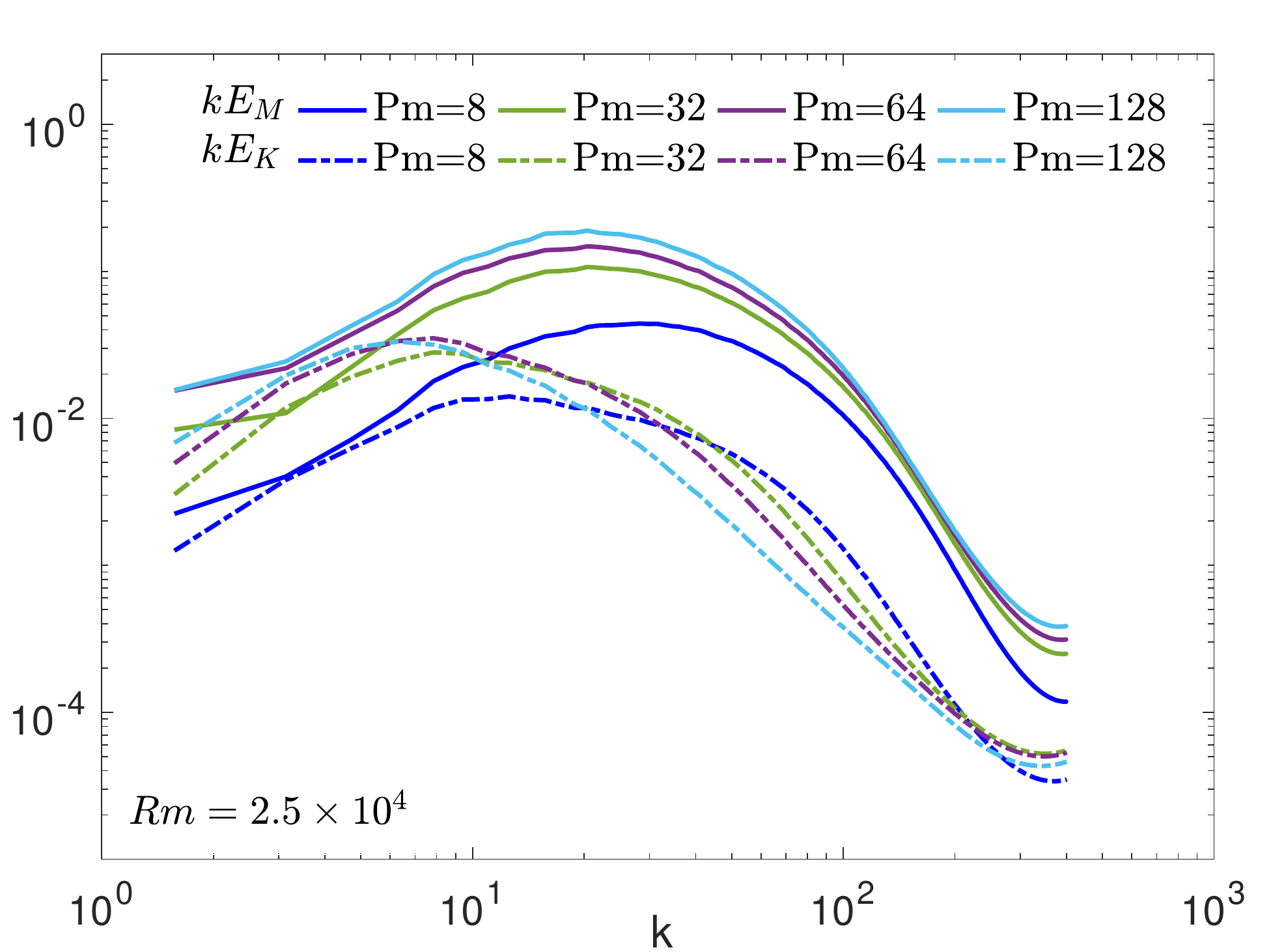}
\caption{Comparison of the compensated shell-averaged kinetic and magnetic energy spectra at different ${\rm Pm}=8,32,64,128$ and fixed ${\rm Rm} = 2.5\times10^4$. As ${\rm Pm}$ increases, the spectral kinetic energy increases at small $k$, but decreases at intermediate and higher $k$ due to increased viscosity. The spectral magnetic energy, however, increases at all $k$, with the peak shifting to lower $k$, till ${\rm Pm}\approx 64$ and then appears to converge at higher ${\rm Pm}=128$ corresponding to the plateau in the $\alpha-{\rm Pm}$ relationship (Figure \ref{FIGURE_AlphaPmRelationship}).}
\label{FIGURE_1DShellAveragedSpectra}
\end{figure}

\subsection{Governing equations in spectral space}
\label{RESULTS_GoverningEquationsInSpectralSpace}
To gain a better understanding of the behaviour of MRI-turbulence with viscous and resistive dissipation described in the last section, we perform a detailed study of its spectral properties and dynamics in 3D Fourier (${\bf k}$-)space by computing and analysing the individual linear and non-linear terms in the main equations at different ${\rm Pm}$ and ${\rm Rm}$, using the data from select simulations described in the previous section (see also Table \ref{TABLE_LargeBoxSimulations}). For this purpose, we employ the approach of \cite{gogichaishvili2017} and \cite{mamatsashvili2020zero} and first decompose the velocity and magnetic field perturbations into spatial Fourier modes
\begin{equation}\label{Fourier}
f({\bf r},t)=\int \bar{f}({\bf k},t)\exp\left({\rm i}{\bf
    k}\cdot{\bf r} \right)d^3{\bf k}, 
\end{equation}
where $f\equiv (\bar{\bf u}, \bar{\bf B})$ and $\bar{f}$ are the corresponding Fourier transforms. Here and below, we consider the velocity deviation from the background Keplerian flow, ${\bf u}\rightarrow {\bf u}+q\Omega x{\bf e}_y$. For the finite simulation box, the grid in Fourier space is determined by the box sizes $L_i$ and numerical resolution $N_i$,
with $i=x,y,z$, so that the cell sizes are given by $\Delta k_i=2\pi/L_i$ and hence the wavenumbers run through values $k_i=n_i\Delta k_i$, where $n_i=0, ±1, ±2,..,±N_i/2$ is integer. 

\begin{figure*}
\centering
\includegraphics[scale=0.295]{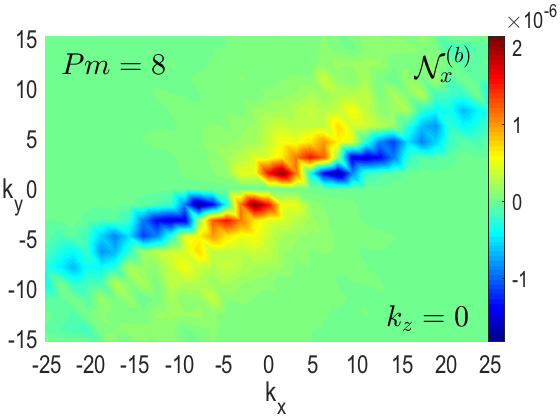}
\includegraphics[scale=0.295]{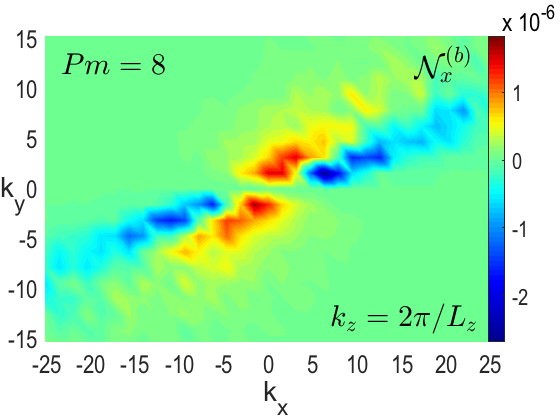}
\includegraphics[scale=0.295]{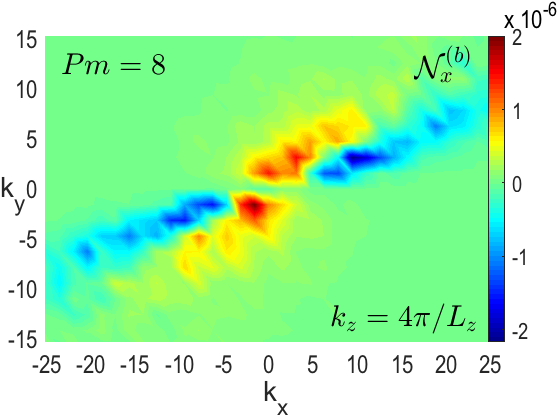}
\includegraphics[scale=0.295]{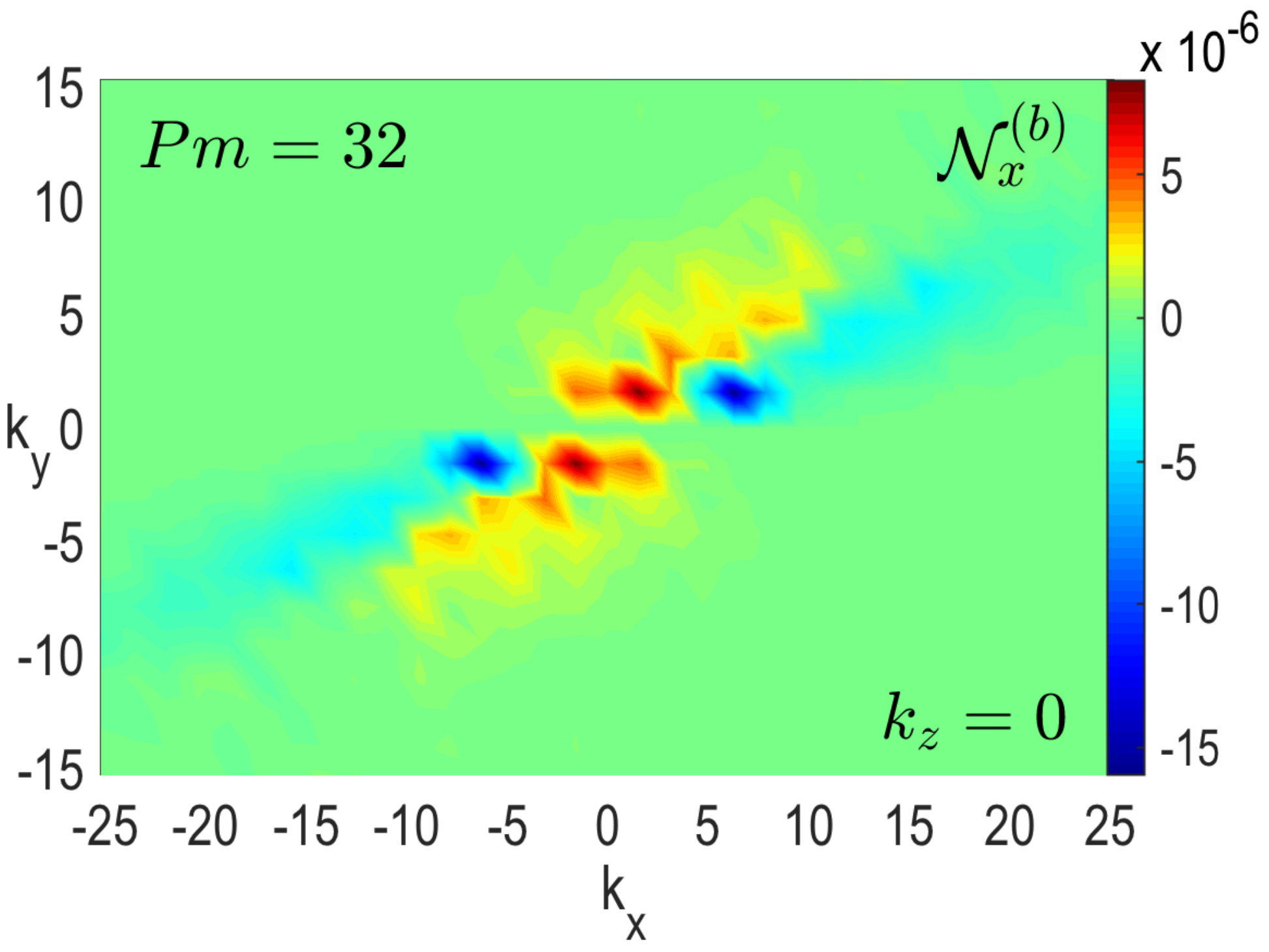}
\includegraphics[scale=0.295]{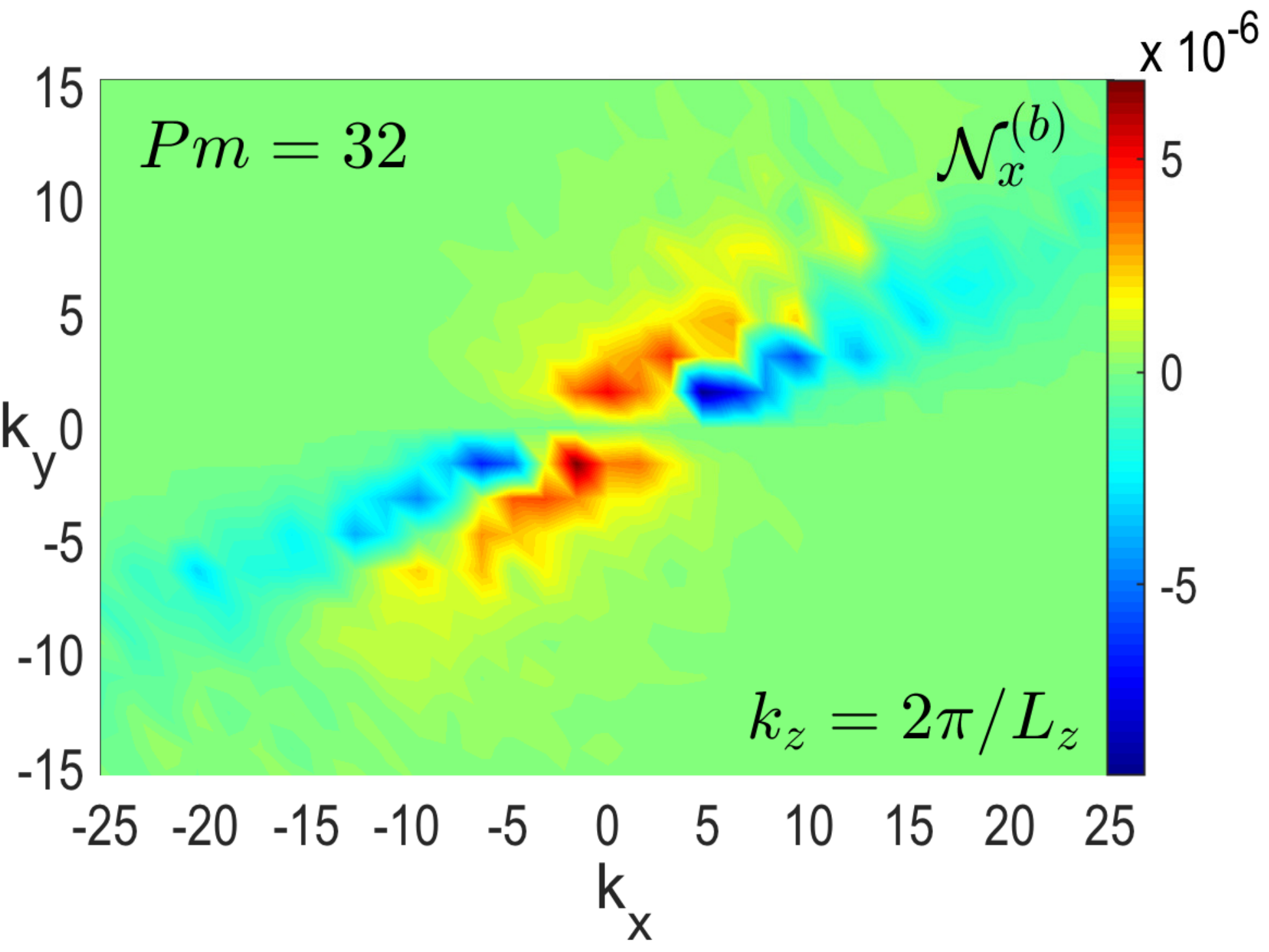}
\includegraphics[scale=0.295]{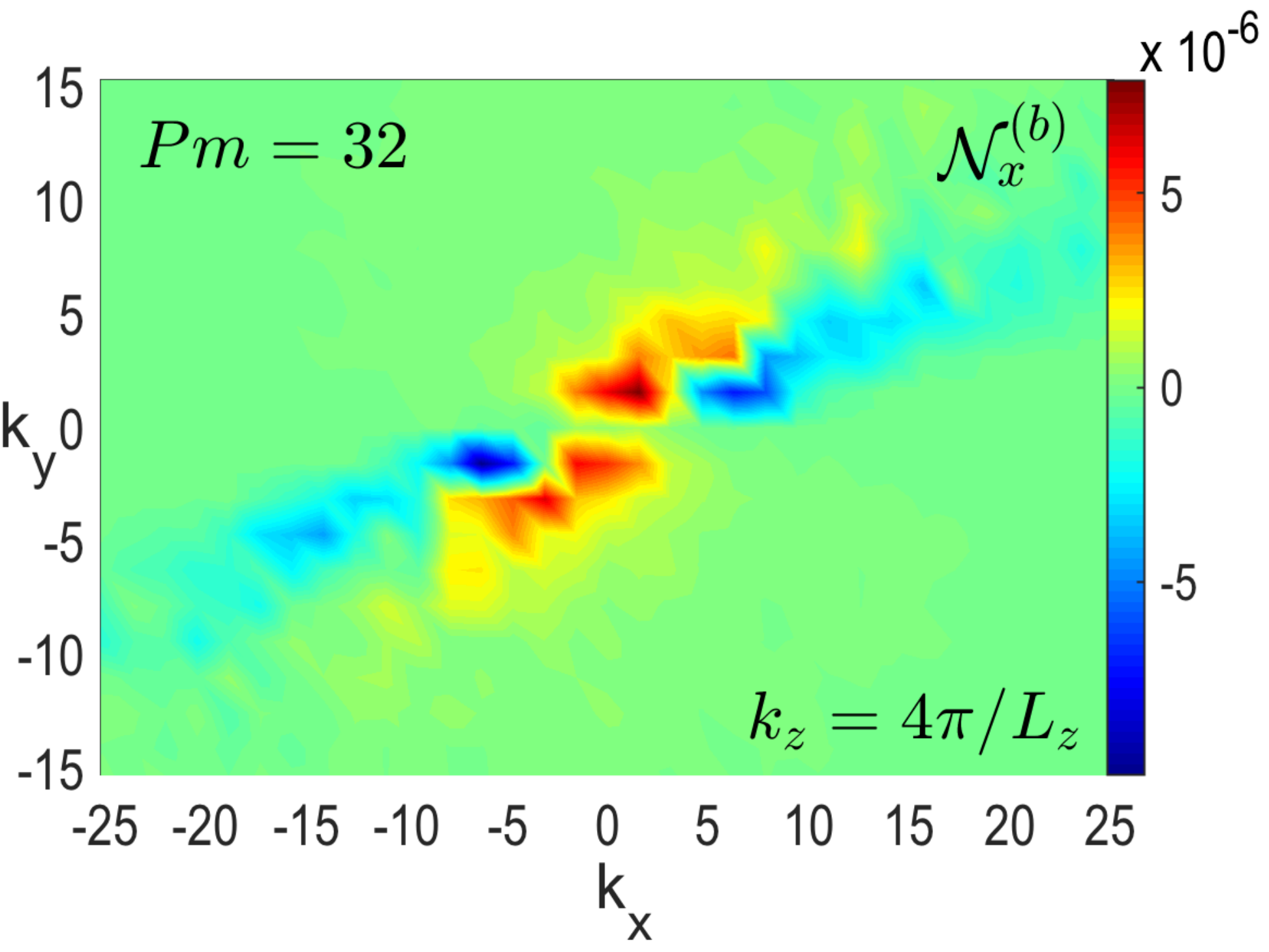}
\includegraphics[scale=0.295]{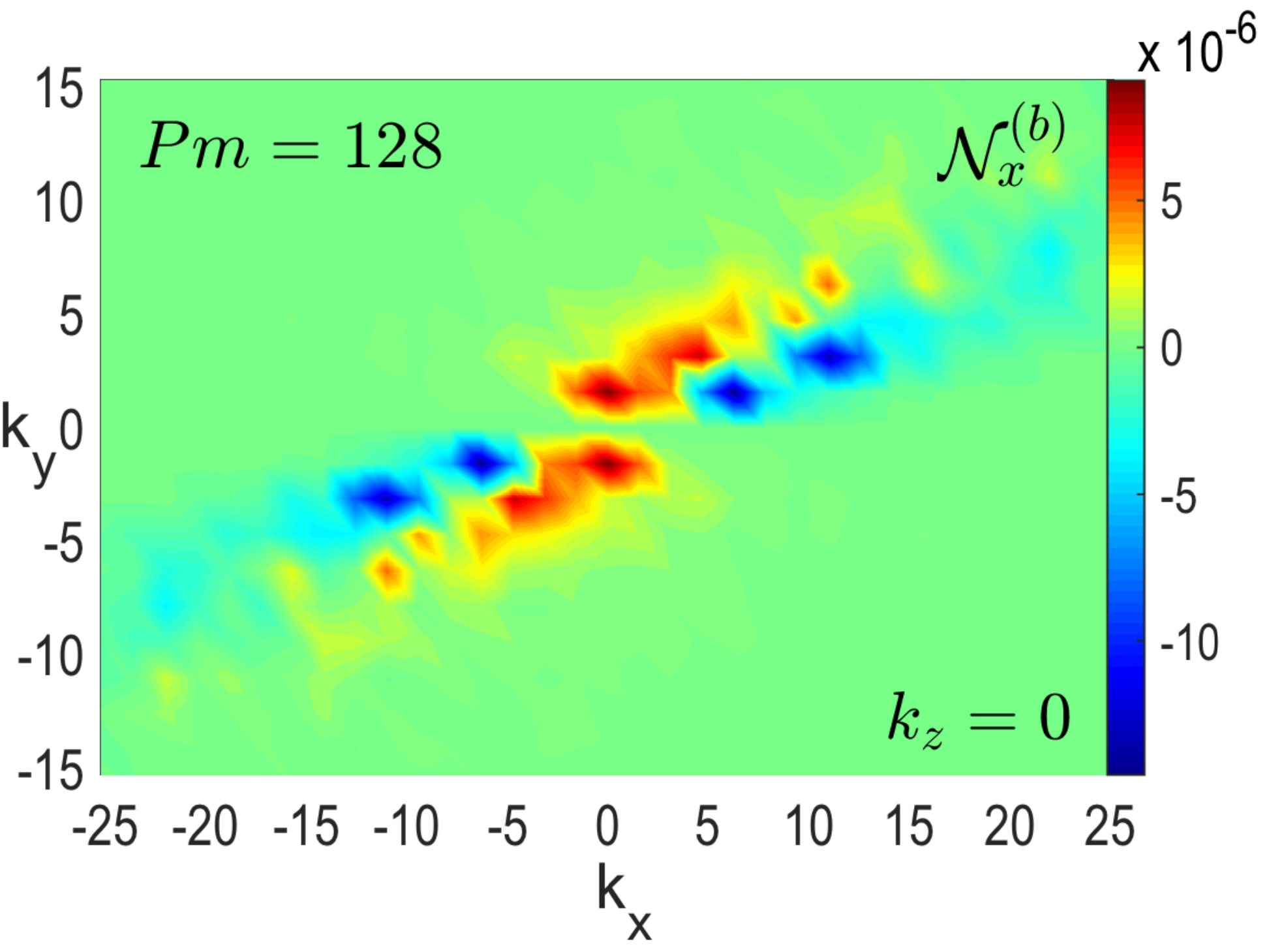}
\includegraphics[scale=0.295]{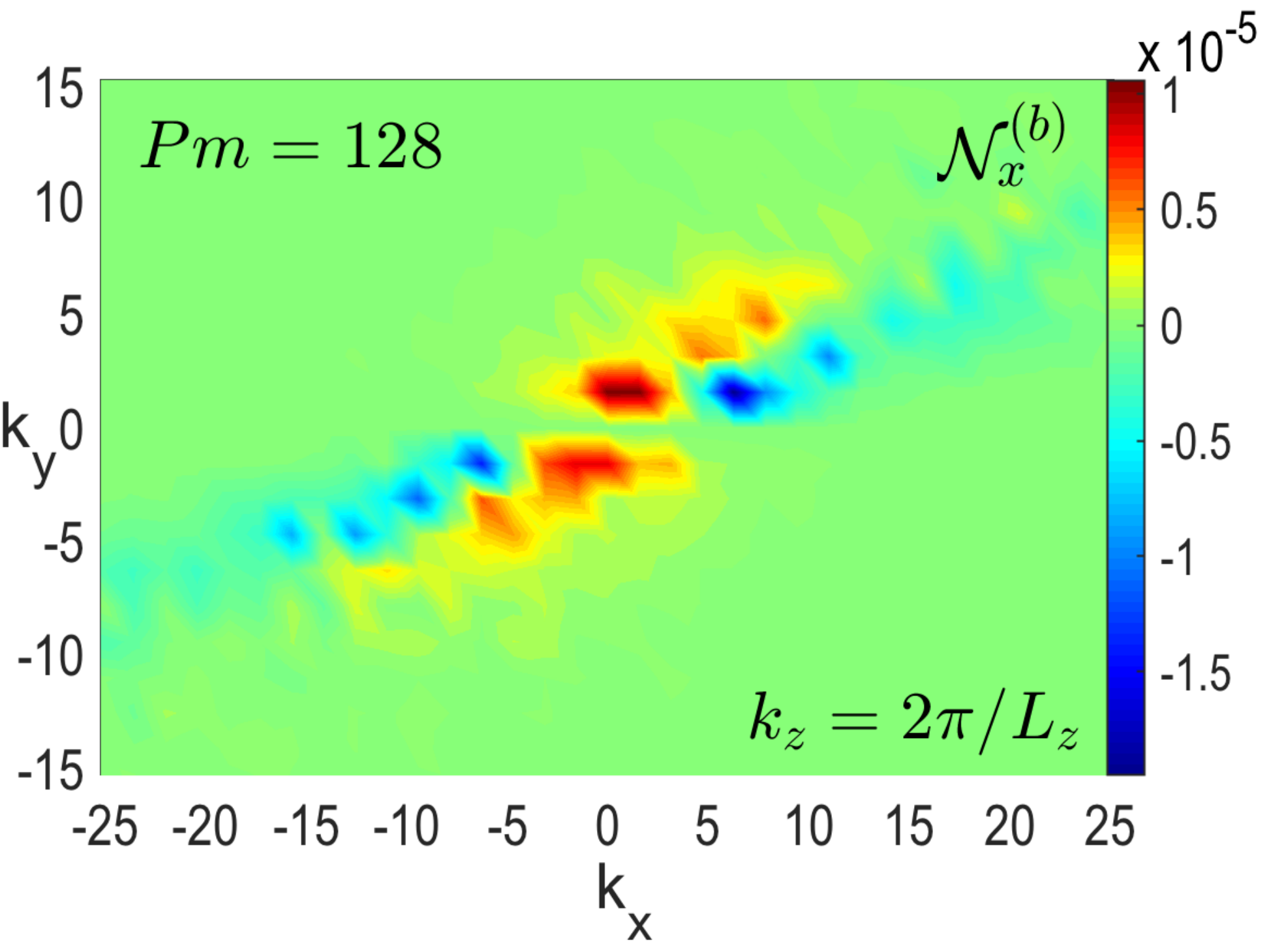}
\includegraphics[scale=0.295]{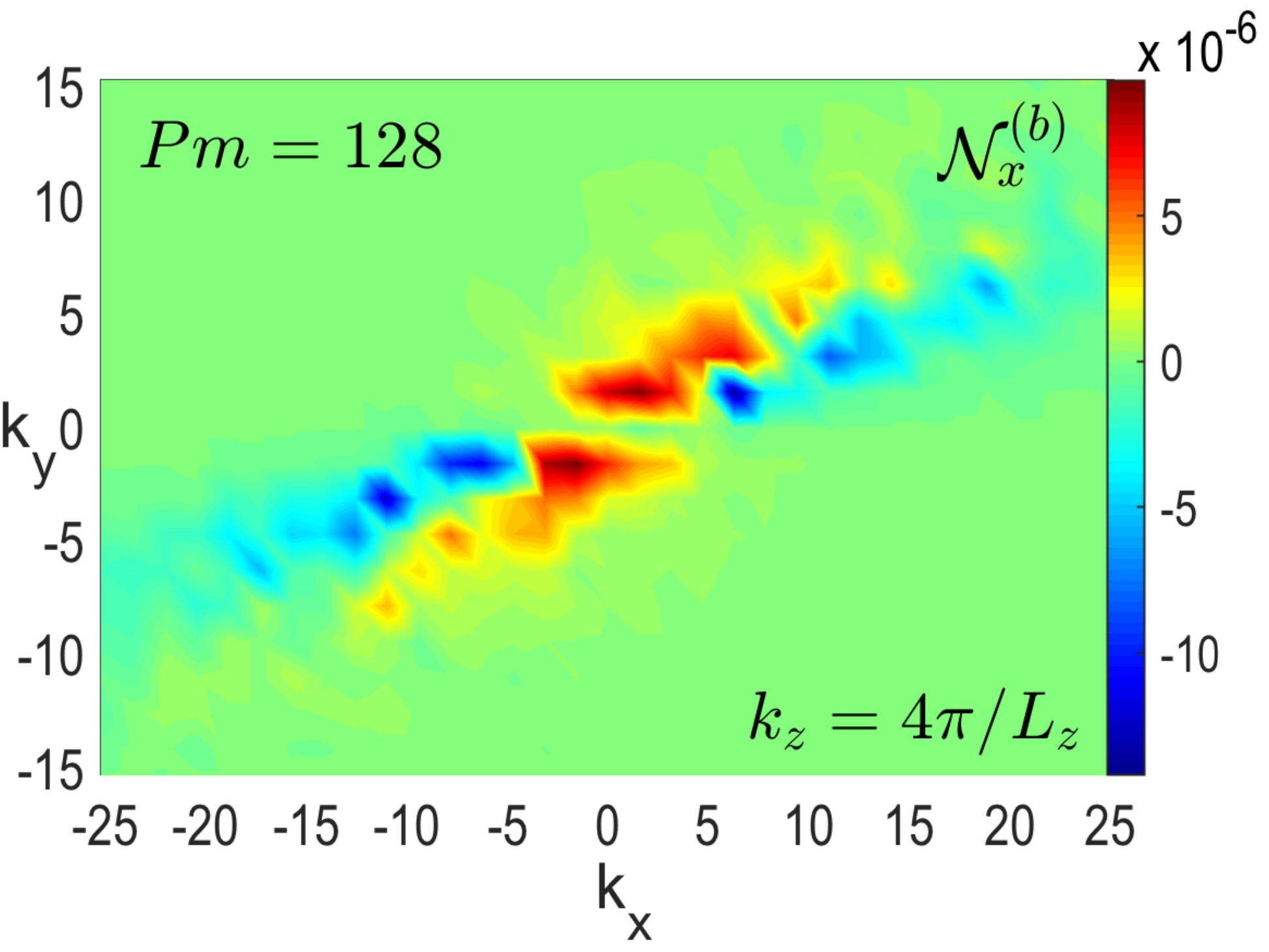}
\caption{Spectrum of the non-linear transfer term ${\cal N}_x^{(b)}$ for the radial magnetic field in the $(k_x, k_y)$-plane at $k_z=0$ (left), $2\pi/L_z$ (middle), $4\pi/L_z$ (right) and different ${\rm Pm}=8$ (top), 32 (middle), 128 (bottom) and a fixed ${\rm Rm}=2.5\times10^4$. It exhibits a characteristic anisotropy (i.e. variation over the wavevector orientation/angle) due to the disc flow shear -- the transverse cascade, which primarily redistributes/transfers $|B_x|^2$ from ``giver'' wavenumbers where ${\cal N}_x^{(b)}<0$ (blue) to ``receiver'' wavenumbers where ${\cal N}_x^{(b)}>0$ (yellow and red), draining and supplying the radial field energy in these regions, respectively. This non-linear transverse redistribution in the $(k_x, k_y)$-plane plays a key role, since it continually replenishes (seeds) the radial field at wavenumbers within the yellow/red regions, which thus represent the growth, or vital area. Note that with increasing ${\rm Pm}$, the action of this term becomes somewhat more concentrated at lower wavenumbers.}
\label{FIGURE_2DSpectraOfNxb}
\end{figure*}

The spectral kinetic, $E_K$, and magnetic, $E_M$, energy densities are defined in the usual way \citep[e.g.,][]{simon2009},
\begin{equation}
E_K=\frac{1}{2}\left(|\overline{(\sqrt{\rho} u_x})|^2+|\overline{(\sqrt{\rho} u_y)}|^2+|\overline{(\sqrt{\rho} u_z)}|^2\right),
\label{EQN_SpectralKE}
\end{equation}
\begin{equation}
E_M=\frac{1}{2}\left(|\bar{B}_x|^2+|\bar{B}_y|^2+|\bar{B}_z|^2\right).
\label{EQN_SpectralME}
\end{equation}
Since MRI-turbulence is primarily magnetically driven, we focus on the dynamics of the magnetic field and, in particular, on its radial, $\bar{B}_x$, and azimuthal, $\bar{B}_y$, components. These components are the most important ones in the turbulent dynamics as they make up the Maxwell stress, which in turn acts to sustain turbulence by extracting energy from the disc flow by means of the MRI. We derive evolution equations for these two components in Fourier space. Substituting equation (\ref{Fourier}) into the induction equation (\ref{SB4}) and multiplying the left- and right-hand sides by the complex conjugate $\bar{\bf B}^{\ast}$, we obtain spectral equations governing the evolution of the quadratic forms of the Fourier amplitudes of the radial and azimuthal fields in the non-dimensional form \citep{mamatsashvili2020zero},
\begin{equation}\label{eq:bxk2} 
\frac{\partial}{\partial
    t}\frac{|\bar{B}_x|^2}{2}=-qk_y\frac{\partial}{\partial k_x}
\frac{|\bar{B}_x|^2}{2} +{\cal D}_x^{(b)}+{\cal
    N}^{(b)}_x
\end{equation}
\begin{equation}\label{eq:byk2}
\frac{\partial}{\partial
    t}\frac{|\bar{B}_y|^2}{2}=-qk_y\frac{\partial}{\partial k_x}
\frac{|\bar{B}_y|^2}{2}+{\cal M}+{\cal
    D}_y^{(b)}+{\cal N}^{(b)}_y.
\end{equation}
These equations, which are central to our analysis, contain different terms of linear and non-linear origin. The linear terms on the right hand side of equations (\ref{eq:bxk2}) and (\ref{eq:byk2}) are:
\begin{enumerate}
\item 
The shear-induced drift, $q k_y\partial/\partial k_x$, which merely advects the magnetic field amplitudes for non-axisymmetric ($k_y\neq 0$) modes along the $k_x$-axis (for $k_y>0$) or opposite the $k_x$ axis (for $k_y<0$) at a speed $q |k_y|$ without producing new energy for these modes,
\item
The Maxwell stress spectrum multiplied by the shear parameter $q$,
\begin{equation}\label{eq:M}
{\cal M}=-\frac{q}{2}(\bar{B}_x\bar{B}_y^{\ast}+\bar{B}_x^{\ast}\bar{B}_y),
\end{equation}
which describes, in general, the energy exchange process between perturbed modes and the background disc flow, and, in the present case, growth of magnetic field perturbations due to the MRI, which is driven by the flow shear. This linear exchange process represents the main energy supplier for the MRI-turbulence,
\item
the negative-definite standard Ohmic dissipation,
\begin{equation}\label{eq:Dbi}
{\cal D}_i^{(b)}=-\frac{k^2}{\rm Rm}|\bar{B}_i|^2, ~~~i=x,y.
\end{equation}
The last terms, ${\cal N}_x^{(b)}$ and ${\cal N}_y^{(b)}$, on the right-hand side of Equations \ref{eq:bxk2} and \ref{eq:byk2}, respectively, represent the \textit{transfer functions} of the magnetic field components via non-linear triadic interactions of modes
\begin{eqnarray}\label{eq:Nbx}
{\cal N}^{(b)}_x=\frac{\rm
    i}{2}\bar{B}_x^{\ast}[k_y\bar{F}_z-k_z\bar{F}_y]+c.c., \\
{\cal N}^{(b)}_y=\frac{\rm
    i}{2}\bar{B}_y^{\ast}[k_z\bar{F}_x-k_x\bar{F}_z]+c.c., 
\end{eqnarray}
where $\bar{F}_x, \bar{F}_y, \bar{F}_z$ are the Fourier transforms of the respective components of the perturbed electromotive force ${\bf F}={\bf u}\times {\bf B}$, which are given mathematically by convolutions in Fourier space. These convolutions describe the net effect of the non-linear interactions of a given mode ${\bf k}$ with two other modes ${\bf k'}$ and ${\bf k-k'}$
\citep{Alexakis2018}, and hence we refer to the interactions as being `triadic'. The explicit expressions for the electromotive force components in Fourier space are
\begin{equation*}\label{eq:App-Fxk}
\bar{F}_x({\bf k},t)=\int d^3{\bf k'}\left[\bar{u}_y({\bf
k'},t)\bar{B}_z({\bf k}-{\bf k'},t)-\bar{u}_z({\bf
k'},t)\bar{B}_y({\bf k}-{\bf k'},t)\right],
\end{equation*}
\begin{equation*}\label{eq:App-Fyk}
\bar{F}_y({\bf k},t)=\int d^3{\bf k'}\left[\bar{u}_z({\bf
k'},t)\bar{B}_x({\bf k}-{\bf k'},t)-\bar{u}_x({\bf
k'},t)\bar{B}_z({\bf k}-{\bf k'},t)\right],
\end{equation*}
\begin{equation*}\label{eq:App-Fzk}
\bar{F}_z({\bf k},t)=\int d^3{\bf k'}\left[\bar{u}_x({\bf
k'},t)\bar{B}_y({\bf k}-{\bf k'},t)-\bar{u}_y({\bf
k'},t)\bar{B}_x({\bf k}-{\bf k'},t)\right].
\end{equation*}
Note that these non-linear transfers do \textit{not} produce net new energy for the turbulence (instead, energy is extracted from the background shear by the MRI by means of the Maxwell stress term $\mathcal{M}$ as mentioned above). Rather, the non-linear transfer terms redistribute power among modes in Fourier space, as well as between the magnetic field and the velocity field, while leaving the total (magnetic+kinetic) spectral energy integrated over all wavenumbers unchanged. Nevertheless, these non-linear transfers play a key role in the self-sustenance and dynamics of MRI-turbulence: they continually replenish MRI-active modes via the transverse cascade process \citep{gogichaishvili2017, mamatsashvili2020zero}, which we will describe in more detail in the next subsection. The behavior of these non-linear transfer terms with ${\rm Pm}$ and ${\rm Rm}$, in turn, determines the dependence of the characteristics of MRI-turbulence on these numbers \citep{riols2015dissipative,riols2017magnetorotational,mamatsashvili2020zero}.
\end{enumerate}

\begin{figure*}
\centering
\includegraphics[scale=0.295]{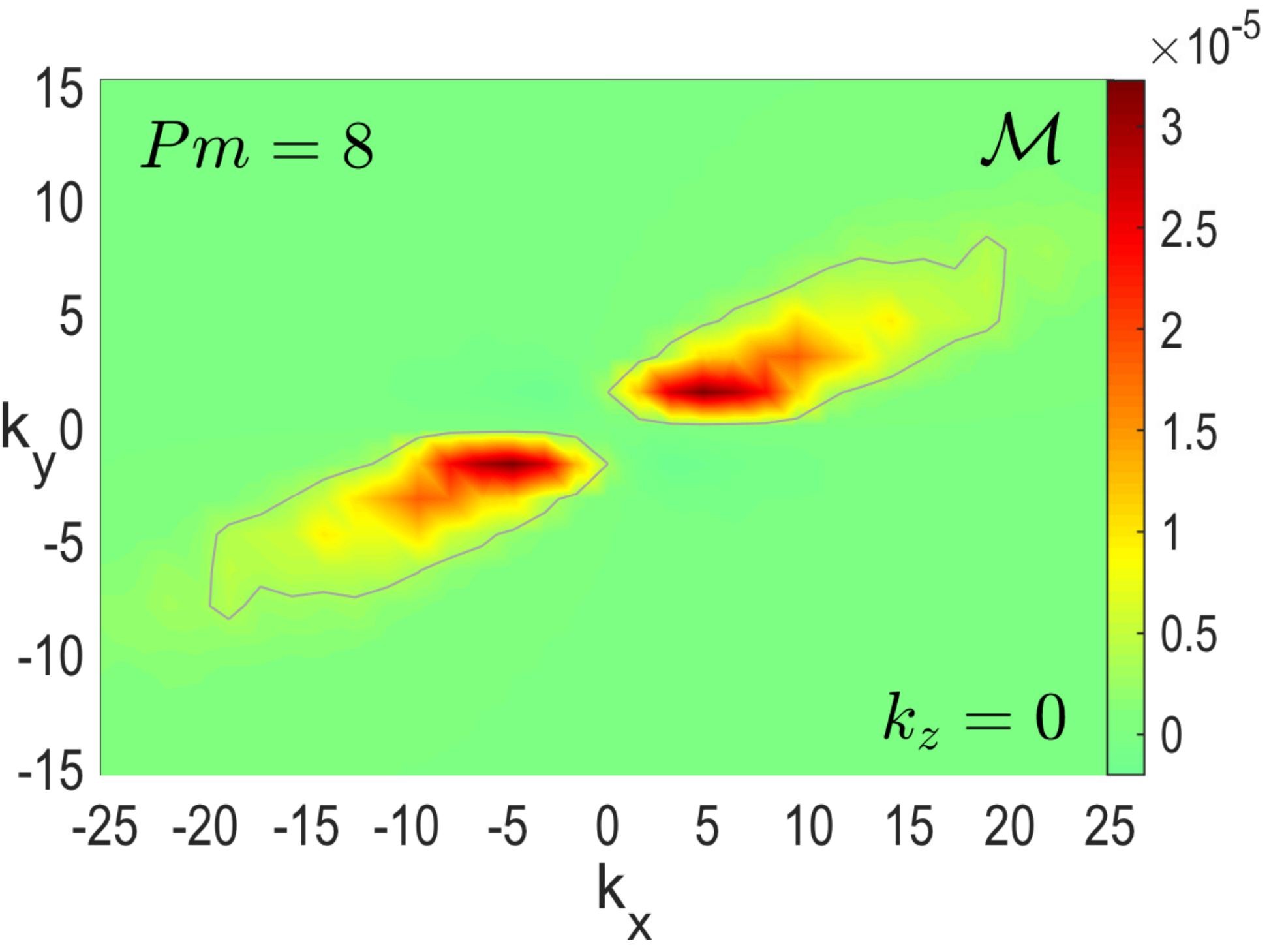}
\includegraphics[scale=0.295]{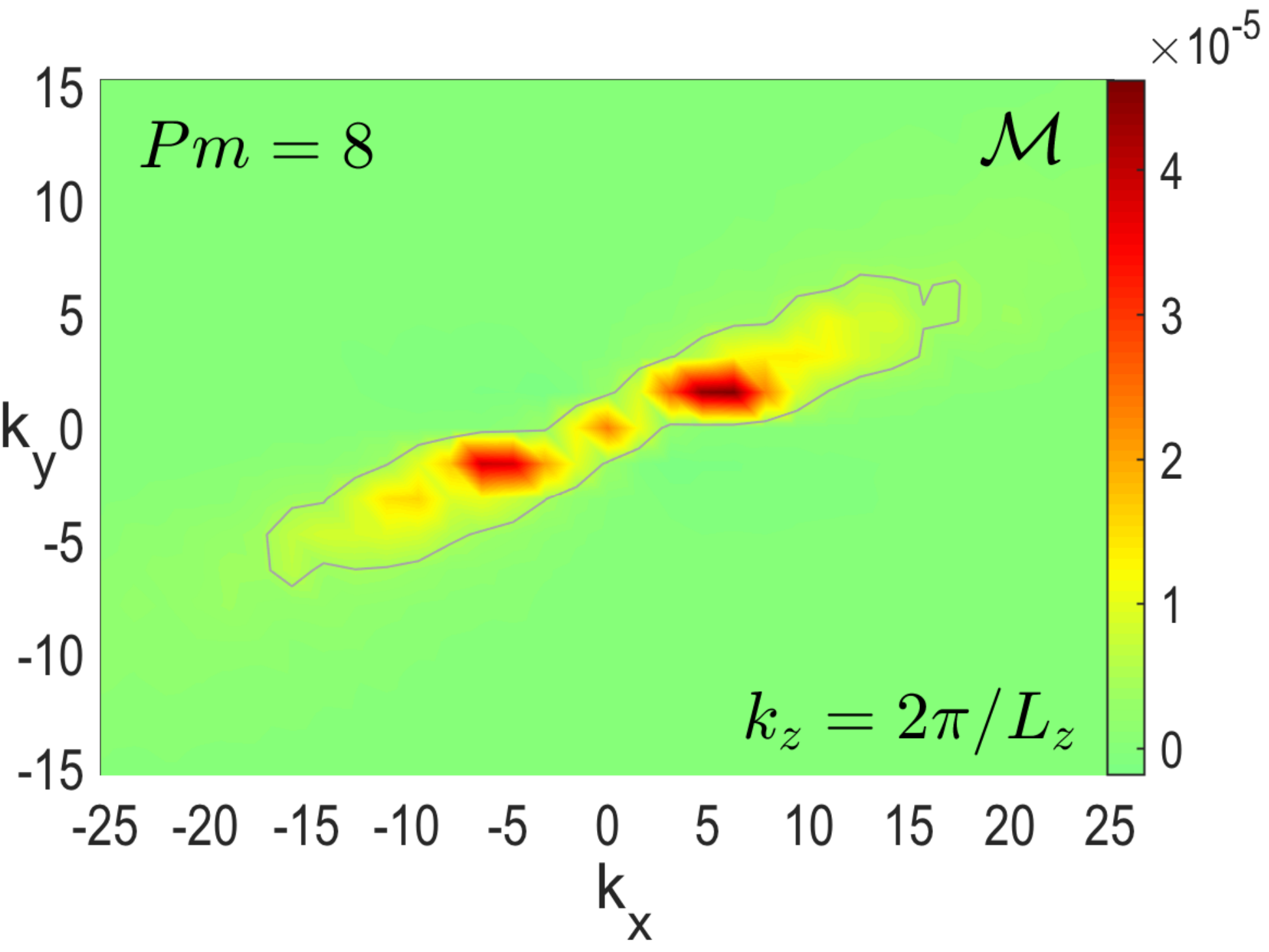}
\includegraphics[scale=0.295]{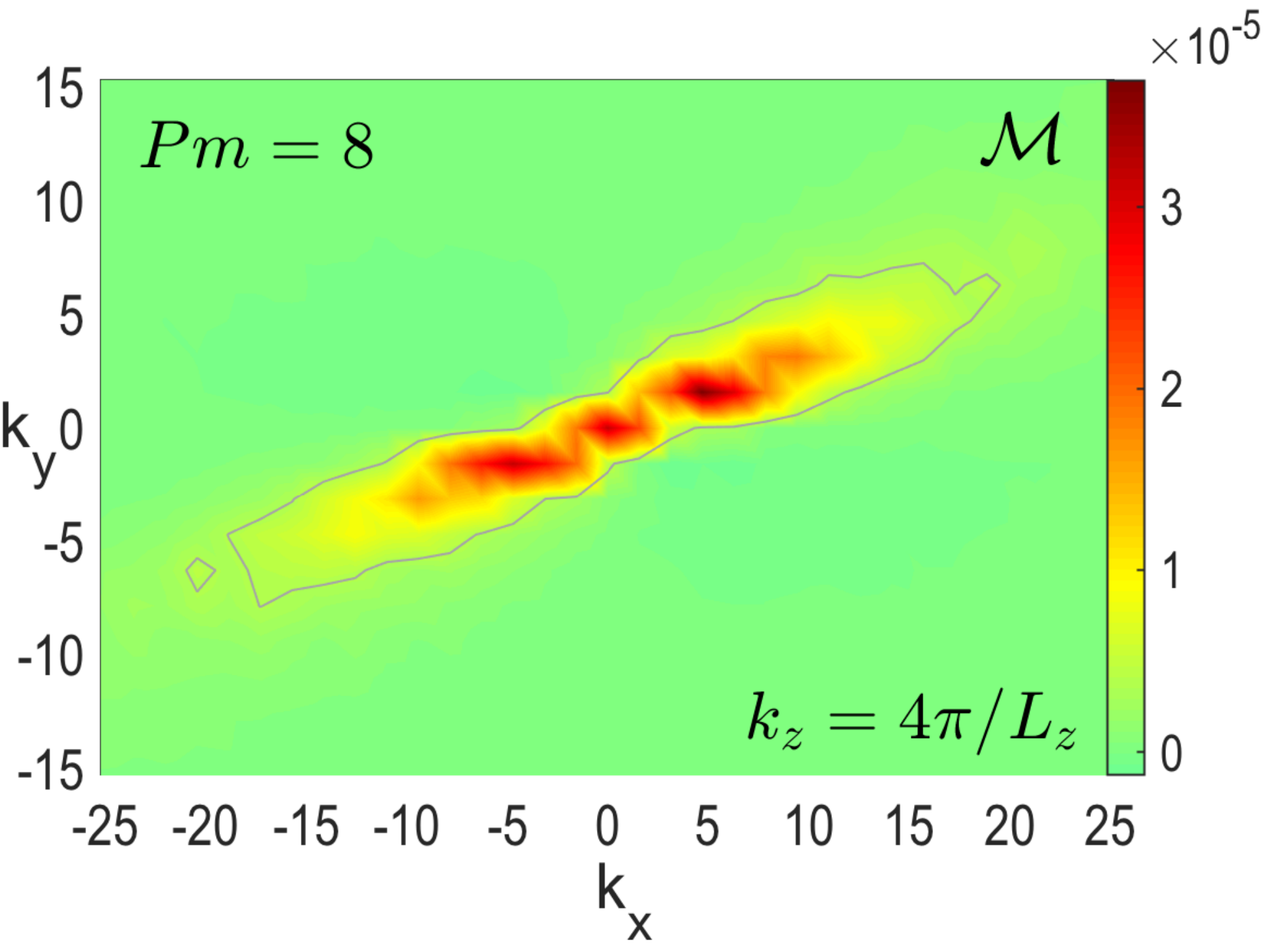}
\includegraphics[scale=0.295]{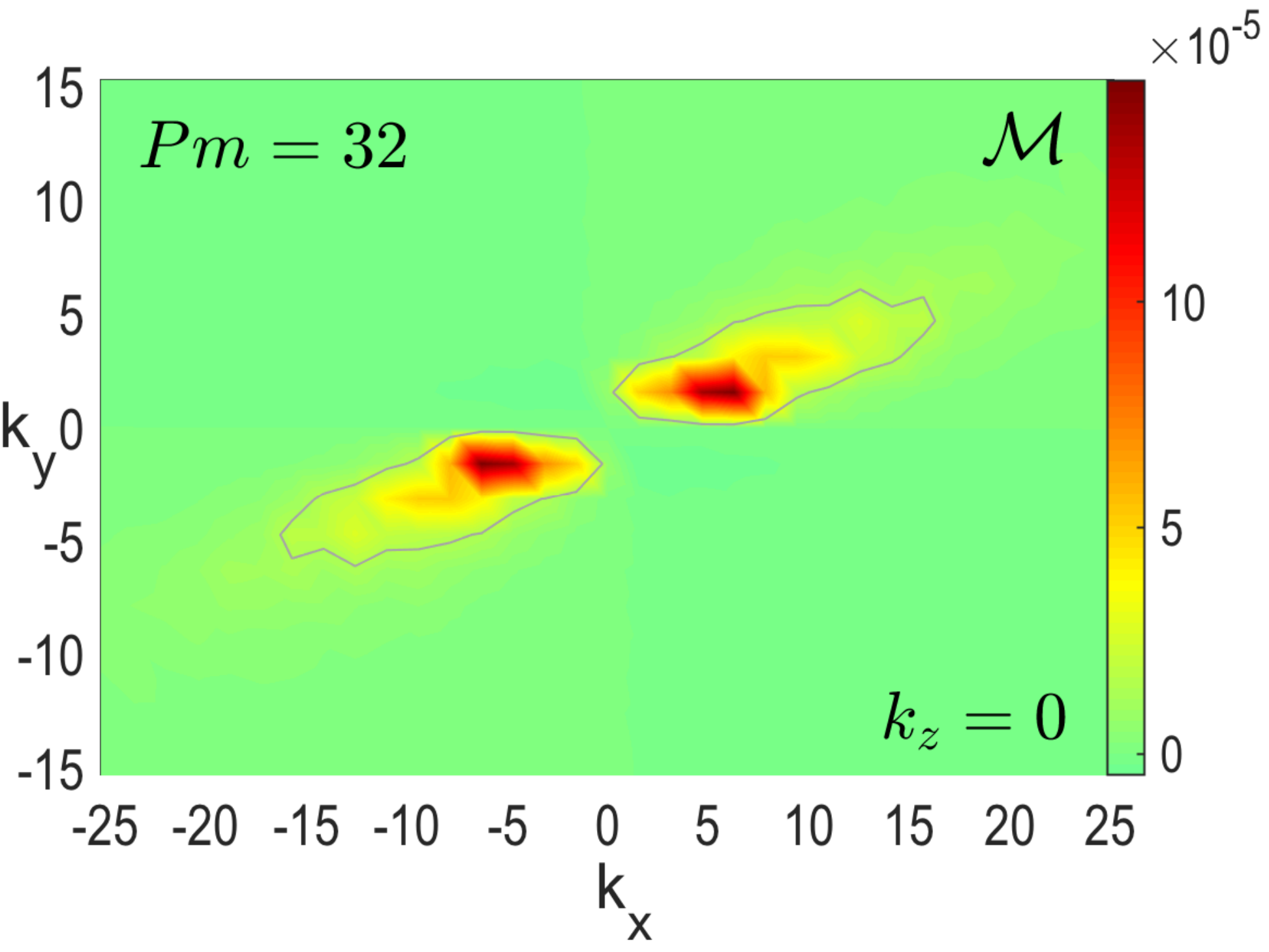}
\includegraphics[scale=0.295]{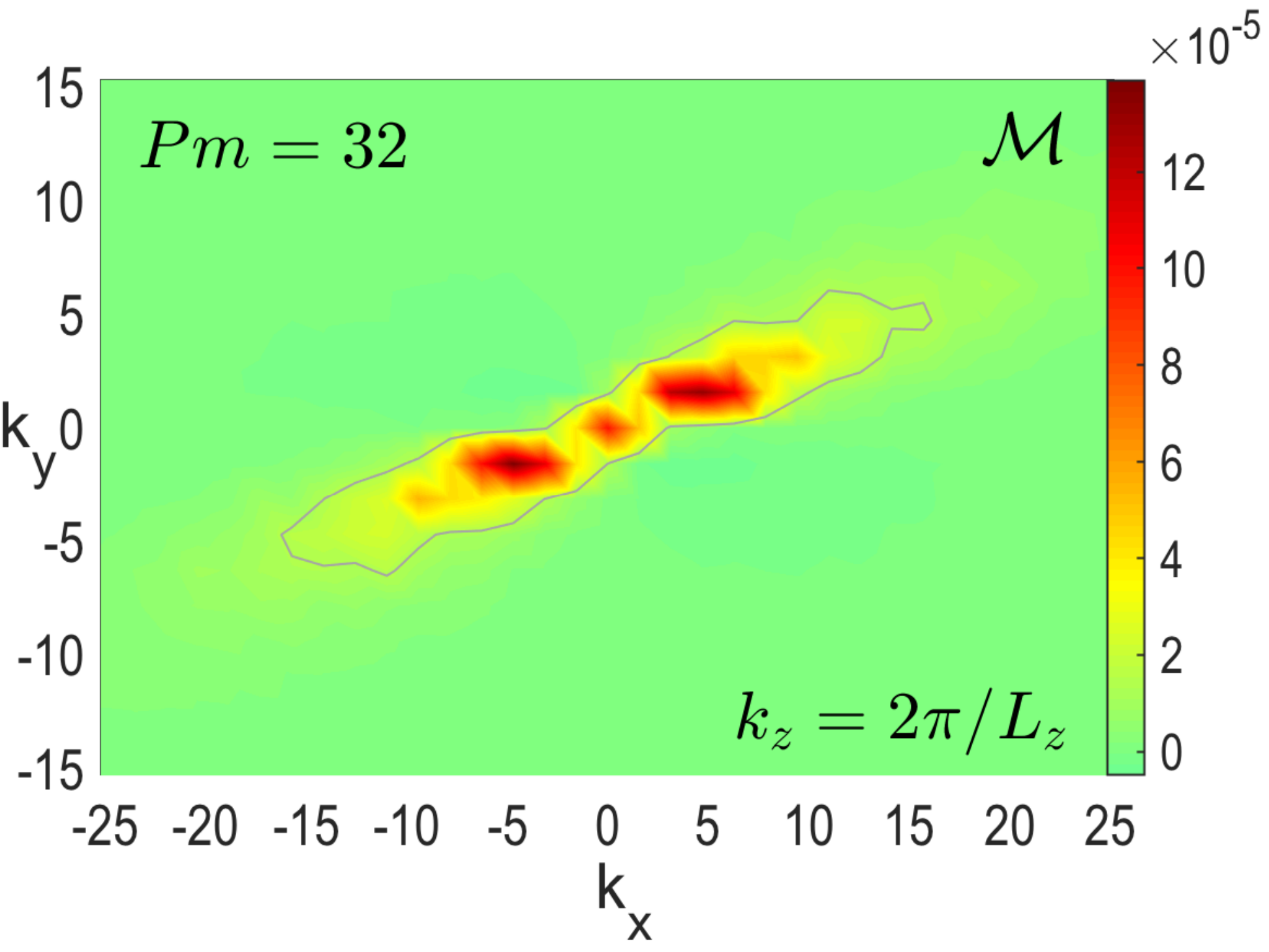}
\includegraphics[scale=0.295]{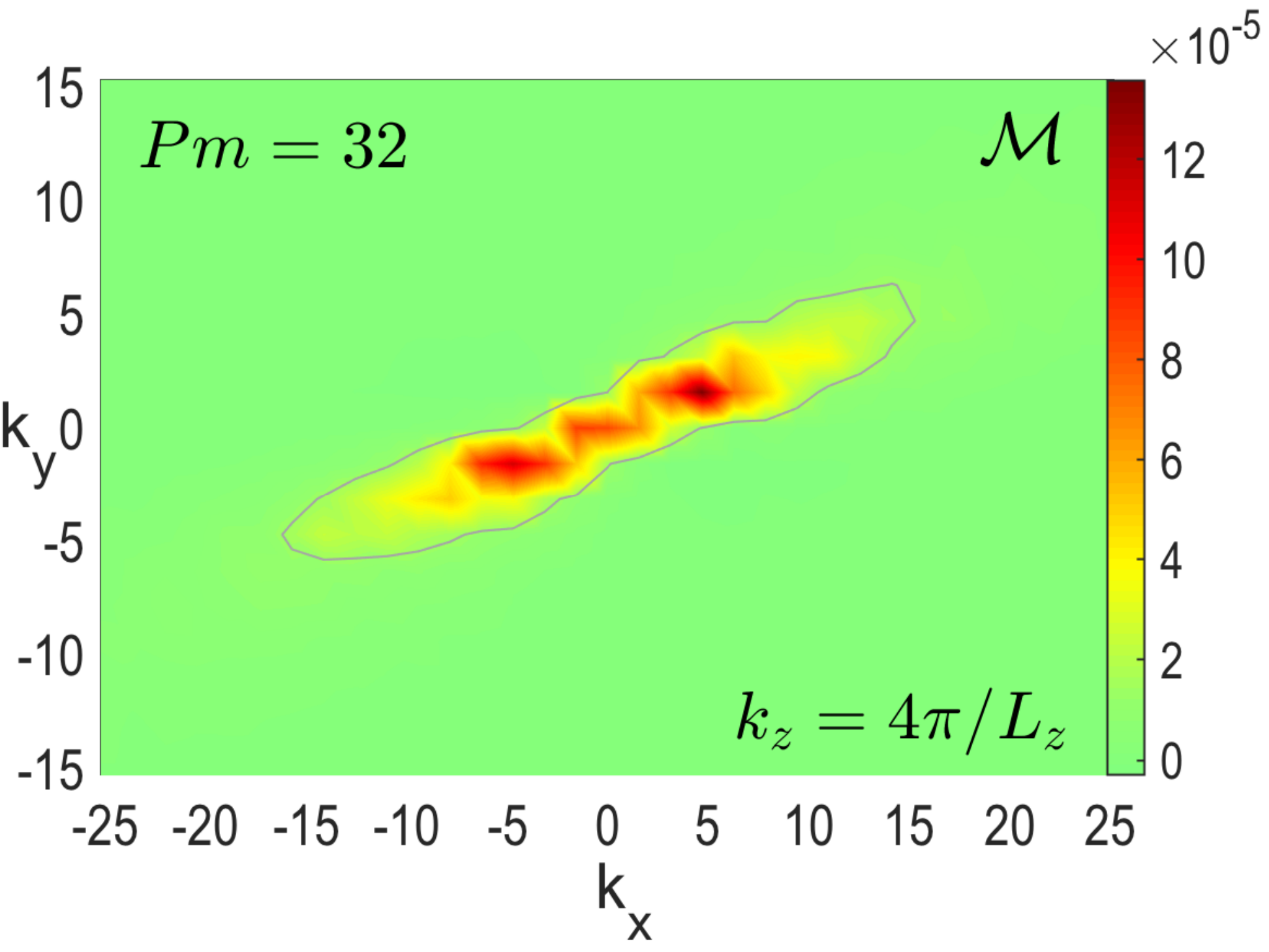}
\includegraphics[scale=0.295]{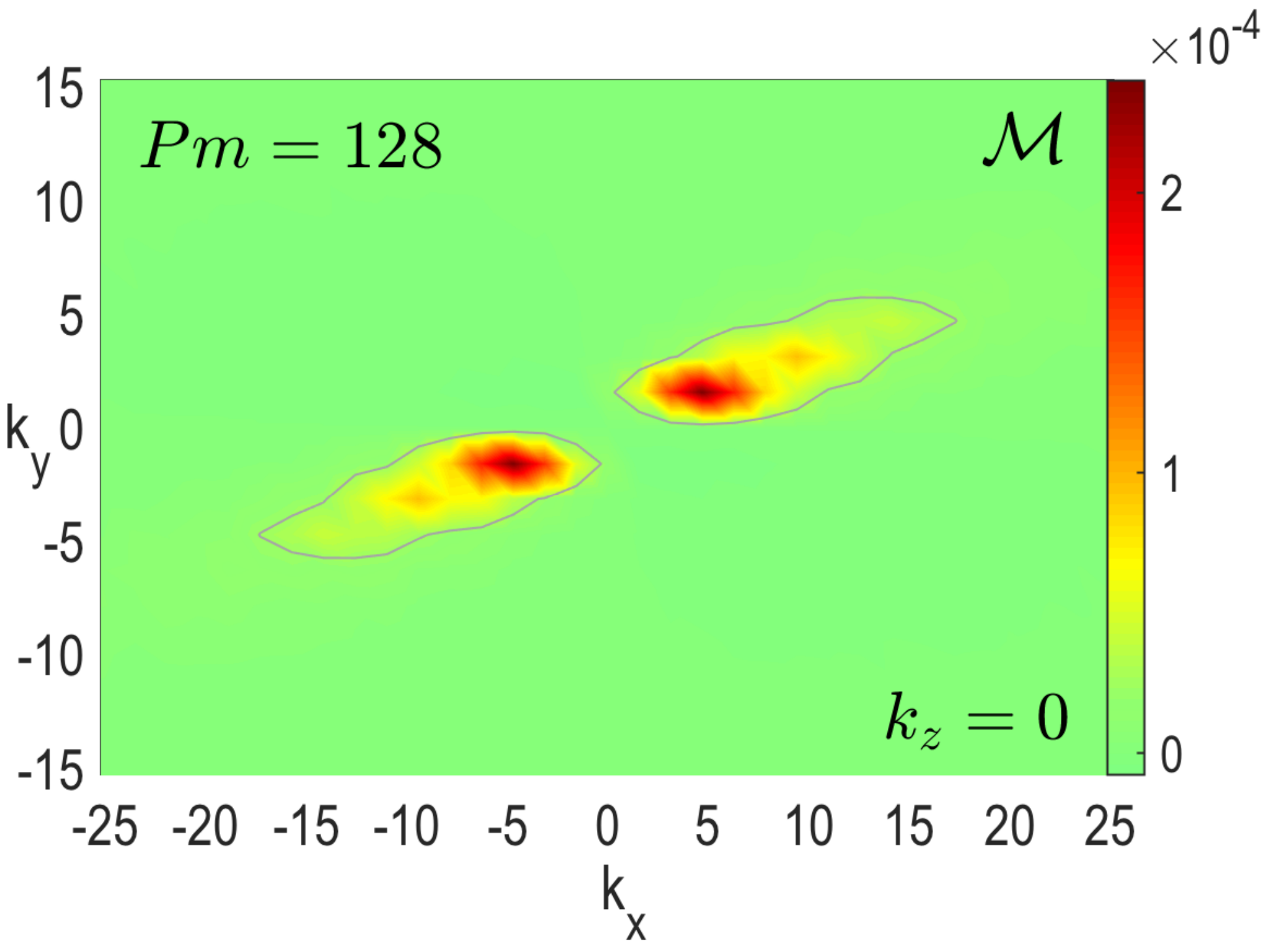}
\includegraphics[scale=0.295]{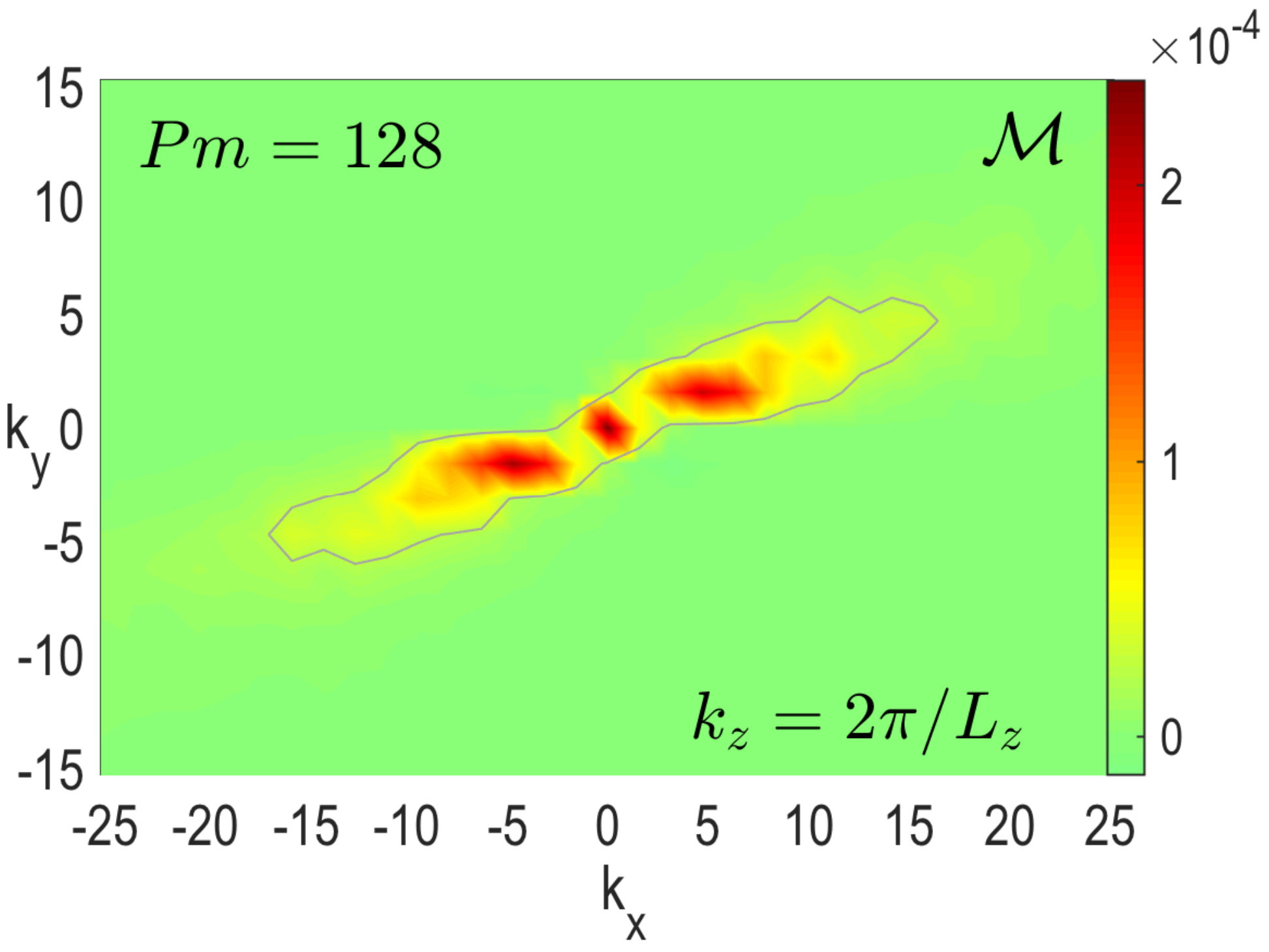}
\includegraphics[scale=0.295]{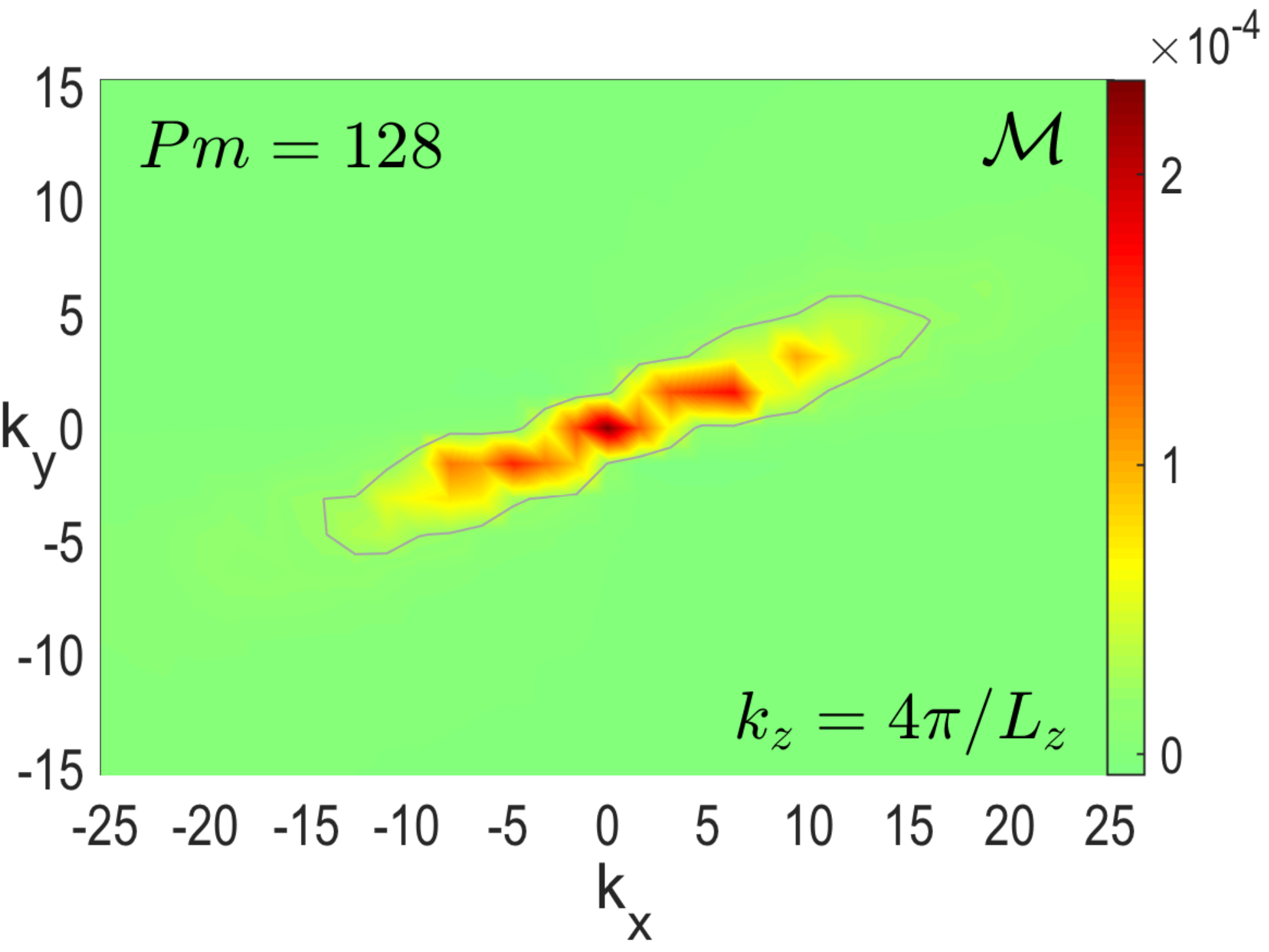}
\caption{Spectrum of Maxwell stress ${\cal M}$ in the $(k_x,k_y)$-plane at  $k_z=0$ (left), $2\pi/L_z$ (middle), $4\pi/L_z$ (right) and different ${\rm Pm}=8$ (top), 32 (middle), 128 (bottom) and a fixed ${\rm Rm}=2.5\times10^4$. Like the spectra of the non-linear terms and energies shown above, this spectrum also exhibits a similar anisotropy due to shear. Grey isocontours, delineating the central part of the vital area, correspond to the level $0.1{\cal M}_{max}$ for a given $k_z$. In this area, the positive Maxwell stress (yellow and red) ensures energy injection and the amplification of azimuthal field energy $|\bar{B}_y|^2$ from the radial field $\bar{B}_x$ due to the nonmodal MRI-growth process. This radial field, in turn, has been newly regenerated by ${\cal N}_x^{(b)}$ due to the non-linear transverse cascade (Figure \ref{FIGURE_2DSpectraOfNxb}). The mode wavenumbers undergo drift through the vital area where ${\cal M}>0$ and eventually leave it; hence, the amplification of the azimuthal field for them is of transient character. Note also that while the vital area shrinks with increasing ${\rm Pm}$, the Maxwell stress inside it increases.}\label{FIGURE_2DSpectraOfMaxwellStress}
\end{figure*}

\subsection{Energy spectra}
\label{RESULTS_EnergySpectra}
We begin with an analysis of the energy spectra (Equations \ref{EQN_SpectralKE} and \ref{EQN_SpectralME}) (the spectra of the dynamical terms will be discussed in Section \ref{RESULTS_SpectraOfDynamicalTerms}). We average all results in time over the last 100 orbits, by which time the simulation solutions have already settled into quasi-steady state (see Figure \ref{FIGURE_TimeSeriesRes128Rm25k}). Figure \ref{FIGURE_2DEkin_Emag_Pm32_kz1} shows the typical form of the 3D kinetic and magnetic energy spectra in the 2D $(k_x,k_y)$-plane. The 2D slice is taken at the first vertical wavenumber $k_z=2\pi/L_z$, at which these energies are larger than at smaller $k_z$. From the plots it is evident that the spectra are strongly anisotropic (i.e. they vary over wavevector orientation). This anisotropy is due to the flow shear, with trailing ($k_x/k_y > 0$) modes having larger energy for a given $k_y$ than leading ($k_x/k_y < 0$) ones do. This shear-induced anisotropy of the energy spectra is a distinctive feature of MRI-turbulence that has also been demonstrated in previous studies \citep{Lesur_Longaretti2011,Murphy_Pessah2015,gogichaishvili2017}. Note that with varying Pm and Rm the inclination of these spectra relative to the $k_x$-axis remains the same, while the extent of the spectra along the $k_x$- and $k_y$-directions changes (not shown). 

Figure \ref{FIGURE_1DShellAveragedSpectra} shows spherical shell-averages of the 3D kinetic and magnetic energy spectra in Fourier space as a function of the total wavenumber magnitude $k=|{\bf k}|$ at different ${\rm Pm}$ (keeping ${\rm Rm}$ fixed). Note that due to the anisotropy of the spectra (Figure \ref{FIGURE_2DEkin_Emag_Pm32_kz1}), the shell-averaged 1D spectra do not display a clear power-law dependence, as was also pointed out in other related studies of MRI-turbulence \citep{fromang2010,Lesur_Longaretti2011,Murphy_Pessah2015}. From this figure we see that with increasing ${\rm Pm}$ the kinetic energy (dashed curves) increases at smaller wavenumbers (i.e. at large scales in physical space), but decreases at intermediate and large wavenumbers due to the increased effect of the viscosity. The viscosity-dominated range, which is located initially at $k\gtrsim k_{\nu}$ (where $k_{\nu}\sim \sqrt{\rm Re}=56$ is the viscous wavenumber\footnote{This estimate of the viscous wavenumber follows from equating the dynamical shear rate $q\Omega$ to the viscous decay rate $\nu k^2$, which yields $k_{\nu}\sim \sqrt{q\Omega/\nu}$, or in non-dimensional units $k_{\nu}\sim \sqrt{\rm Re}$ \citep[see also][]{Guilet2022}. This wavenumber is smaller than the usual viscous wavenumber $k_{\nu0}\sim {\rm Re}^{3/4}$ used in turbulence theory \citep{landau1987} and hence represents its lower bound (corresponding to the largest possible viscous length-scale in the flow). We prefer to use $k_{\nu}$ here as it approximately coincides with the maximum of the viscous dissipation term for the kinetic energy, ${\cal D}_k^{(u)}=2k^2E_K/{\rm Re}$.}) for the lowest ${\rm Pm}=8$ in this figure (dashed blue curve), moves to lower $k$ (larger scales) as $k_{\nu}$ decreases with increasing ${\rm Pm}$ (i.e., decreasing ${\rm Re}$). Consequently, the kinetic energy at these wavenumbers decreases and its spectrum becomes less and less steep. This redistribution of the kinetic energy spectrum towards lower $k$, and of viscous-damping of higher $k$-modes, naturally leads to the emergence of dominant large-scale velocity structures in physical space, as seen in Figure \ref{FIGURE_Res128Rm25kSeriesFlowFieldComparison}. 

The magnetic energy spectrum (solid curves in Figure \ref{FIGURE_1DShellAveragedSpectra}), on the other hand, increases at \textit{all} wavenumbers as ${\rm Pm}$ increases, with lower $k$ (large scales) gaining more power than higher $k$ (small scales), and the peak of the spectrum moving towards lower $k$. This is reflected in the growth of the length-scales of magnetic structures in physical space with ${\rm Pm}$, as seen in Figure \ref{FIGURE_Res128Rm25kSeriesFlowFieldComparison}. These are still smaller in size than velocity structures since the viscous wavenumber $k_{\nu}$ is less than the characteristic wavenumber of the magnetic energy spectrum.\footnote{We take the characteristic wavenumber of the magnetic energy spectrum to be the wavenumber corresponding to the peak of the spectrum.} In contrast to the kinetic energy spectra, the magnetic energy spectra appear to converge, i.e., they no longer vary in shape and magnitude at sufficiently high ${\rm Pm} \gtrsim 64$. In this asymptotic, viscosity-dominated regime at ${\rm Pm}\gg 1$, the peak and most of the magnetic energy spectrum (and therefore the MRI-activity) already lie deep in the viscous range, and thus no longer vary as  ${\rm Pm}$ is increased even further (equivalently as  ${\rm Re}$ is decreased), as long as the turbulence is sustained for a given ${\rm Rm}$. In physical space, this behaviour is reflected in the plateau stage in the $\alpha-{\rm Pm}$ relation (see Figure \ref{FIGURE_AlphaPmRelationship}).

\subsection{Spectra of dynamical terms}
\label{RESULTS_SpectraOfDynamicalTerms}
We now investigate the spectra of the dynamical terms and their dependence on magnetic Prandtl number ${\rm Pm}$ (again, keeping the magnetic Reynolds number ${\rm Rm}$ fixed). The dependence of the spectra of the dynamical terms on Pm at fixed Re, and on Re and Rm (at fixed Pm), will be considered in Section \ref{RESULTS_DependenceOnReAtFixedPm}. 

\subsubsection{Self-sustenance mechanism of MRI dynamo}
\label{RESULTS_SelfSustenanceMechanismOfMRIDynamo}

In \cite{mamatsashvili2020zero}, we performed a detailed analysis of the dynamical processes in zero-net-flux MRI-turbulence in Fourier space, and demonstrated that the most important terms in the self-sustaining dynamics of the turbulence are (i) the linear Maxwell stress, which describes (magnetic) energy supply (injection) due to nonmodal MRI growth, and (ii) the non-linear transfer term for the radial magnetic field, ${\cal N}_x^{(b)}$. The latter quantity is anisotropic in Fourier space due to the background shear, and exhibits a {\it transverse cascade} -- a key non-linear process responsible for replenishing new MRI-active modes. The transverse cascade represents an important type of non-linear redistribution of modes over wavevector orientations/angles that is generic to shear flows.\footnote{The non-linear transverse cascade is a universal process in shear flows, operating both in hydrodynamical and MHD situations. It was first revealed in a simplified model of two-dimensional (2D) spectrally-stable constant shear flow in the purely hydrodynamical case by \cite{horton2010}, and then in the MHD case by \cite{mamatsashvili2014}, and was shown to play a key role in the sustenance of finite-amplitude perturbations (turbulence). Since Keplerian discs are particular examples of shear flows, turbulence in discs also exhibits this transverse cascade \citep{gogichaishvili2017,mamatsashvili2020zero}.} It arises solely due to shear and is thus fundamentally (i.e. topologically) different from the usual classical direct/inverse non-linear cascades that characterise classical theories of turbulence (i.e. which model turbulence in flows without a mean shear) such as those of Kolmogorov, Iroshnikov–Kraichnan, and Goldreich–
Sridhar \citep[see e.g.,][for a review]{Biskamp2003}.

Based on the spectral analysis of these two processes (i.e. the linear Maxwell stress and the non-linear transfer term for the radial magnetic field), we proposed the following self-sustaining scheme for zero-net-flux MRI turbulence in \cite{mamatsashvili2020zero}: the non-linear transverse cascade produces an essential seed \textit{radial} magnetic field, and this radial field then grows by means of the MRI, thereby extracting energy from the background flow and amplifying the azimuthal field. The latter process is proportional to the background shear and is mediated by the Maxwell stress. These two processes are the core of the self-sustenance of MRI turbulence. They occur primarily at small wavenumbers (i.e. large scales), with associated wavelength comparable to the flow size. We refer to this wavenumber range as the \textit{vital area}. The amplified azimuthal field in turn (i) contributes back to the non-linear term ${\cal N}_x^{(b)}$, thus closing the feedback loop, and (ii) is drained by the corresponding non-linear term ${\cal N}_y^{(b)}$, which is negative at these small wavenumbers and which thus acts like turbulent diffusion. 
Thus, ${\cal N}_y^{(b)}$, in contrast to ${\cal N}_x^{(b)}$, \textit{opposes} the self-sustaining process. It is a negative (sink) term for dynamically important active modes located in the vital area (at large scales), and transfers energy from these modes to dynamically passive large wavenumber modes (smaller scales). 

Note that an analogous self-sustaining scheme for the zero-net-flux MRI-dynamo problem was proposed by \cite{herault2011periodic} based on a low-order model involving only a few large-scale active modes rather than on fully developed MRI turbulence as we have here, and this model was later extended by \cite{riols2015dissipative, riols2017magnetorotational} to include the effects of viscosity and resistivity in order to understand the physical nature of the ${\rm Pm}$-dependence of the MRI.

\begin{figure}
\centering
\includegraphics[scale=0.45]{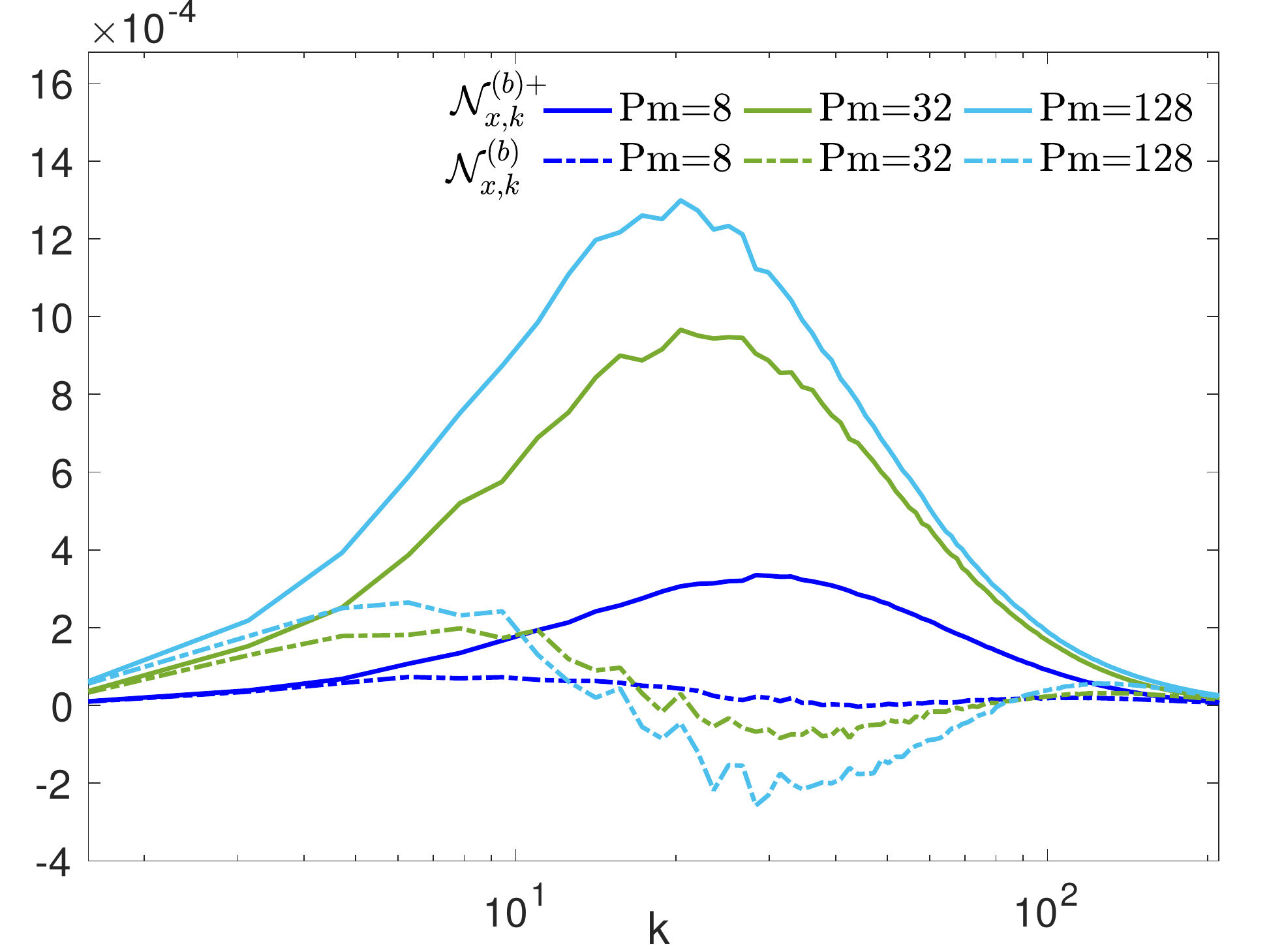}
\includegraphics[scale=0.45]{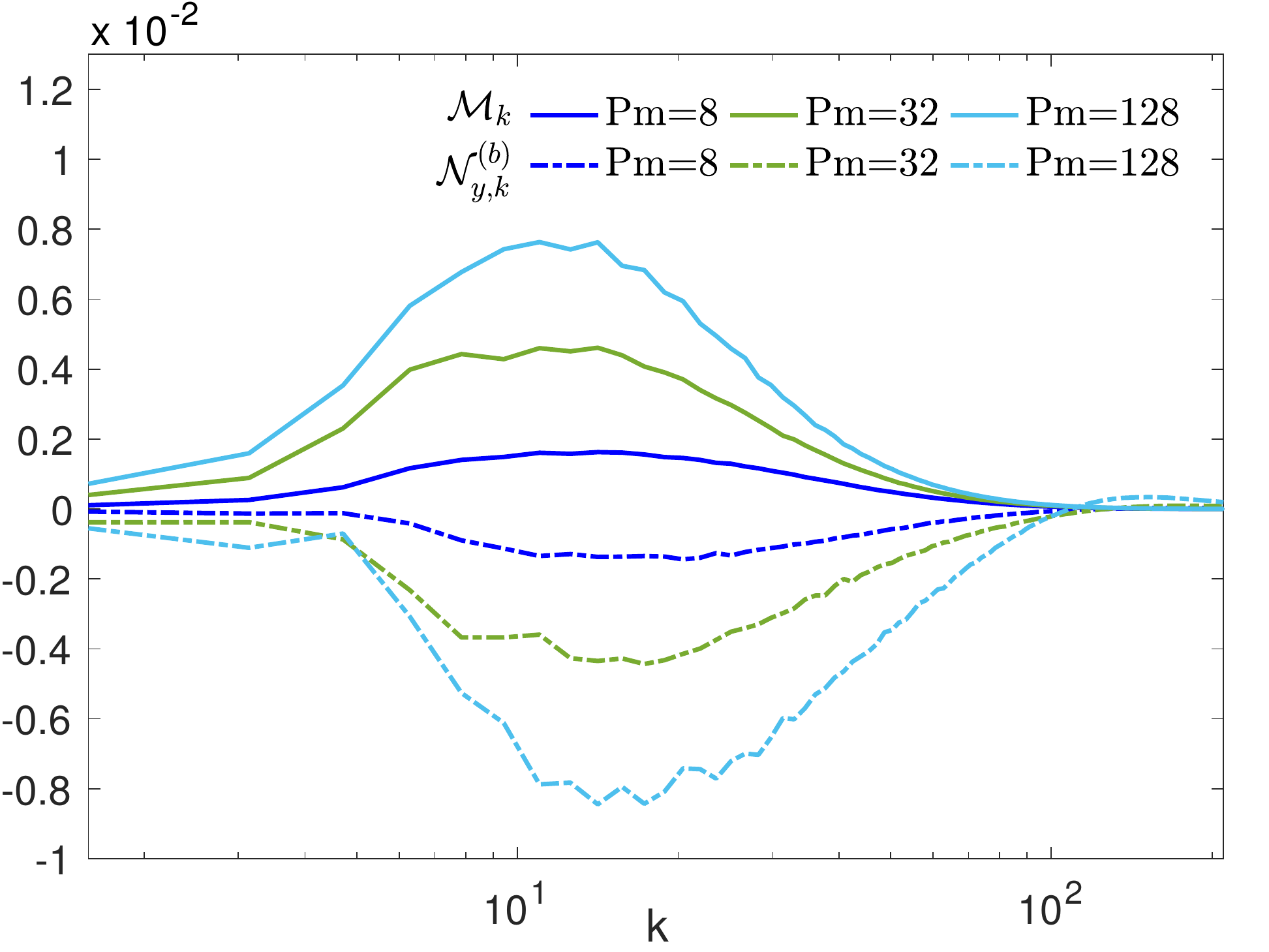}
\caption{Top: shell-averaged non-linear term ${\cal N}_{x,k}^{(b)}$ (dashed lines) and its positive part ${\cal N}_{x,k}^{(b)+}>0$ (solid lines), describing the action of the transverse cascade, which maintains the turbulence. Bottom: shell-averaged Maxwell stress ${\cal M}_k$ (solid lines) and ${\cal N}_{y,k}^{(b)}$ (dashed lines). The colors correspond to results at different ${\rm Pm}=8,~32,~128$ and fixed magnetic Reynolds number ${\rm Rm}=2.5\times 10^4$.}
\label{FIGURE_1DMaxwellStressAnd1DNxb}
\end{figure}

\begin{figure}
\centering
\includegraphics[scale=0.45]{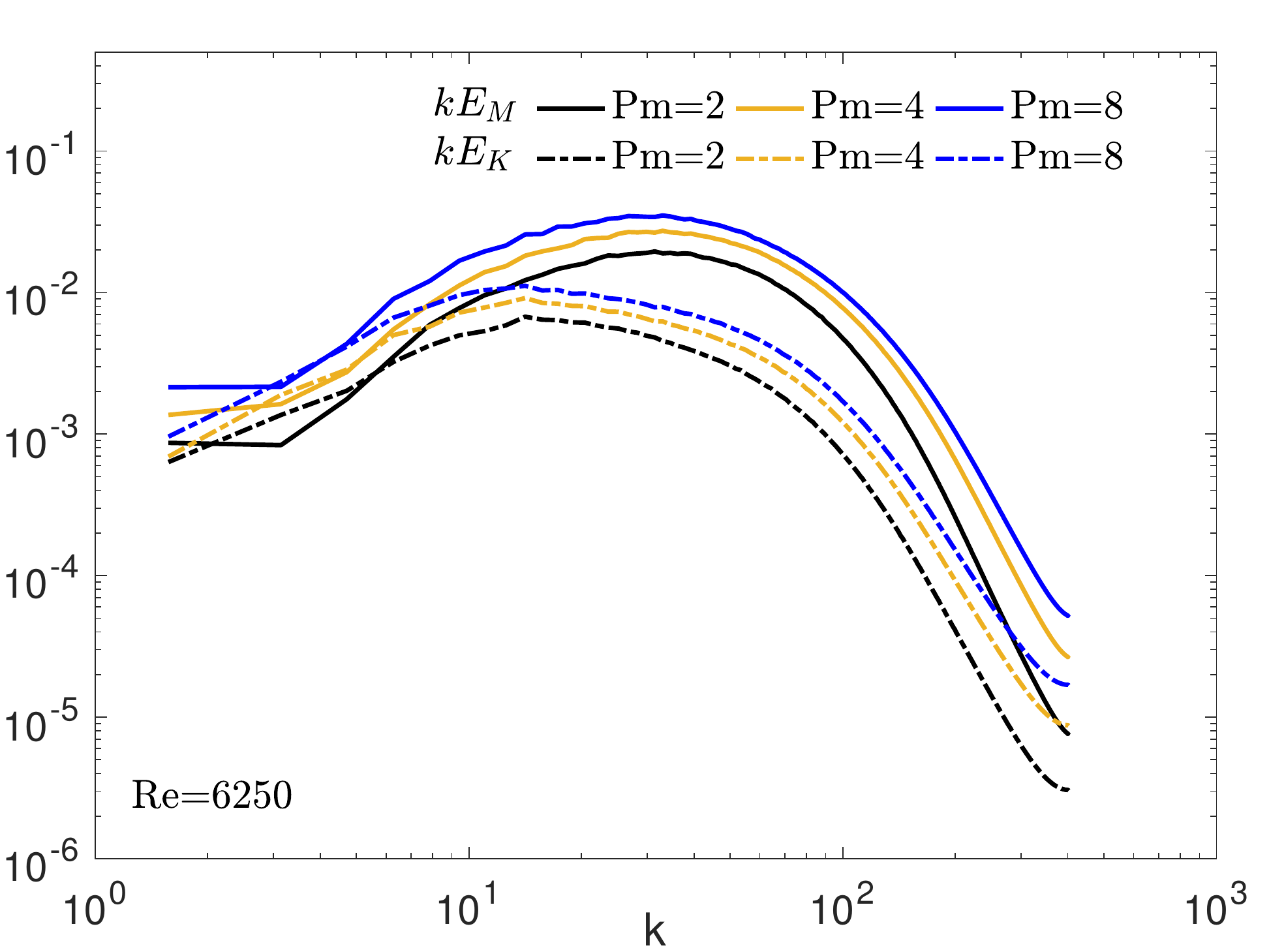}
\includegraphics[scale=0.45]{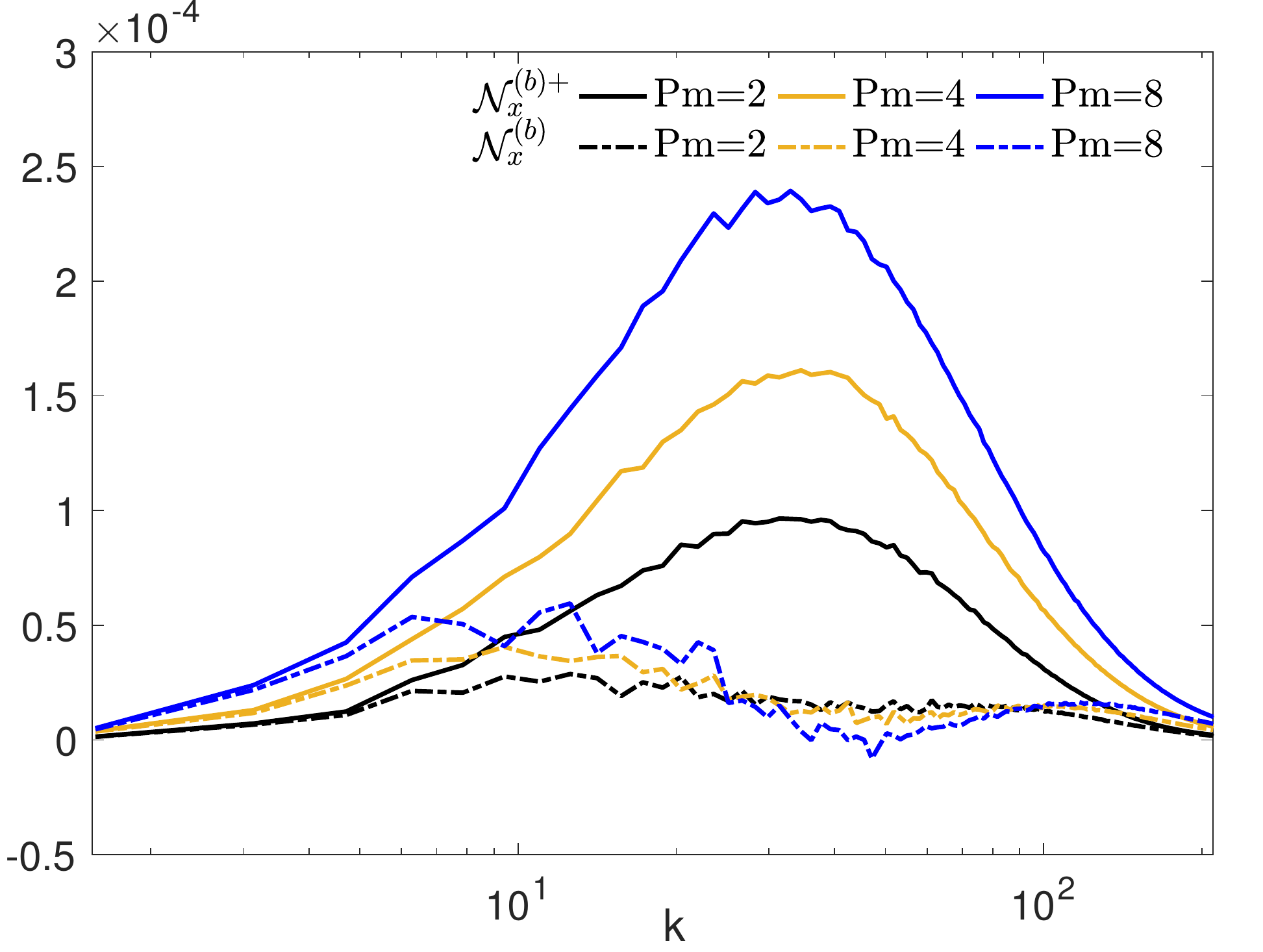}
\caption{Shell-averaged normalized kinetic and magnetic energy spectra (top), and the non-linear transfer term ${\cal N}_{x,k}^{(b)}$ together with its positive part ${\cal N}_{x,k}^{(b)+}>0$ (bottom), at a fixed Reynolds number ${\rm Re}=6250$, but different ${\rm Pm}=2,~4,~8$.}
\label{FIGURE_1DMaxwellStressAnd1DNxb_Re6250}
\end{figure}

\begin{figure}
\includegraphics[scale=0.45]{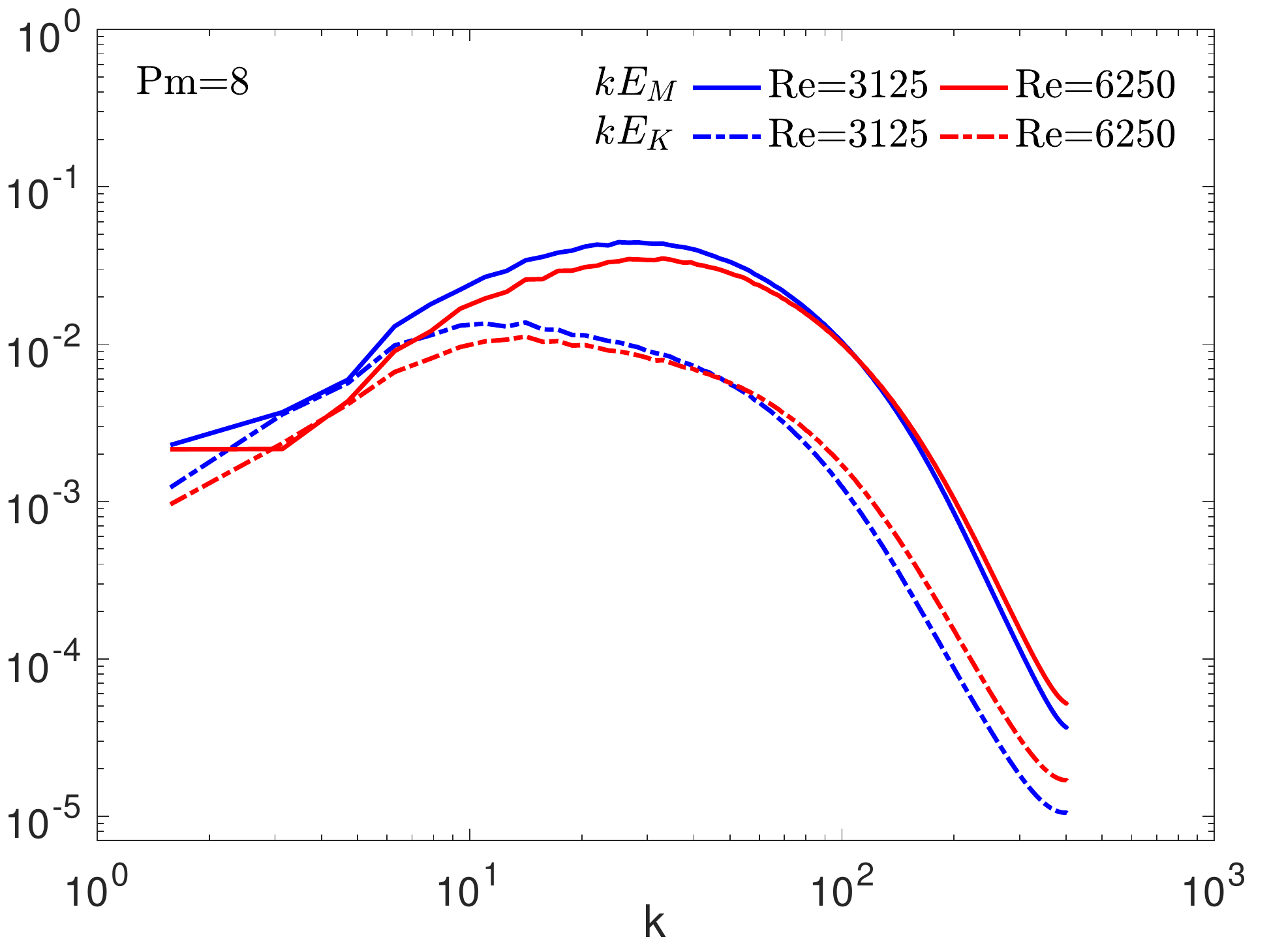}
\includegraphics[scale=0.45]{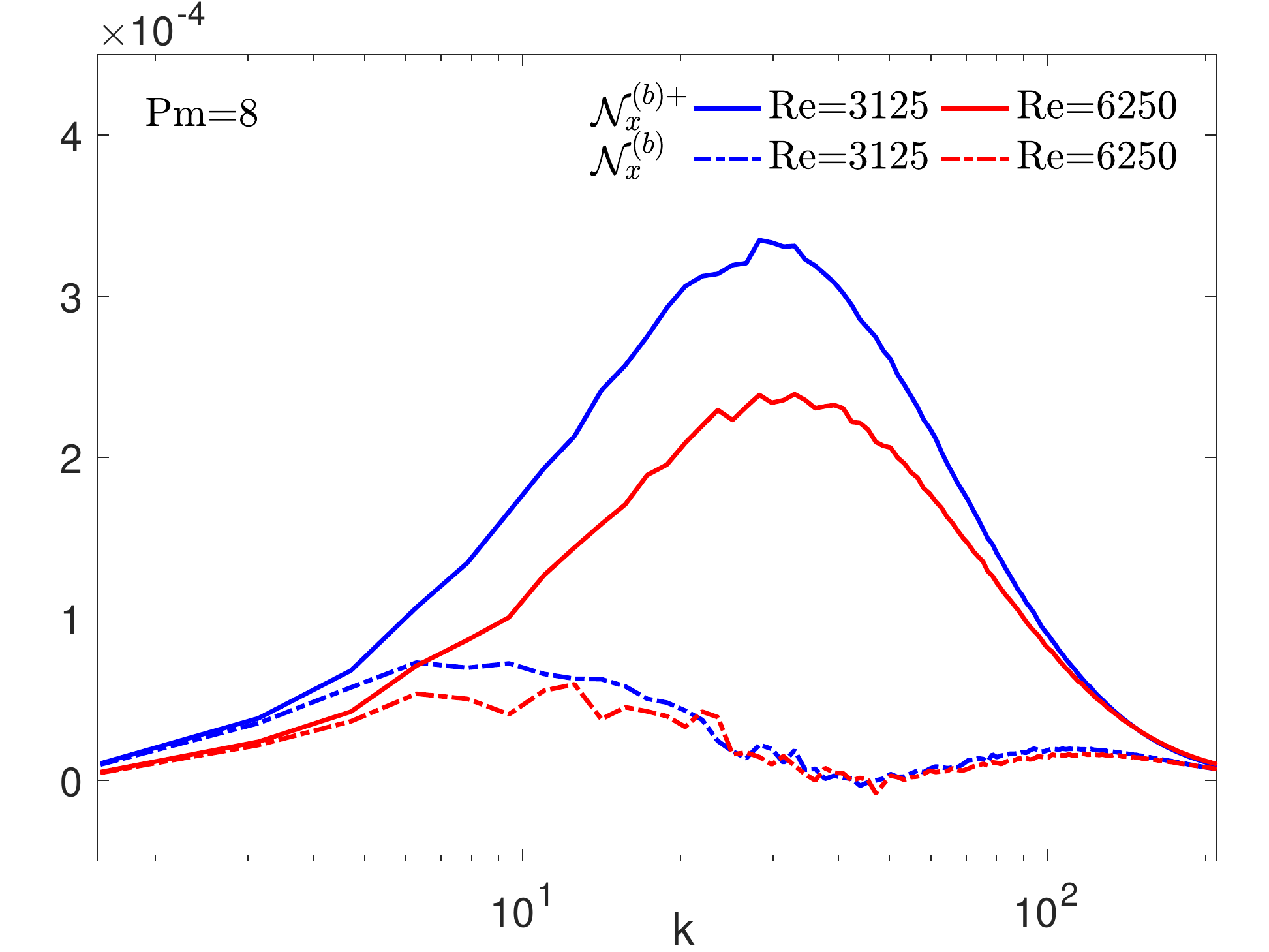}
\caption{Shell-averaged normalized kinetic and magnetic energy spectra (top), and the non-linear transfer term ${\cal N}_{x,k}^{(b)}$ and its positive part ${\cal N}_{x,k}^{(b)+}>0$ (bottom), at a fixed ${\rm Pm=8}$ and different ${\rm Re}=3125$ (blue curves) and $6250$ (red curves). Both energy spectra decrease at lower $k \lesssim 40$ with increasing ${\rm Re}$ due to the weakening of the transverse and inverse cascades at these wavenumbers relative to the direct one.}\label{FIGURE_1DMaxwellStressAnd1DNxb_Pm8}
\end{figure}

\subsubsection{Spectra of dynamical terms}
\label{RESULTS_SpectraOfDynamicalTermsSubsubsection}

The self-sustaining scheme outlined in the previous subsection is a general one for zero-net-flux MRI turbulence in discs, and therefore is at work in the present case of ${\rm Pm}>1$. The behavior of these main dynamical terms (i.e., the Maxwell stress ${\cal M}$ and the non-linear transfer term ${\cal N}_x^{(b)}$) with ${\rm Pm}$, in turn, determines the dependence of turbulent dynamics and transport properties on this parameter. In \cite{riols2017magnetorotational} and \cite{mamatsashvili2020zero}, the opposite regime ${\rm Pm} \lesssim 1$ was considered, whereas here we investigate the underlying dynamical processes in the regime ${\rm Pm} \gtrsim 1$, adopting a similar method of analysis in Fourier space. This will allow us to understand the physics behind the increase of turbulent intensity and transport with ${\rm Pm}$, and in particular the dynamical peculiarities of the new interesting saturation regime -- i.e. the plateau -- which occurs at ${\rm Pm} \gg 1$ (see Figure \ref{FIGURE_AlphaPmRelationship}). The latter regime has already been hinted to exist by \cite{lesur2007} and has been observed for the first time in the fully compressible Keplerian shear simulations presented in this paper and simultaneously in the incompressible sub-Keplerian shear simulations of \cite{Guilet2022}, although their plateau regime is rather different from ours since it occurs for moderately low ${\rm Re}=800$ and large ${\rm Rm}=\mathcal{O}(10^5)$, compared to our simulations where ${\rm Re} \sim 200$ and ${\rm Rm} = \mathcal{O}(10^4)$ in the plateau region.

Figures \ref{FIGURE_2DSpectraOfNxb} and \ref{FIGURE_2DSpectraOfMaxwellStress} show the distribution of the time-averaged non-linear transfer term $N_x^{(b)}$ and of the Maxwell stress ${\cal M}$, respectively, in the $(k_x,k_y)$-plane. Each row shows 2D slices of the 3D spectra of these dynamical terms at a different ${\rm Pm}$, from ${\rm Pm}=8$ (top row) to ${\rm Pm}=128$ (bottom row). We show the spectra of the first three dominant vertical wavenumbers $k_z=(0,1,2)2\pi/L_z$ (corresponding to the left-hand, middle, and right-hand columns, respectively). These wavenumbers carry the largest magnetic energies, and are thus more important in the turbulent dynamics than higher $k_z$ modes. It seen from these figures that, like the energy spectra above, both the Maxwell stress and non-linear transfer function are notably anisotropic in the $(k_x,k_y)$-plane, varying strongly over the polar angle $\phi = arcsin[k_y/(k_x^2 +k_y^2)^{1/2}]$ with inclination towards the $k_x$-axis. This anisotropy is nearly the same for all ${\rm Pm}$ values that we investigate, since, as mentioned above, the extent of the anisotropy is determined by the magnitude of the flow shear. 

The main consequence of this anisotropy for ${\cal N}_x^{(b)}$ is the transverse (i.e., over wavevector angle $\phi$) redistribution/transfer of power, which is evident from its structure in the $(k_x,k_y)$-plane. This term primarily transfers the spectral energy of the radial field, $|\bar{B}_x|^2$, from ``giver'' wavenumbers for which ${\cal N}_x^{(b)}<0$ (blue) acts as a \textit{sink} for the radial magnetic field, to ``receiver'' wavenumbers for which ${\cal N}_x^{(b)}>0$ (yellow and red) acts as a \textit{source}, as well as among different components of the velocities and magnetic field. These two main areas, which are concentrated at lower wavenumbers in Fourier space (corresponding to the flow system scales), are crucial to the self-sustaining dynamics of MRI turbulence and jointly form the vital area of turbulence. 

In this vital area, the positive ${\cal N}_x^{(b)}>0$ (yellow and red in Figure \ref{FIGURE_2DSpectraOfNxb}) continually regenerates and amplifies the radial field mostly for non-axisymmetric modes as a result of non-linear triad interactions with nearby modes. When the flow is in quasi-steady state, the non-linear transverse cascade is in balance with the linear drift term (first term on the right-hand side of equation \ref{eq:bxk2}) inside the vital area. The drift term, acting opposite to the non-linear transfer term, advects $|\bar{B}_x|^2$ for each  non-axisymmetric mode along the $k_x$-axis from the growth (yellow/red) region back into the blue region, where it decreases. Thus, the non-linear transverse cascade operates primarily inside the vital area, replenishing the radial field energy there, while part of this energy is transferred via a direct cascade from the vital area to higher $k$-modes, up to dissipation scales. At these large wavenumbers (small scales), the direct cascade usually dominates over the transverse cascade (see below). 

The dynamics of the azimuthal field is governed by the Maxwell stress ${\cal M}$, and the non-linear transfer term ${\cal N}_y^{(b)}$. The latter, as mentioned above, is negative and hence \textit{opposes} the turbulence sustenance: rather than replenishing energy to modes, as ${\cal N}_x^{(b)}$ does for the radial magnetic field, ${\cal N}_y^{(b)}$ drains azimuthal field energy from the vital area \cite[see][for details]{mamatsashvili2020zero}. The Maxwell stress, on the other hand, is positive in the $(k_x, k_y)$-plane at different $k_z$ and reaches its highest values in the vital area (note the yellow and red regions in Figure \ref{FIGURE_2DSpectraOfMaxwellStress}). In other words, the Maxwell stress injects most of the power and amplifies the azimuthal magnetic field energy $|\bar{B}_y|^2/2$ most efficiently in this low wavenumber area compared to higher wavenumbers. This amplification is a direct result of the MRI acting on modes in the vital area: the instability acts on a seed radial field $\bar{B}_x$ generated by ${\cal N}_x^{(b)}>0$ due to the non-linear transverse cascade, and this leads, in turn, to the growth of the azimuthal field. Consequently, the Maxwell stress spectrum, which is given mathematically by the product of the radial and azimuthal fields, follows (correlates with) the distribution of these two field components and reaches maximum values in the vital area where these components are also highest. Since this azimuthal field is the dominant field component, it provides the main contribution (positive feedback) to ${\cal N}_x^{(b)}$, which, in turn, produces the seed radial field, thus closing the main subcycle of the turbulence self-sustenance mechanism. 

Due to the linear drift in Fourier space, individual non-axisymmetric modes finally leave the vital (growth) area ${\cal M}>0$ in the $(k_x,k_y)$-plane, in other words their amplification is transient and decreases afterwards due to the non-linear transfer of magnetic energy to larger $k$ modes where it is dissipated by resistivity. Given such a transient, or nonmodal character of the MRI amplification \citep[see also][]{mamatsashvili2013,squire2014}, the continual replenishment of the radial field (which in turn triggers MRI-growth) by means of the transverse cascade is crucial for the long-term maintenance of the turbulence. Thus, the concerted interplay of linear MRI growth and the non-linear transverse cascade, which feeds back in a positive manner on the former, is the basis of the self-sustenance of MRI turbulence. This self-sustaining scheme was first proposed and described in depth for a net-toroidal-flux magnetic field configuration in \cite{gogichaishvili2017}, and for zero-net-flux in \cite{mamatsashvili2020zero}, which we refer the reader to for details.

\subsection{Dependence of the spectral dynamics on ${\rm Pm}$ at fixed ${\rm Rm}$}
\label{RESULTS_DependenceOfTurbulentDynamicsInSpectralSpaceOnPm}

Having outlined the self-sustaining scheme of the turbulence, let us now analyse how its main dynamical players -- the Maxwell stress ${\cal M}$ and the non-linear transverse cascade described by the transfer function ${\cal N}_x^{(b)}$ -- change with ${\rm Pm}$. This can provide a clue to the dependence of the strength of MRI turbulence on magnetic Prandtl number which we observe in our simulations. We have seen in Figures \ref{FIGURE_2DSpectraOfNxb} and \ref{FIGURE_2DSpectraOfMaxwellStress} that with increasing ${\rm Pm}$ the transverse cascade, the Maxwell stress, and therefore the whole self-sustaining process, become more concentrated and stronger at smaller wavenumbers (larger scales) and hence the vital area shrinks. 

The shift of these dynamical terms towards smaller wavenumbers can be better characterised by considering their 1D spherically shell-averaged spectra denoted, respectively, as ${\cal M}_k$ for the Maxwell stress and ${\cal N}_{x,k}^{(b)}$ for the non-linear transfer term. We plot these shell-averaged spectra in Figure \ref{FIGURE_1DMaxwellStressAnd1DNxb}. Note that in shear flows, such shell-averaging of spectra should be done with caution, since it tends to smear any spectral anisotropy of the injection and non-linear transfers (the transverse cascade) in Fourier space which arise due to shear \citep{Murphy_Pessah2015}, and instead captures only direct/inverse non-linear cascades along wavenumber ${\bf k}$ \citep[][]{Lesur_Longaretti2011}. Thus, to take into account the specific anisotropic structure of ${\cal N}_x^{(b)}$ in our shell-averaging procedure, we have also shell-averaged this term only over its \textit{positive} values ${\cal N}_x^{(b)}>0$, i.e. over the yellow and red regions in Figure \ref{FIGURE_2DSpectraOfNxb}), 

\begin{equation}
{\cal N}_{x,k}^{(b)+}\equiv\int {\cal N}_x^{(b)}|_{>0}k^2d\Omega,
\label{EQUN_non-linearShellAveragePositive}
\end{equation}
with $d\Omega$ denoting an element of solid angle in Fourier space. We plot equation (\ref{EQUN_non-linearShellAveragePositive}) together with the shell-average of the total ${\cal N}_x^{(b)}$ term in the upper panel of Figure \ref{FIGURE_1DMaxwellStressAnd1DNxb}. Here $N_{x,k}^{(b)+}$ characterizes the action of the non-linear transverse cascade for the radial field $\bar{B}_x$, while its direct/inverse cascade is described by the shell-average ${\cal N}_{x,k}^{(b)}$ of the total transfer term.

\subsubsection{Increase in action of transverse cascade with Pm}
\label{RESULTS_IncreaseInActionOfTransverseCascadeWithPm}

At sufficiently large $k\gtrsim 100$, where the effect of shear is negligible, ${\cal N}_x^{(b)}$ is positive and more isotropic in Fourier space, so ${\cal N}_{x,k}^{(b)}$ (dashed curves) and ${\cal N}_{x,k}^{(b)+}$ (solid curves) coincide. As seen in Figure \ref{FIGURE_1DMaxwellStressAnd1DNxb}, these two quantities behave differently with ${\rm Pm}$ at smaller $k \lesssim 100$. ${\cal N}_{x,k}^{(b)+}$ and, consequently the Maxwell stress determined by it, increases and becomes more and more concentrated (peaked) at lower $k$ with increasing ${\rm Pm}$, consistent with the behavior of these terms in the $(k_x,k_y)$-plane as shown in Figure \ref{FIGURE_2DSpectraOfNxb}. This results in the increased level of turbulence and dominance of larger scales, thus explaining why $\alpha$ and magnetic energy increase with ${\rm Pm}$ as seen in Figure \ref{FIGURE_AlphaPmRelationship}. By contrast, as ${\rm Pm}$ increases, ${\cal N}_{x,k}^{(b)}$, \textit{decreases} at intermediate wavenumbers $10\lesssim k \lesssim 100$, becoming more and more negative, implying that these wavenumbers  lose power overall (of course, this is due to the large anisotropic negative ${\cal N}_x^{(b)}<0$ areas shown in blue in Figure \ref{FIGURE_2DSpectraOfNxb}). On the other hand, ${\cal N}_{x,k}^{(b)}$ is positive and tends to increase more with Pm at small wavenumbers $k \lesssim 10$, indicating the development of an inverse cascade of power towards these wavenumbers compared to its relatively small increase at higher wavenumbers $k \gtrsim 100$, where it exhibits the direct cascade. 

Thus it is evident that with increasing ${\rm Pm}$ the inverse cascade becomes more and more dominant over the direct one, which implies that small wavenumber modes receive increasingly more power from the intermediate wavenumber ones. The result is the reduction of power outflow towards higher wavenumbers, i.e. a reduction in non-linear diffusion. Since in our simulations we increase ${\rm Pm}$ by keeping $\rm Rm$ fixed, i.e. by \textit{decreasing} ${\rm Re}$, the diminishing of the non-linear transfers to high $k$ can alternatively be attributed to the suppression (``arrest'') of these small-scale modes by increased viscosity, as is evident from the shell-averaged kinetic energy spectrum in Figure \ref{FIGURE_1DShellAveragedSpectra}. Thus these small-scale modes can no longer participate in the non-linear interactions/transfers that sustain the dynamo. This also leads to a reduction of resistive dissipation. Indeed, in quasi-steady state, it follows from Equation (\ref{eq:bxk2}) that the total action of the (shell-averaged) non-linear term ${\cal N}_{x,k}^{(b)}$ should balance total resistive dissipation, i.e. $\sum_k{\cal N}_{x,k}^{(b)}\approx -\sum_k{\cal D}_{x,k}^{(b)}$. As a result of this rearrangement of the non-linear term ${\cal N}_{x,k}^{(b)}$ along $k$ as ${\rm Pm}$ is varied, its sum over all wavenumbers is $\sum_k{\cal N}_{x,k}^{(b)}=[2.5,~3.1,~2.7,~2.3,~0.73]\times10^{-3}$ at ${\rm Pm}=8, 16, 32, 64, 128$, respectively, i.e. the total resistive dissipation mostly decreases with Pm (except at ${\rm Pm}=8$), with the smallest value at ${\rm Pm}=128$ corresponding to the plateau in the $\alpha-{\rm Pm}$-relation (Figure \ref{FIGURE_AlphaPmRelationship}). Thus the plateau in fact corresponds to states of \textit{minimal} resistive dissipation.    

The spectral energy of the radial field $|\bar{B}_x|^2$ produced by the transverse cascade (which is described in the shell-averaged case by ${\cal N}_{x,k}^{(b)+}$), and of the associated shell-averaged Maxwell stress ${\cal M}_k$, remain concentrated mostly at low and intermediate wavenumbers within the vital area, and increase with ${\rm Pm}$. By contrast, the fraction of power which is transferred to high wavenumber modes due to the direct cascade, and which eventually gets dissipated via resistivity, decreases compared to its regeneration by ${\cal N}_{x,k}^{(b)+}$. One can say that the transverse cascade prevails over the direct cascade and sustains the turbulence. This is in agreement with the general picture of the ${\rm Pm}$-dependence of the MRI dynamo proposed in \cite{mamatsashvili2020zero}, which is based on the competition between the transverse and direct cascades -- the former dominates over the latter with increasing ${\rm Pm}$, leading to the rise of turbulent intensity. This picture is also in line with the interpretation of the ${\rm Pm}$-effect for non-linear MRI-states by \cite{riols2015dissipative, riols2017magnetorotational}, who, based on the analysis of a low-order model, attributed the decrease of turbulent stress and energy values with decreasing ${\rm Pm}$ to the increase of non-linear magnetic diffusion of large-scale dynamically active modes due to efficient transfer of their energy to small-scale (``slaved'') modes via the direct cascade. Yet another explanation of the ${\rm Pm}$-dependence of MRI turbulence based on the reconnection rate was put forward by \cite{balbus2008}, and elaborated on by \cite{simon2009} and \cite{potter2017} (whose analysis was in physical space). In their picture, increasing ${\rm Pm}>1$ (i.e., increasing viscosity over resistivity) means that the resistive scale-scale becomes smaller and smaller than the viscous one, and, since velocity fluctuations are increasingly damped by viscosity at the resistive scale, velocity fluctuations cannot bring magnetic field lines close enough to each other for their reconnection to occur efficiently. The absence of a sufficient magnetic reconnection rate leads, in turn, to the reduction of the resistive dissipation rate. This interpretation is entirely consistent with our results, discussed in the last paragraph, that resistive dissipation decreases as ${\rm Pm}$ increases.

\subsubsection{Saturation of transverse cascade in plateau region}
\label{RESULTS_SaturationInPlateauRegion}

The magnetic energy and turbulent stress increase with ${\rm Pm}$, as we have seen above, until they eventually reach a plateau stage at about ${\rm Pm_{pl}} \approx 64$ (for ${\rm Rm} =2.5\times10^4$), after which they stay nearly constant (see Figure \ref{FIGURE_AlphaPmRelationship}). Figure \ref{FIGURE_1DShellAveragedSpectra} shows that the dissipation range of the kinetic energy spectrum, $k \gtrsim k_{\nu}$ where $k_{\nu}=\sqrt{\rm Re}$ is the viscous wavenumber defined above, moves to lower $k$. By contrast, the magnetic field spectrum, although also shifted towards lower $k$ as ${\rm Pm}$ is increased, starts to converge with ${\rm Pm}$ for ${\rm Pm}\gtrsim32$ and continues to retain almost the same form even in the plateau stage at ${\rm Pm} \gtrsim {\rm Pm_{pl}}$, with the maximum of the spectral magnetic energy and Maxwell stress becoming concentrated at the intermediate, dynamically active wavenumbers $1 \lesssim k \lesssim 40$ which comprise the vital area. The viscous wavenumber $k_{\nu}$ decreases as ${\rm Pm}$ is increased, and becomes equal to the wavenumber corresponding to the peak of the magnetic energy spectrum at around ${\rm Pm}={\rm Pm_{pl}}$, i.e. just when the plateau stage is first reached. After that, within the plateau region ${\rm Pm} > {\rm Pm_{pl}}$, the viscous wavenumber decreases even further and extends down to the minimum wavenumber in the box $2\pi/L_x$. At this point the the dynamical range as well as all wavenumbers in the box appear to lie fully in the viscosity-dominated regime. In physical space this is manifested in the noticeable increase of the length-scales of velocity structures with ${\rm Pm}$, almost filling the box, in contrast to that of magnetic field ones, as seen at ${\rm Pm}=128$ in Figure \ref{FIGURE_Res128Rm25kSeriesFlowFieldComparison}. 

In this plateau regime, (at ${\rm Pm}=128$, for example) velocity perturbations are already damped at the dynamically active wavenumbers $1 \lesssim k \lesssim 40$ due to viscosity, in contrast to magnetic perturbations (Figure \ref{FIGURE_1DShellAveragedSpectra}). As a result, non-linear direct transfers towards higher (i.e. up to resistive) wavenumbers $k \gtrsim 100$ are suppressed and dominated by the transverse cascade as well as partly by the inverse cascade towards smaller $k\lesssim 10$ within the vital area, as is evident in Figure \ref{FIGURE_1DMaxwellStressAnd1DNxb}. In other words, within the plateau regime, the dominant viscous dissipation efficiently reduces the direct cascade by damping velocities at the small-scales, which in turn results in a suppression of non-linear (turbulent) diffusion of active modes in the vital area \citep[i.e. the loss of energy by non-linear diffusion from large-scale ``master/active'' modes to small-scale ``slaved'' modes is suppressed,][]{riols2017magnetorotational}. In this case, the replenishment rate of these active modes by the transverse cascade operating in the vital area dominates over the weakened direct cascade of power out of this area. As a result, the resistive dissipation at higher $k$ is also greatly reduced, as we have discussed quantitatively in Section \ref{RESULTS_IncreaseInActionOfTransverseCascadeWithPm}. The overall effect is that the spectral energy of the radial field $|\bar{B}_x|^2$, and hence the entire self-sustaining dynamics of the turbulence, are confined to the small and intermediate wavenumbers within the vital area with almost no leakage of energy to high wavenumber modes. This explains why the turbulent dynamics at the plateau stage ${\rm Pm} > {\rm Pm_{pl}}$ is no longer subject to non-linear diffusion and hence maintains the same level of magnetic energy and stress as Pm is increased even further. 

\subsection{Dependence of the spectral dynamics keeping ${\rm Pm}$ or ${\rm Re}$ fixed}
\label{RESULTS_DependenceOnReAtFixedPm}

Above we have explored the dependence of MRI turbulence dynamics, the spectra of the energies, Maxwell and non-linear terms as a function of ${\rm Pm} \equiv {\rm Rm}/{\rm Re}$ at fixed magnetic Reynolds number Rm. Now we explore how the spectral dynamics of MRI turbulence behave with increasing ${\rm Pm}$ and ${\rm Rm}$ at fixed \textit{Reynolds number} ${\rm Re}=6250$, and also with increasing ${\rm Re}$ (and ${\rm Rm}$) at fixed Prandtl number ${\rm Pm}=8$. We do so by comparing the spectra of the energies and the dynamical terms from select simulations. 

\begin{figure}
\centering
\includegraphics[scale=0.25]{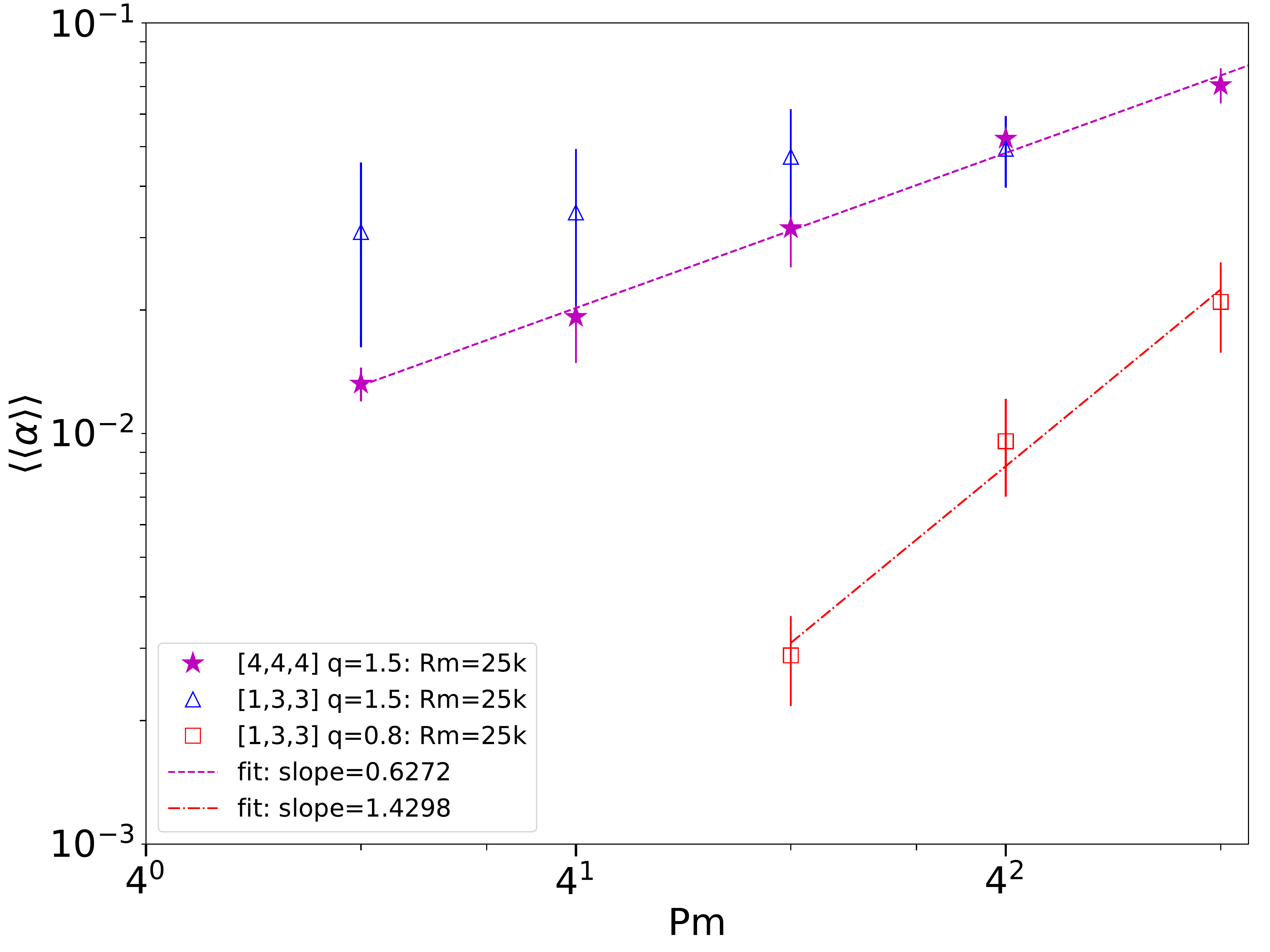}
\caption{Alpha-Pm scaling at fixed $\text{Rm}=2.5\times10^{4}$ showing the dependence of the scaling on the box vertical-to-radial aspect ratio (compare blue triangles to purple stars) and on the shear parameter $q$ (compare blue triangles to red squares).}
\label{FIGURE_AlphaPmScalingBoxAspectRatioShearParameter}
\end{figure}

Different trends are seen depending on which two of the three parameters ($\text{Rm},\text{Re},\text{Pm}$) are varied. In Figure \ref{FIGURE_1DMaxwellStressAnd1DNxb_Re6250} we show the result of keeping ${\rm Re}$ fixed and varying ${\rm Pm}$ by increasing ${\rm Rm}$. Both the kinetic and magnetic energy spectra (top panel) as well as the non-linear transfer terms ${\cal N}_{x,k}^{(b)+}$ and ${\cal N}_{x,k}^{(b)}$ (bottom panel), which describe the action of the transverse and the inverse/direct cascades, respectively, have basically similar shapes with the peaks occurring at the same wavenumbers, and differing only by a simple rescaling of the magnitude, which increases with ${\rm Pm}$ (only the ${\rm Pm}=2$ energy spectra [black curves in the top panel] are slightly steeper at large $k$ due to the smaller ${\rm Rm}$). This implies that the spectral dynamics of the turbulence is similar at fixed ${\rm Re}$, so the only effect of increasing ${\rm Rm}$ in this case is just an increase in the overall energy content (i.e. intensity) of the turbulence due to reduced resistive dissipation. Note, however, that the transverse cascade (bottom panel of Figure \ref{FIGURE_1DMaxwellStressAnd1DNxb_Re6250}) appears to become more efficient than the direct cascade as ${\rm Pm}$ increases [keeping ${\rm Re}$ fixed] in a similar manner to the case where we kept ${\rm Rm}$ fixed and increased ${\rm Pm}$ by decreasing ${\rm Re}$ (Figure \ref{FIGURE_1DMaxwellStressAnd1DNxb}). This suggests that a similar mechanism is likely at work in the simulations of \cite{Guilet2022} where ${\rm Re}$ is fixed and ${\rm Pm}$ varied.

\begin{figure*}
\centering
\includegraphics[scale=0.45]{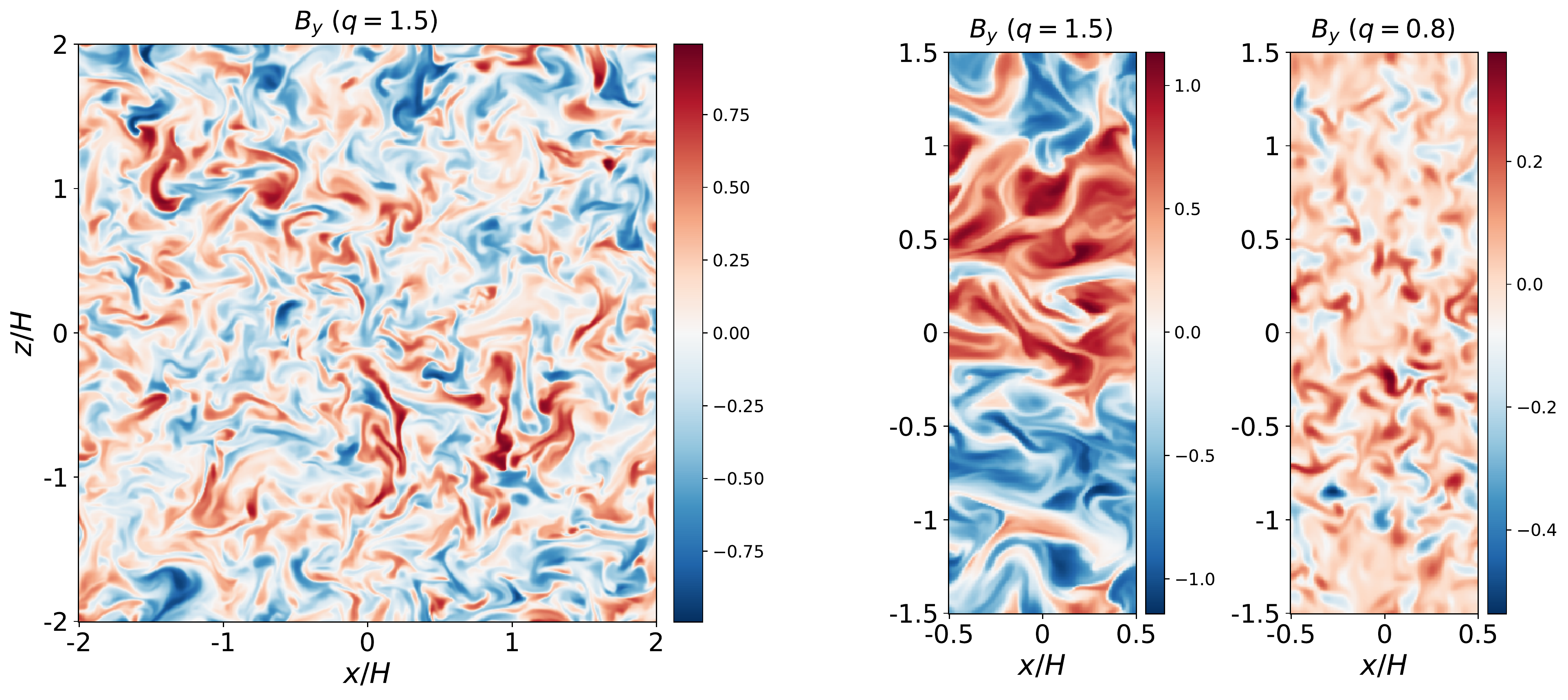}
\caption{Comparison of azimuthal magnetic field $B_y$ in the $xz$-plane in simulations with different vertical-to-radial box size $L_z/L_x$ and different shear parameter $q$. Left: $[L_x,L_y,L_z]=[4,4,4]\,(q = 1.5)$. Middle: $[1,3,3]\,(q = 1.5)$. Right: $[1,3,3]\,(q = 0.8)$. All simulations were run with $\text{Pm}=8$ and $\text{Rm}=2.5\times10^{4}$.}
\label{FIGURE_ByFlowFieldShearParaComparison}
\end{figure*}

\begin{figure}
\centering
\includegraphics[scale=0.23]{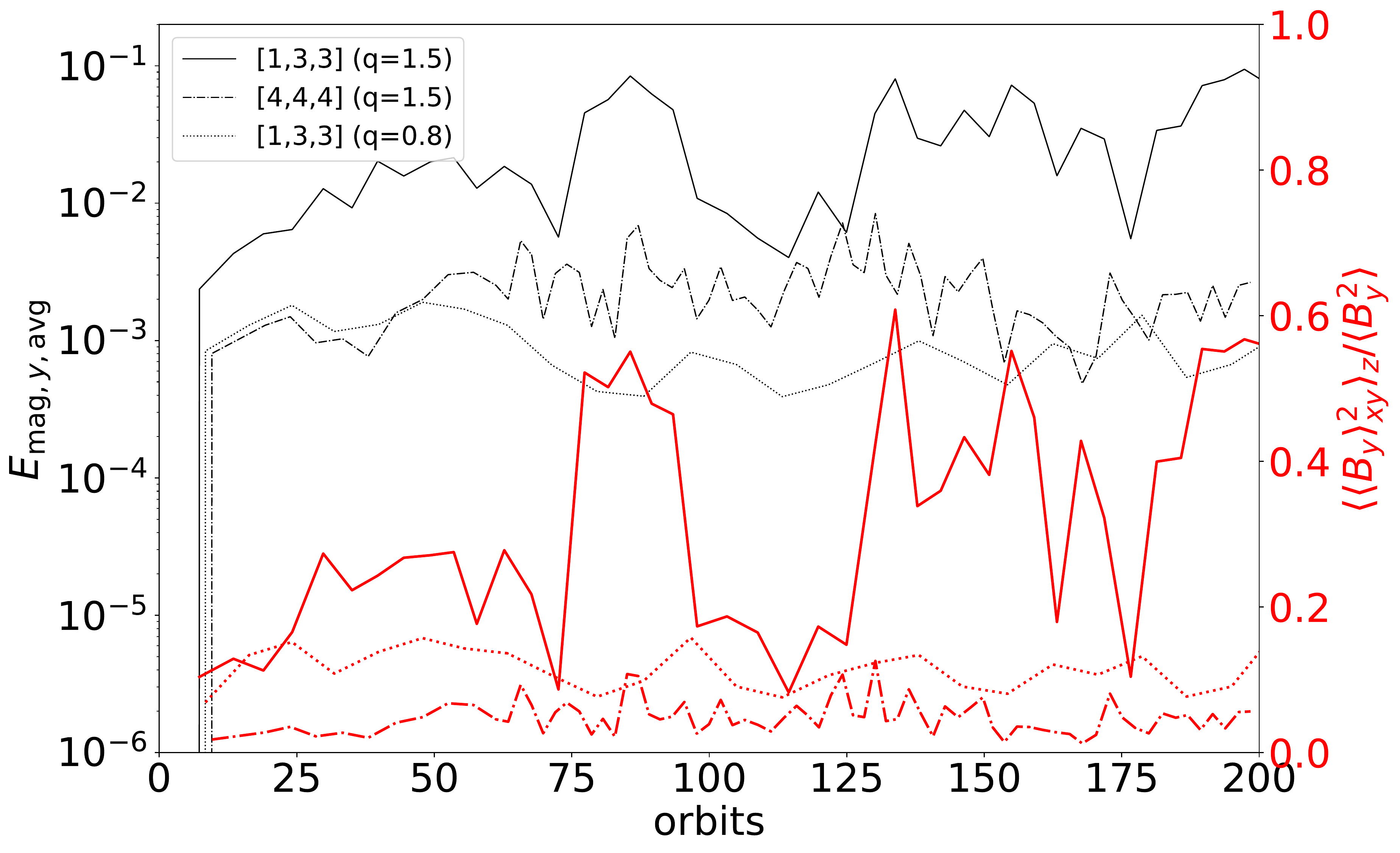}
\caption{Comparison of magnetic energy in the background toroidal field (black curves) and of the ratio of magnetic energy in the background toroidal field to the total energy in toroidal field (red curves) for three different simulations. Solid curves: $[1,3,3]\,(q = 1.5)$. Dot-dashed curves: $[4,4,4]\,(q = 1.5)$. Dotted curves: $[1,3,3]\,(q = 0.8)$.}
\label{FIGURE_EmagyavgTimeSeriesComparison}
\end{figure}

The situation is different when we keep ${\rm Pm}$ fixed and increase ${\rm Re}$ (and thus also ${\rm Rm}$). See Figure \ref{FIGURE_1DMaxwellStressAnd1DNxb_Pm8}. In this case, increasing ${\rm Re}$ (and simultaneously ${\rm Rm}$) leads to a \textit{decrease} of the action of the inverse and and transverse cascades, with this decrease occurring mostly at the lower $k \lesssim 40$ that lie within the vital area. The peak of the non-linear transfer term ${\cal N}_{x,k}^{(b)+}$ (solid curves, bottom panel) slightly shifts towards larger $k$ (smaller scales), while the direct cascade at higher wavenumbers remains unchanged (compare dashed curves in the bottom panel). As a result, both the kinetic and magnetic energy spectra (shown in the top panel) decrease as $\rm{Re}$ and $\rm{Rm}$ are increased at these lower wavenumbers, but the spectral energies both increase at higher wavenumbers, with the kinetic energy spectra increasing somewhat more than the magnetic one. The behavior of the normalized magnetic energy $kE_M$ with ${\rm Re}$ at ${\rm Pm}=8$ that we observe here is qualitatively consistent with that shown by \cite{fromang2010} at ${\rm Pm}=4$ (see their Figure 3), although the exact shape of the spectra differ between their simulations and ours, perhaps due to the different set-up used in that study (\cite{fromang2010} employ a `finger-box' domain of size $[1,\pi,1]$ and a higher resolution $512/H$, compared to our cubic box domain of size $[4,4,4]$ at resolution $128/H$). 

When we compare the shell-averaged spectra in Figure \ref{FIGURE_1DMaxwellStressAnd1DNxb_Pm8} (simulations at fixed Pm and varying Re) with the results from the simulations where we kept Rm fixed (and varied Re) (Figures \ref{FIGURE_1DShellAveragedSpectra} and \ref{FIGURE_1DMaxwellStressAnd1DNxb}), we see qualitatively similar behavior both in the energy spectra, and also of the non-linear transfer term ${\cal N}_{x,k}^{(b)}$ and its positive part ${\cal N}_{x,k}^{(b)+}$ (the latter of which maintains the MRI turbulence, as discussed above). This implies that for the magnetic field dynamics, the Reynolds number turns out to be a more important parameter. Increasing ${\rm Re}$ (at fixed Pm, or at fixed Rm), allows (frees) small-scale modes up to viscous wavenumbers to come into play in the dynamical processes. This initiates an efficient direct cascade of magnetic energy from smaller wavenumbers in the vital area to these newly-released high-wavenumber modes (turbulent magnetic diffusion). Thus magnetic energy from large-scales is converted mostly into kinetic energy at small-scales. At the same time, the transverse cascade, the inverse cascade (at smaller wavenumbers), and hence the overall turbulent level all decrease, as can be seen in Figure \ref{FIGURE_AlphaPmRelationship} where turbulent stress and magnetic energy clearly decrease when Re is increased (whether by keeping Pm fixed [i.e. moving along a vertical line from top to bottom], or by keeping Rm fixed [i.e. moving diagonally from top-right to bottom-left]). \

\section{Discussion}
\label{RESULTS_Discussion}

\subsection{Effect of changing the box aspect ratio}
\label{RESULTS_EffectOfChangingBoxAspectRatio}

\cite{shi2016} carried out several ideal and non-ideal zero-net-flux magnetohydrodynamic shearing box simulations of large vertical-to-radial ratio ($L_z/L_x \gtrsim 2.5$). They found (i) stronger angular momentum transport compared to boxes with smaller aspect ratio, and, in their non-ideal runs,  (ii) weak alpha-Pm scaling (see their Figure 11). To check this we have also carried out several simulations in tall boxes ($H\times3H\times3H$) at different ${\rm Pm}$ and shear parameters $q$ (Table \ref{TABLE_TallBoxSimulations}). \footnote{To reduce computational cost, our tall box runs are somewhat shorter ($H\times3H\times3H$) in the azimuthal and vertical directions than those employed in the non-ideal simulations of \cite{shi2016} ($H\times4H\times4H$), but the resolution per scale height is the same in our runs as it is in theirs ($128/H$).} We carried out four simulations for this tall box at ${\rm Pm}= 2, 4, 8$ and $16$, all keeping the magnetic Reynolds number $\text{Rm} = 25000$ and Keplerian shear $q=1.5$ constant (see blue triangles in Figure \ref{FIGURE_AlphaPmScalingBoxAspectRatioShearParameter}). We found that the turbulent transport is larger in these tall boxes compared to otherwise identical cubic box runs of size $4H\times4H\times4H$ (denoted by the purple stars in Figure \ref{FIGURE_AlphaPmScalingBoxAspectRatioShearParameter}). In our tall box runs we also find a much weaker alpha-Pm scaling between $\text{Pm} = 2$ and $\text{Pm} = 8$, and little to no scaling for $\text{Pm}> 8$. Hence our tall box runs are consistent with those of \cite{shi2016}.

The weaker scaling in boxes of large vertical-to-radial aspect ratio can be understood in terms of the large-scale, vertically-dependent mean toroidal field that arises in these boxes, even when they are initialized with zero net-magnetic-flux. This coherent, mean field component is clearly seen by comparing the snapshots of the total $B_y$ in the $xz$-plane from the cubic [4,4,4] and tall [1,3,3] boxes both at $\text{Pm}=8$ and $\text{Rm}=25000$ (left-hand and middle panels of Figure \ref{FIGURE_ByFlowFieldShearParaComparison}). The contribution to the total toroidal magnetic energy density from this mean toroidal field can also be quantified by equation (\ref{EQUN_EmagyMean}), which we plot in Figure \ref{FIGURE_EmagyavgTimeSeriesComparison}. The black curves show the magnetic energy density of the mean toroidal field, while the red curves show the \textit{fraction} of the total toroidal magnetic energy density which is contained in the mean toroidal field. The latter diagnostic, in particular, reveals that $20\%$ to $60\%$ of the energy in the total toroidal field comes from this mean field in tall boxes (solid red curve), compared to about only $5\%$ in cubic boxes (dot-dashed red curve). 

\begin{figure}
\centering
\includegraphics[scale=0.35]{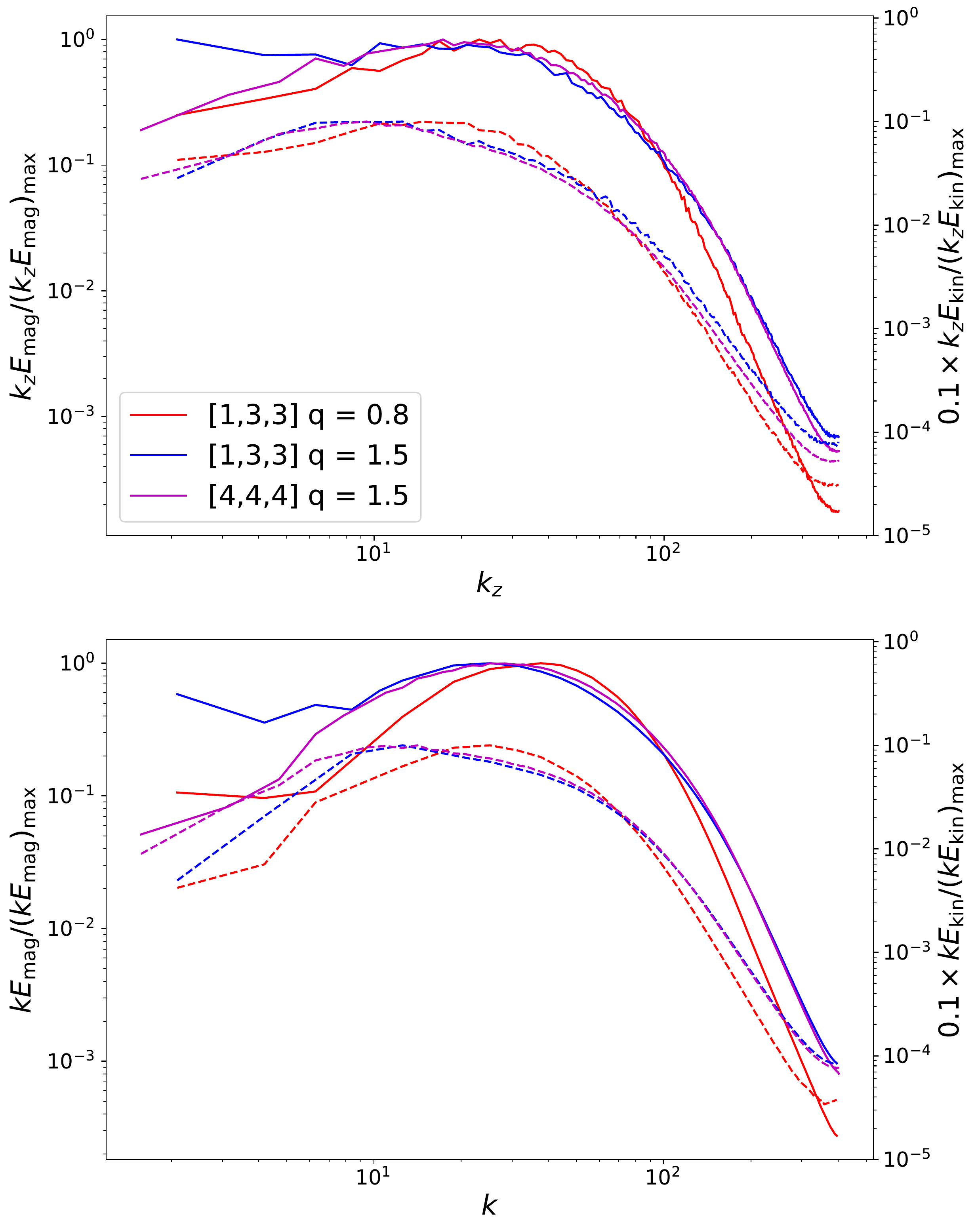}
\caption{Top panel: 1D spectra (times $k_z$) of magnetic energy (solid lines) and kinetic energy (dashed lines) as a function of vertical wavenumber $k_z$, showing dependence on vertical-to-radial box aspect ratio (compare blue and magenta curves) and on the shear parameter $q$ (compare blue and red curves). Magnetic energy spectra are plotted against the left-hand $y$-axis, while kinetic energy spectra are plotted against the right-hand $y$-axis. The spectra have been normalized by their maximum values. Bottom panel: same as top panel, but showing shell-averaged spectra. All simulations were run with $\text{Pm}=8$ and $\text{Rm}=25000$.}
\label{FIGURE_1DShellAveragedSpectraComparingBoxSizeAndShearParameter_Rm25k}
\end{figure}

This mean toroidal field arises because in tall boxes the total toroidal field changes more slowly than the orbital period, resulting in a regular mean toroidal field, which varies on time-scales larger than the dynamical time $\sim \Omega^{-1}$ \citep{lesur2008localized,shi2016}. Thus, the properties and dynamics of MRI turbulence in tall boxes is more similar to net-toroidal-flux MRI rather than to zero-net-flux MRI. Because of the presence of this large-scale toroidal field, more of the energy lies at large scales compared to the pure ZNF case. Consequently, the turbulence in tall boxes is less sensitive to the dissipative scales, which is reflected in the smaller sensitivity of large vertical-to-radial aspect ratio simulations on the magnetic Prandtl number. This can be seen by comparing shell-averaged spectra of kinetic and magnetic energies taken from the cubic and tall boxes, respectively: compare the blue ([1,3,3] box) and magenta ([4,4,4] box) curves in the bottom panel of Figure \ref{FIGURE_1DShellAveragedSpectraComparingBoxSizeAndShearParameter_Rm25k}. The spectra in the [1,3,3] box are flatter at large scales (small $k$) in contrast to the spectra in the [4,4,4] box. Thus, there is more energy in the largest scales in the tall box and the flow is less sensitive to the dissipation scales (and therefore to the magnetic Prandtl number).

\subsection{Effect of changing the shear parameter}
\label{RESULTS_EffectOfChangingShearParameter}
So far we have concentrated on exploring the dependence of MRI turbulence on magnetic Prandtl number in the regime of Keplerian shear (with shear parameter $q=1.5$) which is relevant to (thin) accretion discs. However, a second astrophysically relevant regime in which one expects the MRI to be active at high magnetic Prandtl numbers is the interior of protoneutron stars, mainly due to a large neutrino viscosity \citep{guilet2015neutrino,Guilet2022}. The angular frequency within the outer layers of the protoneutron star decreases with radius and should thus be MRI-unstable, but the shear is weaker than in Keplerian discs. \cite{reboul2021global} carried out 3D global MHD simulations in a spherical shell of protoneutron stars, and found a shear parameter of $q \sim 0.8$ in the outer, MRI-unstable regions of the PNS (see their Figure 6). Thus, we carry out three simulations in the [1,3,3] boxes with a shear parameter of $q = 0.8$, keeping the magnetic Reynolds number fixed at $\text{Rm}=25000$ (Table \ref{TABLE_TallBoxSimulations}). These are to be compared to Keplerian shear $q = 1.5$ simulations run with the same box size [1,3,3] and at the same ${\rm Rm}$. Compare the red and blue data points in Figure \ref{FIGURE_AlphaPmScalingBoxAspectRatioShearParameter}, which correspond to the runs at $q=0.8$ and $q=1.5$, respectively. At a given magnetic Prandtl number, the turbulent transport (and also magnetic energy; not shown) is considerably weaker in the sub-Keplerian shear case ($q=0.8$), which is not surprising given that the shear is smaller and therefore there is less orbital energy for the MRI to tap into (see also \cite{pessahchanpsaltis2008}). The $\alpha-$Pm power-law scaling, on the other hand, is significantly stronger in the $q=0.8$ runs compared to the $q = 1.5$ runs, with $\delta \sim 1.43$.

Despite the $q=0.8$ simulations being run in tall boxes ($H\times3H\times3H$), we do not observe a similar large scale toroidal field at $q=0.8$ which we observe in the case of Keplerian shear. Compare the middle panel of Figure \ref{FIGURE_ByFlowFieldShearParaComparison}, which shows the toroidal field $B_y$ from a snapshot taken from the tall-box $q=1.5$ run at $\text{Pm}=8$ and $\text{Rm}=25000$, to the the right-hand panel, which shows the same quantity from the tall-box $q=0.8$ run at the same ${\rm Pm}$ and ${\rm Rm}$. We observe only small-scale turbulence in the toroidal magnetic field in the sub-Keplerian shear simulation. Furthermore, the mean toroidal field in the $q=0.8$ run comprises only around $10\%$ of the total toroidal magnetic field energy (see dotted red curve in Figure \ref{FIGURE_EmagyavgTimeSeriesComparison}), a considerably smaller fraction than that found at $q=1.5$ (solid red curve in Figure \ref{FIGURE_EmagyavgTimeSeriesComparison}), for which between $20\%$ and $60\%$ of the total toroidal magnetic field energy is in the mean toroidal field. Turning to the spectra for kinetic and magnetic energy (see Figure \ref{FIGURE_1DShellAveragedSpectraComparingBoxSizeAndShearParameter_Rm25k}), the spectra in this tall box [1,3,3] for the $q=0.8$ run (red curves) are steeper at large scales ($k \lesssim 40$) than those for the $q=1.5$ run (blue curves). Thus, the turbulence in the sub-Keplerian shear runs is more sensitive to the dissipation scales (and therefore to the magnetic Prandtl number), which explains the steeper $\alpha-$Pm scaling compared to the runs with Keplerian shear. 

\subsection{Comparison to previous work}
\label{RESULTS_ComparisonPreviousWork}
\begin{figure}
\centering
\includegraphics[scale=0.24]{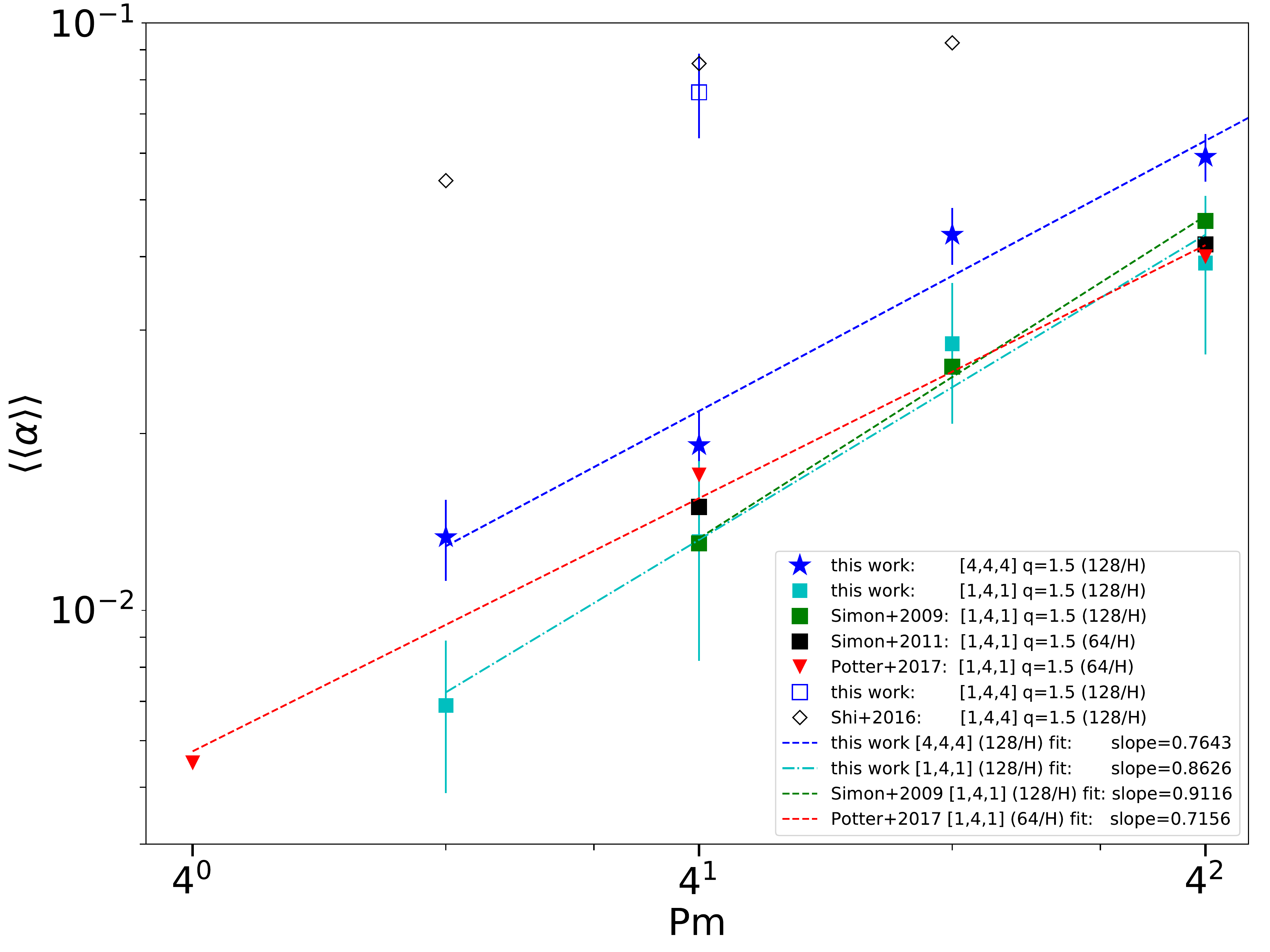}
\caption{Comparison of $\alpha-\text{Pm}$ scaling between select zero-net-flux MRI simulations from the literature and from this work. The box size $[L_x,L_y,L_z]$ (in units of the initial scale height $H$) and resolution (in terms of number of cells per $H$) are indicated in the legend. All simulations employ zero-net-magnetic-flux with a shear parameter of $q = 1.5$ (Keplerian shear).}
\label{FIGURE_AlphaPmComparisonPreviousWork}
\end{figure}

Relatively few authors have investigated the saturation of the MRI dynamo (zero-net-flux MRI) and associated turbulence in the regime $\text{Pm} > 1$. The best points of comparison to our work are select simulations from \cite{simon2009}, \cite{simon2011}, \cite{potter2017}, and \cite{shi2016}.\footnote{\cite{Guilet2022} have recently investigated the $\alpha-$Pm effect in shearing box simulations at large ${\rm Pm}$, but their simulations are restricted to sub-Keplerian flow characteristic of the interiors of protoneutron stars.} The zero-net-flux simulations described by the first three of these authors were carried out in `finger-boxes' with a configration $[1,4,1]$, while \cite{shi2016} carried out `tall-box' simulations with a configuration of $[1,4,4]$. \cite{simon2009}, \cite{simon2011}, and \cite{shi2016} carried out their simulations at a resolution $128$ cells per $H$, whereas \cite{potter2017} carried out their (zero-net-flux) simulations at $64$ cells per $H$. For convenience we have plotted select simulations from these papers and from our own runs in Figure \ref{FIGURE_AlphaPmComparisonPreviousWork}.

In our cubic box $[4,4,4]$ runs at $\text{Rm}=12500$ (blue stars) we measured an $\alpha$-Pm scaling with power-law index $\delta \sim 0.76$. To facilitate better comparison with the literature, we repeated select runs at $\text{Pm} = 2,4,8$ and 16 (cyan squares) in finger boxes [1,4,1] (also at fixed $\text{Rm}=12500$). The stress in these simulations is lower and the scaling is somewhat stronger at $\delta \sim 0.86$ compared to our cubic box runs, but the results of our finger box simulations are in excellent agreement with those of \cite{simon2009} (green squares) and \cite{potter2014} (red triangles).

As already discussed in detail in the previous two subsections, we also find that the dependence of turbulent saturation on ${\rm Pm}$ is sensitive to the vertical-to-radial box aspect ratio and to the shear parameter. Tall boxes have weaker scaling than cubic boxes, while boxes with a larger shear parameter have weaker scaling than boxes with a smaller shear parameter. We carried out four simulations in tall boxes $[1,3,3]$ at $\text{Rm}=25000$ and found very weak scaling between $\text{Pm}=2$ and $\text{Pm} = 16$ (see blue triangles in Figure \ref{FIGURE_AlphaPmScalingBoxAspectRatioShearParameter}) in agreement with \cite{shi2016}. Note that these runs are in slightly shorter boxes, and at higher Rm, compared to the runs of \cite{shi2016}. We have therefore repeated our tall box run at $\text{Pm}=4$ in a box of size $[1,4,4]$ and at $\text{Rm}=12500$ (see blue square at $\text{Pm}=4$ in Figure \ref{FIGURE_AlphaPmComparisonPreviousWork}. The turbulent stress from this run is within one standard deviation of that measured by \cite{shi2016}.

Both the box vertical-to-radial aspect ratio and shear parameter dependence of the the scaling of turbulent stress with ${\rm Pm}$ can be explained by the strength of the mean toroidal field that arises due to the MRI dynamo. The weaker this toroidal field is, the stronger the $\alpha$-Pm scaling becomes, as discussed in detail in Section \ref{RESULTS_EffectOfChangingBoxAspectRatio}. Although our simulations are all initialized with zero-net-magnetic-flux, the weaker scaling that we observe when a stronger toroidal mean field arises is qualitatively consistent with the simulations which include a mean background field ab initio, i.e. the net-toroidal-flux simulations of \cite{simon2009} (see their Figure 7), and the net-toroidal-flux and net-vertical-flux simulations of \cite{potter2017} (see their Figure 2). Both authors find (i) weaker scaling in simulations initialized with a mean field compared to the zero-net-flux case, and (ii) stronger scaling the weaker this background toroidal or vertical background field. Note that this dependence of the scaling on the strength of a background field was already foreshadowed by \cite{lesur2007} (for the net-vetical-flux case), who similarly found that turbulent transport becomes more sensitive to magnetic Prandtl number the weaker the background field.

\section{Conclusions}
\label{CONCLUSIONS}

We have carried out 3D non-ideal unstratified isothermal shearing box MHD simulations of the zero-net-magnetic-flux magnetorotational instability (MRI) in order to investigate the self-sustaining dynamics of MRI turbulence in the sparsely explored regime of large magnetic Prandtl numbers ${\rm Pm}\gg 1$. Such conditions are expected to occur in the neutrino-cooled discs resulting from binary neutron star or neutron star-black hole mergers \citep{rossi2008}, in AGN and low mass X-ray binaries \citep{balbus2008}, and also in the interiors of protoneutron stars \citep{Guilet2022}. We have carried out a large number of simulations of zero-net-flux MRI turbulence, exploring different regimes of the main parameters of the flow system, such as Reynolds, ${\rm Re}$, and magnetic Reynolds, ${\rm Rm}$, numbers, magnetic Prandtl number ${\rm Pm}={\rm Rm}/{\rm Re}$, shear parameter, $q$, and box aspect ratio, $L_z/L_x$.

Our main results are presented in Figure \ref{FIGURE_AlphaPmRelationship}. We have shown, consistent with previous studies in the literature, that turbulent saturation level (intensity) increases with magnetic Prandtl number for $\text{Pm}\geq 2$, as a power-law $\sim Pm^\delta$ both for magnetic energy and turbulent stress $alpha$ parameter. Furthermore, we extended the range of ${\rm Pm}$ to 128 (in certain cases up to 256) and showed that this scaling still holds up to $\text{Pm} \sim 128$. The scaling, however, appears to be sensitive to the magnetic Reynolds number: for cubic boxes of size $4H\times4H\times4H$ (where $H$ is the scale height) we find that $\delta \sim 0.6$ at $\text{Rm} = 2.5\times10^{4}$, but there is a stronger scaling of $\delta \sim 0.76$ at lower $\text{Rm} = 1.2\times10^{4}$, and a weaker scaling of $\delta \sim 0.5$ at higher $\text{Rm} = 5\times10^{5}$.

To better understand the dynamics of the zero-net-flux MRI (i.e. MRI dynamo) in this high ${\rm Pm}$ regime, we have performed a detailed analysis of the dynamical processes in Fourier ${\bf k}$-space, in particular their interplay and dependence on ${\rm Pm}$ and ${\rm Rm}$, following the procedure laid out in \cite{mamatsashvili2020zero}. As demonstrated in that paper using incompressible simulations with the spectral code \textsc{SNOOPY}, as well as in the fully compressible finite-volume simulations with \textsc{PLUTO} which we have presented in this work, a key non-linear process ensuring turbulent sustenance is the \textit{transverse cascade} -- non-linear anisotropic redistribution of power over wavevector orientations in Fourier space, which is an intrinsic non-linear process in shear flows. Specifically, the self-sustenance of zero-net-flux MRI turbulence is determined by the interplay of nonmodal MRI growth, which is of linear nature, and the aforementioned non-linear transverse cascade, which exerts positive feedback on the former. (As a result, the peak of the Maxwell stress spectrum, which describes the action of the MRI, is well correlated with the positive (i.e. source) parts of the spectrum of the non-linear transfer term, which corresponds to the replenishment of active modes by the transverse cascade.) 

The increase (i.e. scaling) of turbulent intensity with ${\rm Pm}$ can be understood in terms of the competition between a direct cascade (which moves energy from small wavenumber active modes to large wavenumbers) and the non-linear transverse cascade which moves energy around over wavevector angles from small to intermediate wavenumbers, thereby producing (seeding) new modes in those regions of Fourier space where they, in turn, can undergo MRI growth. Since these active modes contribute most to the energetics of the MRI turbulence, the dynamically important region in Fourier space where these modes are  located is referred to as the \textit{vital area} of the turbulence. As the magnetic Prandtl number is increased this self-sustaining process becomes concentrated into ever smaller wavenumbers (larger scales) and the transverse cascade becomes more efficient at putting energy into the vital area, dominating over the direct cascade which transfers energy out of this area to larger wavenumber modes, thus acting like a turbulent diffusion for active modes. The end result is that less energy is transported to -- and dissipated at -- the resistive scale, and thus the magnetic energy (and turbulent stress) grow with increasing $\text{Pm}$. As we showed, this process is also at work when  the turbulence level decreases with increasing ${\rm Re}$ at a fixed magnetic Prandtl number because of increasing dominance of the direct cascade over the transverse transverse one. 

A novel effect that we observe is that the power-law increase of the turbulent energy and stress with ${\rm Pm}$ eventually saturate to a plateau at high enough ${\rm Pm} \gtrsim 100$ for all ${\rm Rm}$ which we considered, and thereafter remain nearly constant with respect to these numbers. A similar plateau regime of zero-net-flux MRI turbulence has also been reported recently by \cite{Guilet2022}, albeit it at fixed ${\rm Re}$ and increasing ${\rm Rm}$. As we showed, this plateau stage corresponds to the minimum of resistive dissipation. In physical space, the sizes of turbulent structures (correlation length) both in the velocity and magnetic field accordingly increase with ${\rm Pm}$, with the former being always larger than the latter and comparable to the box size at the plateau stage. In this highly viscosity-dominated regime all MRI modes lie entirely within the viscous subrange, the direct cascade transporting energy out to high wavenumber modes is almost suppressed, or very inefficient due to high viscosity, whereas the non-linear transverse cascade is maximally efficient in replenishing the large-scale modes in the vital area and hence sustaining the turbulence. Thus, the small-scale modes are basically placed outside the self-sustaining dynamics, no longer draining energy from large-scale active modes. This explains the independence of the saturated energy and stress from the dissipation coefficients in this limiting, high ${\rm Pm}\gg 1$ plateau regime.

Finally, we have further found that the $\alpha-\text{Pm}$ scaling is sensitive to various numerical and physical parameters, in particular the vertical-to-radial aspect ratio ($L_z/L_x$) as well as the shear parameter $q$. We observe larger saturation levels and weaker scaling in tall boxes ($L_z/L_x \geq 3$) than we do in cubic boxes ($L_z=L_x$), in agreement with the tall box simulations of \cite{shi2016}. This can be explained by the shallower magnetic energy spectra at small wavenumbers in the tall boxes due to the emergence of a regular large-scale mean toroidal magnetic field in contrast to those in the cubic boxes. A novel result is the sensitivity of the $\alpha-\text{Pm}$ scaling to the shear parameter. We find significantly stronger scaling (with $\delta \sim 1.4$) in shear flows characteristic of the interiors of protoneutron stars (with $q = 0.8$) than we do in Keplerian shear flows ($q = 1.5$). These results can be understood in terms of the strength of the mean toroidal field that arises in unstratified MRI dynamo simulations. The stronger this field (relative to the total toroidal magnetic field), the more energy lies at larger scales and the less sensitive the dynamo is to the dissipation scales, which in turn results in a weaker $\alpha$-Pm scaling.

In order to keep the physics as simple as possible we have limited our investigations in this work to unstratified isothermal boxes. A natural next step therefore is to investigate whether our results generalize to conditions resembling those found in real discs in which $\text{Pm}\gg 1$, in particular the neutrino-cooled discs from binary neutron star mergers, e.g. by carrying out simulations of the MRI at high ${\rm Pm}$ including vertical stratification, neutrino cooling, and a tabulated equation of state that models the conditions in these systems. We plan on carrying out such simulations in the near future.

\section*{Acknowledgements}
Simulations were run on the Sakura, Cobra, and Raven clusters at the Max Planck Computing and Data Facility (MPCDF) in Garching, Germany. The authors would like to thank Masaru Shibata, J\'er\^ome Guilet, Alexis Reboul-Salze, and Henrik Latter for helpful discussions. GM also thanks Frank Stefani for carefully reading the manuscript and for useful comments. Finally, we thank Dr John Alan Kennedy and Dr Michele Compostella at MPCDF for assistance in sharing large data files. This work received funding from the European Union's Horizon 2020 research and innovation programme under the ERC Advanced Grant Agreement No. 787544.

\section*{Data availability}
The data underlying this article will be shared on a reasonable request to the corresponding author.




\bibliographystyle{mnras}
\bibliography{2022MRIAtHighPmPaper_Bibliography} 




\appendix

\section{Resolution study}

\begin{figure}
\centering
\includegraphics[scale=0.2]{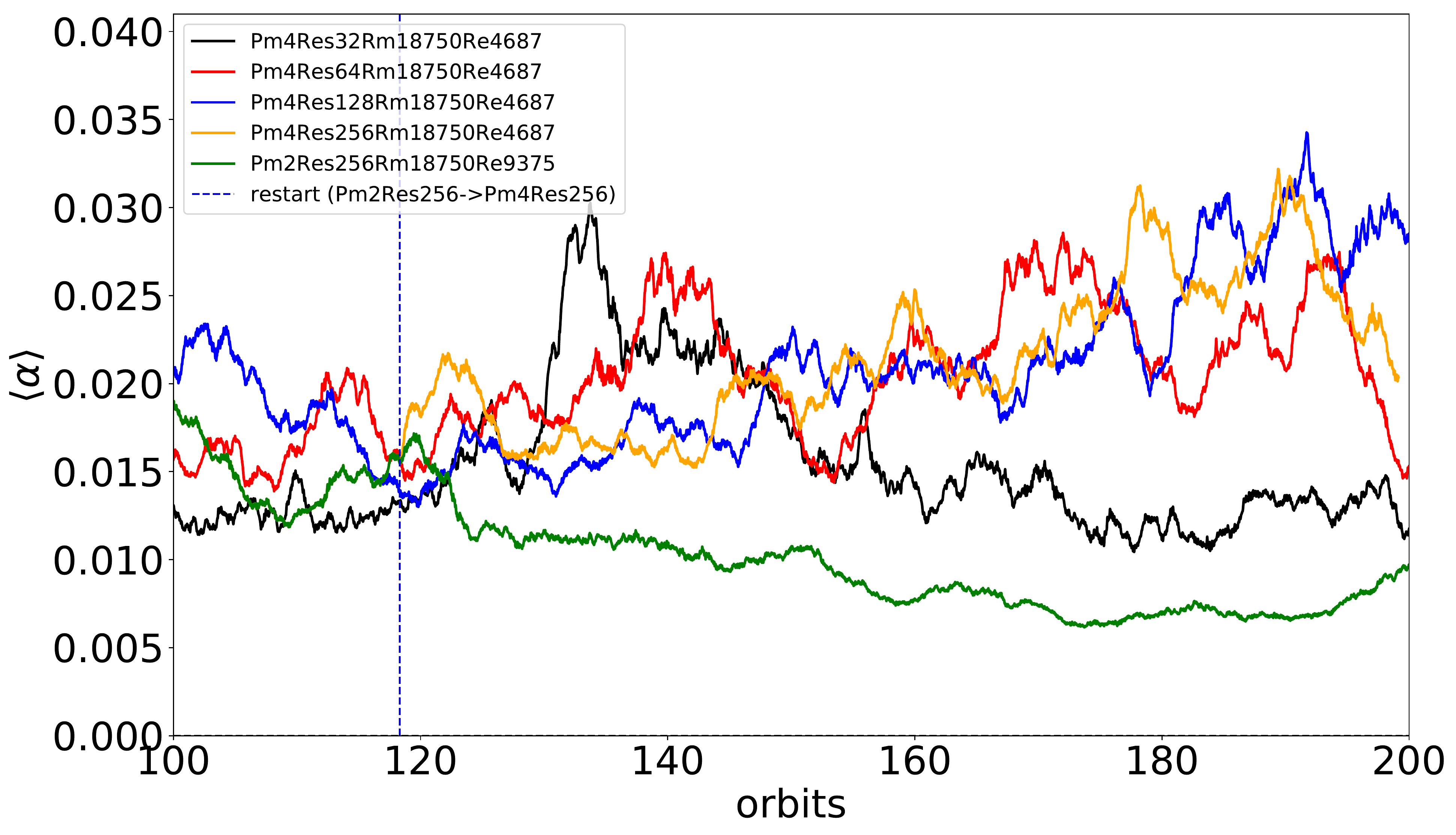}
\caption{Time-series of volume-averaged alpha parameter taken from a resolution study at $\text{Pm}=4$, $\text{Rm}=18750$ and $\text{Re}=4687$. The resolution (cells per scale height) are denoted by the colors: $32/H$ (black), $64/H$ (red), $128/H$ (blue), and $256/H$ (gold). Note that highest resolution $\text{Pm}=4$ run was restarted from orbit 125 of a $256/H$ $\text{Pm}=2$ run (green).}
\label{FIGURE_AlphaTimeSeriesComparisonResolutionStudy}
\end{figure}

\begin{figure}
\centering
\includegraphics[scale=0.27]{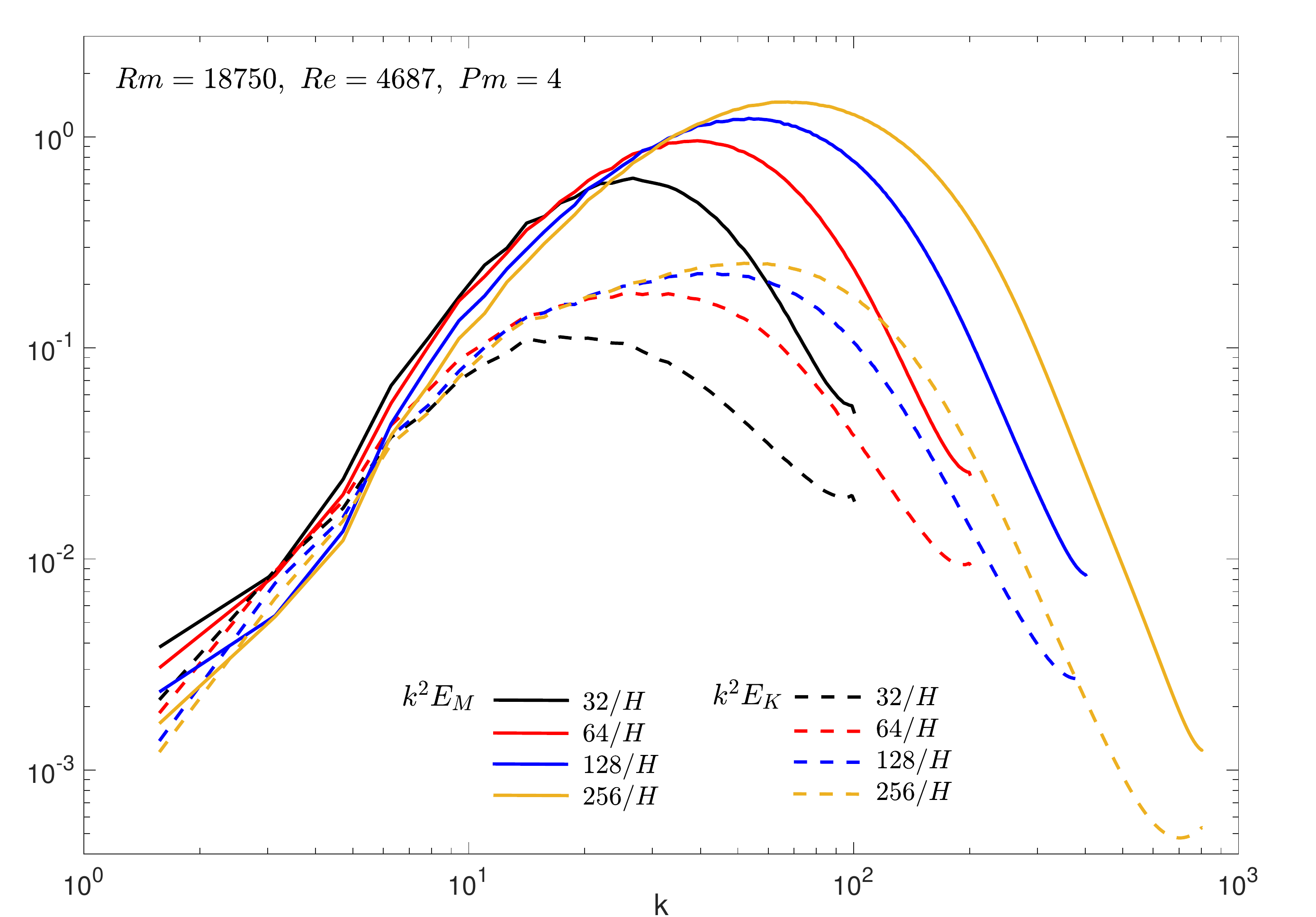}
\caption{Shell-averaged dissipation spectra of magnetic (solid lines) and kinetic (dashed lines) energy from a resolution study at $\text{Pm}=4$, $\text{Rm}=18750$ and $\text{Re}=4687$. The resolution (cells per scale height) are denoted by the colors: $32/H$ (black), $64/H$ (red), $128/H$ (blue) and $256/H$ (gold).}
\label{FIGURE_DissipationSpectraResolutionStudy}
\end{figure}

\label{APPENDIX_ConvergenceStudy}
To check the influence of numerical dissipation on our results, we have repeated the run with $\text{Rm}=18750$, $\text{Pm}=4$ and $\text{Re}=4687$ at four different resolutions $32/H, 64/H, 128/H$, and $256/H$ in the cubic box $4H\times4H\times4H$. The time-series of the volume-averaged alpha parameter from each run is shown in Figure \ref{FIGURE_AlphaTimeSeriesComparisonResolutionStudy}. The stress in the high resolution $\text{Pm}=2$ run (green) gradually decays over the last 100 orbits of the simulation (except for a transient increase over the last 10 orbits), which is likely because $\text{Pm}=2$ lies at the margin of stability, as discussed at the end of Section \ref{INTRO_RoleOfExplicitDissipationCoeffs}. At $\text{Pm}=4$, on the other hand, the stress gradually increases in the highest three resolution runs over the last 100 orbits. Furthermore, the time-series at $\text{Pm}=4$ are in excellent agreement (except for the lowest resolution run with $32/H$): in particular the curves for $128/H$ (blue) and $256/H$ (gold) runs track one another very closely over the last 100 orbits.

To check resolution scale by scale, in Figure \ref{FIGURE_DissipationSpectraResolutionStudy} we also plot the shell-averaged dissipation spectra, $k^2E_K$ and $k^2E_M$, respectively, for kinetic and magnetic energies from the $\text{Pm}=4$, $\text{Rm}=18750$ runs. For the four resolutions $32/H, 64/H, 128/H$, and $256/H$, the wavenumbers corresponding to the maxima (peaks) of these spectra increase with resolution and are $k_{p,M}=26.7,~39.27,~53.41,~67.54$ for the magnetic dissipation, and $k_{p,K}=17.28,~29.84,~42.41,~54.98$ for the kinetic dissipation. At a given resolution, these spectra coincide with those for other resolutions at $k\leq k_{p}$, where corresponding $k_p$ is to be taken for each resolution, implying  that these wavenumbers are increasingly resolved as the peaks move up with increasing resolution in Figure \ref{FIGURE_DissipationSpectraResolutionStudy}. The dissipation spectra differ at larger wavenumbers $k>k_p$, on the other hand, which are thus unresolved, but which tend to converge as the resolution is increased. Thus the wavenumber of the peak $k_p$, which separates resolved wavenumbers (to its left) and unresolved ones (to its right), represents the largest wavenumber (i.e. the smallest wavelength) which is still quite well resolved. The number of grid cells at this wavelength is $2\pi/(k_{p,M}\Delta x) \approx 7, 10, 15, 24$ for the magnetic and $2\pi/(k_{p,K}\Delta x)\approx 11, 13, 19, 29$ for the kinetic dissipation spectra, at resolutions $32/H, 64/H, 128/H, 256/H$, respectively, where $\Delta x=L_x/N_x$ is the cell size. In this sense, one can interpret $k_p$ as the characteristic cut-off wavenumber of numerical viscous and resistive dissipation, which naturally increases with increasing resolution and corresponding decreasing numerical dissipation. 

We can also estimate physical viscous and dissipation scales of MRI-turbulence. The viscous scale $l_{\nu}=2\pi/k_{\nu}=2\pi/\sqrt{\rm Re}$ was introduced in the text, and the resistive scale can be computed as $l_{\eta}=l_{\nu}/\sqrt{\rm Pm}$ at ${\rm Pm}>1$ \citep[e.g.,][]{fromang2010,rincon2019}. At $\text{\rm Pm}=4$ and $\text{Re}=4687$, $l_{\nu}=0.092H$ and $l_{\eta}=0.046H$. The cut-off length, $2\pi/k_{p}$, of numerical dissipation approaches these physical dissipation lengths at higher resolutions $128/H$ and $256/H$ for which $(2\pi/k_{p,K}, 2\pi/k_{p,M})=(0.15, 0.12)H$ and $(0.11, 0.09)H$, respectively. In particular, our fiducial simulation with $128/H$ resolves these dissipation scales with the number of cells per viscous and resistive wavelength being $l_{\nu}/\Delta x\approx 12$ and $l_{\eta}/\Delta x\approx 6$, respectively. The latter lower resolution for the magnetic dissipation scale is not really a problem, because all the key dynamical processes occur at wavenumbers $k < k_{p,M}=53.41$ (for $128/H$) and hence are well resolved. The resolution improves even more in the new and important plateau regime (Figure 1), where $\text{Re}$ is smaller, $\text{Pm}$ is larger and the energy spectra shift to lower wavenumbers. For example, in this regime at $\text{Rm}=18750$ and $\text{Pm}=64$, we find $l_{\nu}/\Delta x\approx 47$ and $l_{\eta}/\Delta x\approx 6$. This together with the fact that the key dynamics shift to lower $k$ (larger scales) as ${\rm Pm}$ is increased (by increasing the explicit viscosity) give us confidence that our fiducial resolution of $128/H$ captures the different linear and non-linear dynamical processes over an extensive range of wavenumbers, from the smallest, corresponding to the largest system length-scale, up to the highest ones comparable to the shortest resistive scale.

\section{Tables of Simulations}
\label{APPENDIX_TablesOfSimulations}

\begin{table*}
\centering
\caption{Unstratified isothermal shearing box MHD simulations at various magnetic Prandtl numbers $\text{Pm}$. All simulations were carried out in a box of size $4H\times4H\times4H$, a resolution (`Res') of $512^3$ (128 cells per scale height), and with a shear parameter of $q=1.5$ (Keplerian shear). Here $\text{Rm}$ denotes the magnetic Reynolds number, Pm the magnetic Prandtl number, and $\text{Re} \equiv \text{Rm}/\text{Pm}$ the Reynolds number (Re has been rounded down to the nearest whole number in the table). Other columns $\langle\langle E_\text{mag}\rangle\rangle$,$\langle\langle\alpha\rangle\rangle$,$\langle\langle M_{xy} \rangle\rangle$, $\langle\langle R_{xy} \rangle \rangle$, and $R$ denote the time- and volume-averaged magnetic energy density, alpha, Maxwell stress, and Reynolds stress, and the ratio of the two, respectively. All simulations were run for 200 orbits ($1257\,\Omega^{-1}$). Time-averages were taken from orbit 100 to orbit 200.}
\label{TABLE_LargeBoxSimulations}
	\begin{tabular}{lccccccccr}
		\hline
		Run	& Res & Rm & Re & Pm &$\langle\langle E_\text{mag}\rangle\rangle$ &$\langle\langle\alpha\rangle\rangle$&$\langle\langle M_{xy} \rangle\rangle$&$\langle\langle R_{xy} \rangle \rangle$ &$R$ \\ 
		\hline
		USTRMRIPm2Res128Re6250    & $512^3$ & $12500$ & $6250$ & $2$ & $0.024654$ & $0.013325$ & $0.010983$ & $0.002342$ & $4.689934$\\
		USTRMRIPm4Res128Re3125    & $512^3$ & $12500$ & $3125$ & $4$ & $0.035726$ & $0.019114$ & $0.015991$ & $0.003124$ & $5.119071$\\
		USTRMRIPm8Res128Re1562    & $512^3$ & $12500$ & $1562$ & $8$ & $0.083441$ & $0.043599$ & $0.036941$ & $0.006658$ & $5.548418$\\
		USTRMRIPm16Res128Re781    & $512^3$ & $12500$ & $781$ & $16$ & $0.116744$ & $0.059179$ & $0.050808$ & $0.008371$ & $6.069633$\\
		USTRMRIPm32Res128Re390    & $512^3$ & $12500$ & $390$ & $32$ & $0.183800$ & $0.087765$ & $0.076170$ & $0.011595$ & $6.569327$\\
		USTRMRIPm64Res128Re195 & $512^3$ & $12500$ & $195$ & $64$ & 0.237740 & 0.104540 & 0.092387 & 0.012153 & 7.601919\\
		\hline
		USTRMRIPm2Res128Re9375   & $512^3$ & $18750$ & $9375$ & $2$ & $0.030083$ & $0.016308$ & $0.013444$ & $0.002864$ & $4.693883$\\
		USTRMRIPm4Res128Re4687    & $512^3$ & $18750$ & $4687$ & $4$ & $0.038566$ & $0.020708$ & $0.017263$ & $0.003445$ & $5.011101$\\
		USTRMRIPm8Res128Re2344    & $512^3$ & $18750$ & $2344$ & $8$ & $0.064677$ & $0.034079$ & $0.028753$ & $0.005325$ & $5.399456$\\
		USTRMRIPm16Res128Re1172    & $512^3$ & $18750$ & $1172$ & $16$ & $0.108701$ & $0.055624$ & $0.047489$ & $0.008135$ & $5.837607$\\
		USTRMRIPm32Res128Re586    & $512^3$ & $18750$ & $586$ & $32$ & $0.147352$ & $0.072422$ & $0.062683$ & $0.009739$ & $6.436292$\\
		USTRMRIPm64Res128Re293    & $512^3$ & $18750$ & $293$ & $64$ & $0.213264$ & $0.098103$ & $0.085934$ & $0.012170$ & $7.061322$\\
		USTRMRIPm90Res128Re208    & $512^3$ & $18750$ & $208$ & $90$ & $0.242010$ & $0.106918$ & $0.094469$ & $0012449$ & $7.588394$\\
		\hline
		USTRMRIPm2Res128Re12500   & $512^3$ & $25000$ & $12500$ & $2$ & $0.024189$ & $0.013232$ & $0.010882$ & $0.002350$ & $4.631550$\\
		USTRMRIPm4Res128Re6250    & $512^3$ & $25000$ & $6250$ & $4$ & $0.035615$ & $0.019237$ & $0.016000$ & $0.003236$ & $4.943757$\\
		USTRMRIPm8Res128Re3125    & $512^3$ & $25000$ & $3125$ & $8$ & $0.059672$ & $0.031634$ & $0.026643$ & $0.004991$ & $5.337965$\\
		USTRMRIPm16Res128Re1562    & $512^3$ & $25000$ & $1562$ & $16$ & 0.100257 & $0.051604$ & $0.043942$ & $0.007662$ & $5.735189$\\
		USTRMRIPm32Res128Re781    & $512^3$ & $25000$ & $781$ & $32$ & $0.141098$ & $0.070569$ & $0.060798$ & $0.009770$ & $6.222866$\\
		USTRMRIPm64Res128Re390    & $512^3$ & $25000$ & $390$ & $64$ & $0.197615$ & $0.093552$ & $0.081633$ & $0.011919$ & $6.849049$\\
		USTRMRIPm90Res128Re277    & $512^3$ & $25000$ & $277$ & $90$ & $0.208069$ & $0.095177$ & $0.083744$ & $0.011432$ & $7.325385$\\
		USTRMRIPm128Res128Re195   & $512^3$ & $25000$ & $195$ & $128$ & $0.243376$ & $0.106686$ & $0.094480$ & $0.012205$ & $7.740826$\\
		\hline	
		USTRMRIPm2Res128Re25000   & $512^3$ & $50000$ & $25000$ & $2$ & $0.027289$ & $0.014875$ & $0.012244$ & $0.002631$ & $4.653349$\\
		USTRMRIPm4Res128Re12500   & $512^3$ & $50000$ & $12500$ & $4$ & $0.025290$ & $0.013836$ & $0.011443$ & $0.002392$ & $4.783425$\\
		USTRMRIPm8Res128Re6250    & $512^3$ & $50000$ & $6250$ & $8$ & $0.046855$ & $0.024965$ & $0.020899$ & $0.004066$ & $5.139953$\\
		USTRMRIPm16Res128Re3125   & $512^3$ & $50000$ & $3125$ & $16$ & $0.067468$ & $0.035377$ & $0.029914$ & $0.005463$ & $5.476027$\\
		USTRMRIPm32Res128Re1562   & $512^3$ & $50000$ & $1562$ & $32$ & $0.100496$ & $0.051498$ & $0.044019$ & $0.007479$ & $5.885534$\\
		USTRMRIPm64Res128Re781    & $512^3$ & $50000$ & $781$ & $64$ & $0.143874$ & $0.071228$ & $0.061569$ & $0.009659$ & $6.374429$\\
		USTRMRIPm128Res128Re390   & $512^3$ & $50000$ & $390$ & $128$ & $0.202365$ & $0.095001$ & $0.083140$ & $0.011861$ & $7.009762$\\
		USTRMRIPm160Res128Re312   & $512^3$ & $50000$ & $312$ & $160$ & $0.230167$ & $0.104682$ & $0.091870$ & $0.012812$ & $7.170490$\\
		USTRMRIPm180Res128Re312   & $512^3$ & $50000$ & $277$ & $180$ & $0.217971$ & $0.099191$ & $0.087409$ & $0.011782$ & $7.418628$\\
		USTRMRIPm210Res128Re238   & $512^3$ & $50000$ & $238$  & $210$ & $0.233012$ & $0.104444$ & $0.092318$ & $0.012126$ & $7.613376$\\
		USTRMRIPm256Res128Re312   & $512^3$ & $50000$ & $195$ & $256$ & $0.259255$ & $0.112683$ & $0.099976$ & $0.012707$ & $7.867968$\\
		\hline
	\end{tabular}
\end{table*}

\begin{table*}
\centering
\caption{Unstratified isothermal shearing box MHD simulations at various
magnetic Prandtl number $\text{Pm}$: dependence on vertical-to-radial box aspect ratio,  shear parameter $q$, and box size. The box size is given in shorthand notation by $[L_x,L_y,L_z]$ (in units of the scale height $H$). The resolution is 128 cells per $H$ in all simulations: $128\times512\times384$ for the tall boxes, and $128\times512\times128$ for the finger boxes. All other columns have the same meaning as in Table \ref{TABLE_LargeBoxSimulations}.}
\label{TABLE_TallBoxSimulations}
	\begin{tabular}{lcccccccccr}
		\hline
		Run	& Box Size & q & Rm & Re & Pm &$\langle\langle E_\text{mag}\rangle\rangle$ &$\langle\langle\alpha\rangle\rangle$&$\langle\langle M_{xy} \rangle\rangle$&$\langle\langle R_{xy} \rangle \rangle$ &$R$ \\ 
		\hline
		USTRMRIPm2Res128Re12500H3  & [1,3,3] & 1.5 & $25000$ & $12500$ & $2$ & $0.066374$ & $0.030970$ & $0.025836$ & $0.005135$ & $5.031343$\\
		USTRMRIPm4Res128Re6250H3   & [1,3,3] & 1.5 & $25000$ & $6250$ & $4$ & $0.073655$ & $0.034560$ & $0.029076$ & $0.005485$ & $5.300971$\\
		USTRMRIPm8Res128Re3125H3   & [1,3,3] & 1.5 & $25000$ & $3125$ & $8$ & $0.110337$ & $0.047218$ & $0.039981$ & $0.007237$ & $5.524667$\\
		USTRMRIPm16Res128Re1562H3  & [1,3,3] & 1.5 & $25000$ & $1562$ & $16$ & $0.110520$ & $0.049407$ & $0.042373$ & $0.007034$ & $6.024036$\\
		\hline
		USTRMRIPm8Res128Re3125H3q  & [1,3,3] & 0.8 & $25000$ & $3125$ & $8$ & $0.007490$ & $0.002882$ & $0.002735$ & $0.000148$ & $18.53935$\\
		USTRMRIPm16Res128Re1562H3q & [1,3,3] & 0.8 & $25000$ & $1562$ & $16$ & $0.022772$ & $0.009580$ & $0.000500$ & $0.009580$ & $18.1508$\\
		USTRMRIPm32Res128Re781H3q & [1,3,3] & 0.8 & $25000$ & $781$ & $32$ & $0.053625$ & $0.020919$ & $0.019899$ & $0.001020$ & $19.503944$\\
		\hline
		USTRMRIPm2Res128Re6250H & [1,4,1] & 1.5 & 12500 & 6250 & 2 & decaying\\
		USTRMRIPm4Res128Re3125H & [1,4,1] & 1.5 & 12500 & 3125 & 4 & 0.022380 & 0.012129 & 0.009822 & 0.002306 & 4.259117 \\
		USTRMRIPm8Res128Re1562H & [1,4,1] & 1.5 & 12500 & 1562 & 8 & 0.046309 & 0.024789 & 0.020411 & 0.004378 & 4.662352\\
		USTRMRIPm16Res128Re781H & [1,4,1] & 1.5 & 12500 & 781 & 16 & 0.077530 & 0.038880 & 0.032815 & 0.006065 & 5.410612\\
		\hline
	\end{tabular}
\end{table*}


\bsp	
\label{lastpage}
\end{document}